\documentclass[rmp,aps,nofootinbib,endfloats]{revtex4}

\usepackage{graphics}
\usepackage{epsfig}
\usepackage{amsmath}
\usepackage{color}

\def\inbar{\,\vrule height1.5ex width.4pt depth0pt}
\def\IR{\relax{\rm I\kern-.18em R}}
\def\IC{\relax\hbox{$\inbar\kern-.3em{\rm C}$}}
\newcommand{\be}{\begin{equation}}
\newcommand{\ee}{\end{equation}}
\newcommand{\bea}{\begin{eqnarray}}
\newcommand{\eea}{\end{eqnarray}}

\newcommand{\caloa}{{\cal{O}}(\alpha)}
\newcommand{\caloasq}{{\cal{O}}(\alpha^2)}
\newcommand{\caloaem}{{\cal{O}}(\alpha E/M)}

\newcommand{\ms}{\overline{\mbox{{\small MS}}}}

\newcommand{\sef}{s^2_{\mbox{{\tiny eff}}}}
\newcommand{\cef}{c^2_{\mbox{{\tiny eff}}}}
\newcommand{\Dref}{\Delta r_{\mbox{{\tiny eff}}}}
\newcommand{\sefl}{\sin^2 \theta_{\mbox{{\tiny eff}}}^{\mbox{{\tiny lept}}}}
\newcommand{\Li}{{\rm Li}}

\newcommand{\noi}{\noindent}
\newcommand{\lsim}{\mathrel{\lower4pt\hbox{$\sim$}}
\hskip-12.5pt\raise1.6pt\hbox{$<$}\;}
\newcommand{\gsim}{\mathrel{\lower4pt\hbox{$\sim$}}
\hskip-11.5pt\raise1.6pt\hbox{$>$}\;}

\newcounter{saveeqn}

\begin{document}
\title{Radiative Corrections in  Precision Electroweak Physics: a Historical Perspective}

%\author{William J. Marciano}
%\email{marciano@bnl.gov}
%\affiliation{Physics Department, Brookhaven National Laboratory, Upton, NY 11973}
\author{Alberto Sirlin}
\email{alberto.sirlin@nyu.edu}
\affiliation{Department of Physics, New York University,  New York, NY 10003}
\author{Andrea Ferroglia}
\email{aferroglia@citytech.cuny.edu}
\affiliation{New York City College of Technology, Brooklyn, NY 11201}

\begin{abstract}  
The aim of this article is to review the very important role played by radiative corrections in precision electroweak physics,  in the framework of both the Fermi Theory of Weak Interactions and the Standard Theory of Particle Physics. Important theoretical developments, closely connected with the study and applications of the radiative corrections, are also reviewed. The role of radiative corrections in the analysis of some important signals of new physics is also discussed.
\end{abstract}

%\date
\maketitle

\newpage
\tableofcontents

\newpage

\section{INTRODUCTION}
\label{sec:intro}
The aim of this article is to review, from a historical perspective, the very important role played by radiative corrections in precision electroweak physics, in the framework of both the original Fermi Theory of Weak Interactions and  in the renormalizable Standard Theory of Particle Physics, usually referred to as the Standard Model. Those two areas are discussed in Sections~\ref{sec:Fermi} and \ref{sec:SM}, respectively.

Studies of such corrections are closely connected with important developments in theoretical particle physics, which are also reviewed.
The role of radiative corrections in the analysis of some important signals of new physics is also discussed.

As shown in the Contents,  six Subsections are based on the Fermi Theory of Weak Interactions and twenty one Subsections are based on the Standard Theory of Particle Physics. They review important and interesting subjects in Electroweak Physics. On the other hand, in view of the magnitude of the area, encompassing more than fifty years of physics, it was found not possible to cover every conceivable subject. Taking this into account, the authors apologize beforehand for the omission of important and interesting developments that lie beyond the scope of this article.

There are a number of excellent reviews of Gauge Theories in general and the Standard Theory of Particle Physics in particular.
Among them: \textcite{Abers:1973R, Taylor:1976R, Faddeev:1980R, Cheng:1984R, Pokorski:1987R, Bailin:1993R, Quigg:1983R, Aitchison:1989R, Gunion:2000R, Einhorn:1991R, Donoghue:1980R, Bardin:1999R, Boehm:2001R, Paschos:2007R, Beg:1974R, Beg:1982R, Sirlin:1993bu, Merritt:1995md, Sirlin:2000R, Sirlin:2003R,  Alexander:1988mw, Ellis:1986jba, Altarelli:1989R, Bardin:1995R, Aoki:1982R,Langacker:2010zza, Langacker:1995R, Jegerlehner:2008R, Jegerlehner:1991R, Hollik:1993R, Weinberg:1974R}.

\section{RADIATIVE CORRECTIONS IN THE FERMI THEORY OF WEAK INTERACTIONS}
\label{sec:Fermi}

The powerful and highly successful relativistic methods developed by Feynman, Schwinger, Tomonaga, Dyson and others to evaluate the radiative corrections in Quantum Electrodynamics\footnote{ See, for example, \textcite{SchwingerQED}, \textcite{FeynmanQED},  \textcite{Kinoshita:1990zz}, \textcite{Schweber:1994xx}.}, were first applied to the Weak Interactions in the mid-fifties. In particular, \textcite{Behrends:1955mb} studied the $\caloa$ electromagnetic corrections to muon decay in the framework of the four-fermion Fermi Theory of Weak Interactions.

We recall that in this theory the interaction Lagrangian density for muon decay is given by
\be
{\cal L} = -g_i [\bar\psi_{e} \Gamma^i\psi_\mu] [ \bar\psi_{\nu_\mu}\Gamma_i\psi_{\nu_e}] + {\rm h. c.}\, , \label{eqone}
\ee
where $i$ runs over the scalar $(S)$, vector $(V)$, tensor $(T)$, axial vector $(A)$ and pseudoscalar $(P)$ interactions.
Explicitly, we have\footnote{In this paper we use the notational conventions and $\gamma$ matrices of  \textcite{bjor}. We also use ``natural units'' $\hbar=c=1$. In  Eq.(\ref{eqone}) it is understood that the contravariant and covariant indices are contracted  and summed from $0$ to $3$ as in $[\gamma^\mu] [\gamma_\mu]$, $[\sigma^{\mu \nu}] [\sigma_{\mu \nu}]$, etc \ldots}:
\be
\Gamma^S = 1\, , \quad \left(\Gamma^V\right)^\mu=\gamma^\mu\, , \quad \left(\Gamma^T\right)^{\mu \nu}=\frac{\sigma^{\mu\nu}}{\sqrt{2}} = \frac{i}{2 \sqrt{2}} \left( \gamma^\mu \gamma^\nu - \gamma^\nu \gamma^\mu\right)\, , \quad \left(\Gamma^A\right)^\mu = i\gamma^\mu\gamma_5\, , \quad \Gamma^P = i \gamma_5\, . \label{eqtwo}
\ee
 Eq.(\ref{eqone}) is the interaction Lagrangian density in the charge-retention order in which leptons of equal charge are placed in the same covariant.
${\cal L}$ can be written also in the  charge-exchange order
\be 
{\cal L} = -\tilde{g_i} [\bar\psi_{\nu_\mu} \Gamma^i \psi_\mu] [\bar\psi_e \Gamma_i\psi_{\nu_e}] + {\rm h. c.} \, ,\label{eqthree}
\ee
 where the $\tilde{g_i}$ are related to $g_i$ by Fierz transformations (Fierz, 1937).

While Eq.(\ref{eqone}) is very convenient for actual calculations in the Fermi theory, Eq.(\ref{eqthree}) conforms more closely with current formulations in which $\mu$ decay arises from charged current interactions.

The $\caloa$ radiative corrections to muon decay in the Fermi theory arise from the interchange of a virtual photon between the $\mu$ and the $e$, the electromagnetic field renormalizations of these particles, and the inner bremsstrahlung contributions.

An important result   is that in the charge-retention order of Eq.(\ref{eqone}), the $\caloa$ corrections to muon decay are ultraviolet (UV) convergent only for the vector and axial vector interactions \cite{Behrends:1955mb}. This can be readily understood by analogy with QED\null. It is well known that in the scattering of an electron by an external potential, the UV divergence of the vertex part cancels against those in the wave function renormalizations of the external legs (by the Ward Identity). For the vector coupling in muon decay in the charge retention order, we have an analogous situation, except for the fact that the muon and electron have different masses. However, as the coefficients of these divergences are independent of the fermion masses, they also cancel in muon decay. The corrections involving the axial vector coupling in the charge-retention order can be obtained from those in the vector case by means of the formal transformation $\psi_e\to \psi^\prime_e = \gamma_5\psi_e$, $m_e\to -m_e$ in the Lagrangian density. Thus, they only differ from the vector case by the change $m_e\to -m_e$ and, consequently, the UV divergences cancel also for the axial vector coupling. In contrast, for the $S$, $T$, and $P$ interactions of the charge-retention order, the analogy with QED is no longer valid and the $\caloa$ corrections are logarithmically ultraviolet divergent.

\subsection{Non-Conservation of Parity. The Two-Component Theory of the Neutrino}
\label{sec:2.1}

In 1956 Lee and Yang proposed the revolutionary idea that parity is not conserved in the weak interactions 
\cite{Lee:1956qn} and this was soon verified by elegant experiments. In order to accommodate parity non-conservation, Eq.(\ref{eqone}) was generalized to 
\be
{\cal L} = -[\bar\psi_e\Gamma^i\psi_\mu] [\bar\psi_{\nu_\mu} \Gamma_i (g_i + g^\prime_i \gamma_5)\psi_{\nu_e}] + {\rm h. c.} \, , \label{eqfour}
\ee
 with an analogous modification of Eq.(\ref{eqthree}).

To lowest order, Eq.(\ref{eqfour}) leads to the following expression for the energy-angle distribution of $e^-(e^+)$ from the decay of a polarized $\mu^-(\mu^+)$ at rest:
\begin{align}
dN(x,\theta) &=   \frac{d^3p}{(2\pi)^4} \frac{m_\mu E_0A}{6} \left\{3(1-x) +2\rho\left[\frac{4}{3}x -1-\frac{1}{3} \frac{m^2_e}{E^2_0x} \right]\right. +3\eta \frac{m_e}{E_0} \frac{(1-x)}{x} 
\nonumber \\
&\mp P\beta\xi\cos\theta \left.\left[\!1-x+2\delta \left(\frac{4x}{3}-1-\frac{1}{3} \frac{m^2_e}{m_\mu E_0}\right)\right]\right\}\, , \label{eqfive}
\end{align}
where the upper and lower signs refer to $\mu^-$ and $\mu^+$, respectively, $\theta$ is the angle between the $e^\mp$ momentum and the spin direction of the $\mu^\mp$; $x=E/E_0$ where $E$ is the $e^\mp$ energy and $E_0=(m^2_\mu+m^2_e)/2m_\mu$ its maximum value; $p$ is the $e^\mp$ momentum, $\beta=p/E$ and $P$ the degree of polarization of $\mu^\mp$. The parameter $\rho$ that describes the energy distribution of $e^\mp$ from unpolarized muons was introduced long ago by \textcite{Michel:1950qe} and is generally referred to as the Michel parameter. The parameters $\xi$ and $\delta$, which are currently employed to describe the effects of parity non-conservation, were introduced by \textcite{Kinoshita:1957zz, Kinoshita:1957za}. Alternative expressions to Eq.(\ref{eqfive}), using different parametrizations, were obtained by \textcite{Bouchiat:1957zz} and \textcite{Larsen:1957zz}.  Since $E_0>> m_e$, the terms proportional to $m^2_e$ in the cofactors of $\rho$ and $\delta$ are very small. For the same reason, the term proportional to $\eta$ is potentially significant only in the very low energy part of the spectrum. 
For a more detailed discussion of theoretical and experimental aspects of muon decay, and the relations between the parameters $A$, $\rho$, $\eta$, $\xi$, $\delta$ and the couplings $g_i$ and $g_i'$ see for example\footnote{In several early papers, including \textcite{Kinoshita:1957zz, Kinoshita:1957za} and \textcite{Sachs:1975R}, $\gamma_5$ was defined with a sign opposite to the one employed in the present article.} \textcite{Kinoshita:1957zz, Kinoshita:1957za}; \textcite{Berman:1962xx}; \textcite{Sachs:1975R};  \textcite{SirlinT}; \textcite{FetscherPDG}.

In 1957, \textcite{Lee:1957qr}, \textcite{Salam:1957st} and \textcite{Landau:1957tp} re-introduced the two-component theory of the neutrino, an elegant formulation that had been long abandoned because it leads to parity non-conservation! This theory can be regarded as a special case of the four-component theory of a massless neutrino, subject to the subsidiary condition
%
%%\alpheqn
%
\be
 a_-\psi_\nu = \psi_\nu \, , \label{eqsixa}  
\ee
 or
\be
a_+\psi_\nu = \psi_\nu \, , \label{eqsixb}
\ee
%
%%\reseteqn
%\setcounter{equation}{6}
%
 where
\be
a_\mp = \frac{1\mp\gamma_5}{2} \, ,\label{eqseven}
\ee
are the left and right chiral projectors.

If Eq.(\ref{eqsixa}) is satisfied, the massless neutrino has helicity $h=-1$ and the corresponding antineutrino has $h=1$. If Eq.(\ref{eqsixb}) is satisfied, the signs are reversed. From measurements of the polarization and angular distribution of high energy positrons in $\mu^+$ decays, it was concluded that $\bar \nu_e$ and $\nu_\mu$ have opposite helicities. Moreover, the helicity of $\bar\nu_e$ in $\beta$ decay was found to be positive. These observations led to the conclusion that both $\bar\nu_e$ and $\bar\nu_\mu$ have $h=+1$, correspondingly $\nu_e$ and $\nu_\mu$ have $h=-1$, and therefore 
Eq.(\ref{eqsixa}) holds.

Comparing Eq.(\ref{eqsixa}) with the Lagrangian density in Eq.(\ref{eqfour}) one readily finds
%\alpheqn
\bea
g_S  &=& g^\prime_S = g_T = g^\prime_T = g_P = g^\prime_{P} =0 \, , \label{eqeighta} \\
g_i^\prime &\equiv& -g_i \qquad (i=V,A). \label{eqeightb}
\eea
%\reseteqn
%\setcounter{equation}{8}
Namely, in the two-component neutrino theory only the vector and axial vector couplings of the charge-retention order survive, precisely the interactions for which the $\caloa$ radiative corrections had been previously found to be convergent \cite{Behrends:1955mb}.

Comparison of Eqs.(\ref{eqeighta},\ref{eqeightb}) with the general expressions relating $\rho$, $\delta$, $\eta$ and $\xi$ to the coupling constants, further leads to the important conclusions:
%
%\alpheqn
%
\begin{align}
\rho & =  \delta = \frac{3}{4} \, ,\label{eqninea} \ \\
\xi & =  -\frac{g_Vg^\ast_A + g_Ag^\ast_V} { |g_V|^2+|g_A|^2}  \, ,\label{eqnineb} \\
\eta & =  \frac{1}{2} \left[\frac{|g_A|^2 - |g_V|^2}{|g_A|^2 + |g_V|^2} \right]. \label{eqninec}
\end{align}
%
%\reseteqn
%
Thus, in the two-component theory of the neutrino the parameters $\rho$ and $\delta$ are sharply predicted, while $\xi$ and $\eta$ depend only on $g_V$ and $g_A$!

\subsection{Radiative Corrections to Muon Decay in the Two-Component Theory of the Neutrino: Cancellation of Mass Singularities in Integrated Observables}

In comparing theory with experiment in muon decay, it is important to evaluate the $\caloa$ corrections since they play a very significant role. Including those corrections in the framework of the two-component theory of the neutrino, one obtains the following expression for the energy-angle distributions of $e^-(e^+)$ in the decay of a polarized $\mu^-(\mu^+)$ at rest \cite{Kinoshita:1958ru}:
%
%\alpheqn
\bea 
dN(x,\theta) &=& \frac{d^3p}{(2\pi)^4} \frac{m_\mu E_0}{3} 2(|g_V|^2
+|g_A|^2) \left\{3-2x-\frac{m^2_e}{E^2_0x} + \frac{6\eta m_e}{E_0} \frac{(1-x)}{x} \right. \nonumber \\
& &+\frac{\alpha}{2\pi} f(x) \pm P\beta\xi\cos\theta \left.\left[ 1-2x+\frac{m^2_e}{m_\mu E_0} + \frac{\alpha}{2\pi} g(x) \right] \right\},   \label{eqtena} 
\eea
\noi where
\begin{align}
f(x) &= (6-4x) R(x) + 6(1-x) \ln x + \frac{(1-x)}{3x^2}  [(5+17x-34x^2) (\omega +\ln x) - 22x + 34x^2] \, , \label{eqtenb}  \\
g(x) & =  (2-4x)R(x) + (2-6x) \ln x - \frac{1-x}{3x^2}  \Biggl[(1+x+34x^2)(\omega +\ln x) + 3-7x-32x^2 
%\nonumber \\&  
+4(1-x)^2 \frac{\ln(1-x)}{x}\Biggr], \label{eqtenc} \\
R(x) & =  2 \Li_2(x) - \frac{\pi^2}{3} - 2 +\omega \left[ \frac{3}{2} + 2\ln \left(\frac{1-x}{x}\right)\right]  -(2\ln x -1) \ln x+ \left(3\ln x-1-\frac{1}{x}\right) \ln (1-x) \, ,  \label{eqtend} \\
\omega & =  \ln\left(\frac{m_\mu}{m_e}\right)\, , \label{eqtene} 
\end{align}
%\reseteqn
%
and
\be
\Li_2(x) = -\int^x_0 dt \frac{\ln (1-t)}{t} \label{eq:2_19}
\ee
is  the dilogarithm function\footnote{See, for example, \textcite{lewin}.}.
 In Eqs.(\ref{eqtenb},\ref{eqtenc}), we have neglected terms of ${\cal{O}}(\alpha m_e/E)$, although all the contributions of $\caloa$ not proportional to $\cos\theta$ have been evaluated exactly \cite{Behrends:1955mb, Grotch:1968zz, Nir:1989xxx}. The terms $m^2_e/E^2_0 x$ and $m^2_e/m_\mu E_0$ in Eq.(\ref{eqtena}) are very small and frequently omitted in the literature.

Integrating Eq.(\ref{eqtena}) over all values of $p$ and $\theta$, one obtains the expression for the muon lifetime $\tau_\mu$, including ${\cal O}(\alpha)$ corrections
\be
\frac{1}{\tau_\mu} = \frac{(|g_V|^2 + |g_A|^2 ) m^5_\mu}{192\pi^3} \left[ 1-\frac{8m^2_e}{m^2_\mu} + 4\eta\frac{m_e}{m_\mu} \right]\left[1+\frac{\alpha}{2\pi} \left(\frac{25}{4} - \pi^2\right)\right], \label{eqeleven}
\ee
 where we have neglected terms of order $(m_e/m_\mu)^4$, $\eta (m_e/m_\mu)^3$, and $\alpha m_e/m_\mu$. 

The $\caloa$ radiative corrections have a large effect on the $e^\mp$ spectrum in $\mu$ decay. In fact, they decrease the decay probability for large $x$ and increase it for small $x$. In order to estimate the magnitude of this effect, it has been pointed out that if the $e^\mp$ spectrum in Eq.(\ref{eqtena}) is fitted with an effective uncorrected formula of the Michel type (cf. Eq.(\ref{eqfive})) over the range $0.3\lsim x\lsim 0.95$, the parameter $\rho_{\rm eff}$ obtained in this manner is $\rho_{\rm eff} \approx 0.71$ rather than the value 3/4 of the two-component theory \cite{Kinoshita:1958ru}. Similar observations hold for the parameter $\delta$ that governs the $x$ dependence of the $\cos\theta$ part of the decay probability. Since current determinations of $\rho$ and $\delta$ agree with the predictions of Eq.(\ref{eqninea}) at the $0.035~\%$ and $0.046\%$ levels, respectively \cite{Bayes:2011T}, it is clear that the radiative corrections play a crucial role in verifying the validity of the two-component theory of the neutrino.

On the other hand, the $\caloa$ corrections to the muon lifetime given in Eq.(\ref{eqeleven}) amount to only $-4.2\times10^{-3}$. The reason why the corrections to the electron spectrum are quite large while the corrections to $\tau_\mu$ are rather small has been traced to the cancellation of ``mass singularities'' in integrated observables, discovered in \cite{Kinoshita:1958ru}. In the case of muon decay, it implies that the corrections to the lifetime and the integrated asymmetry are finite in the mathematical limit $m_e\to0$. The properties discussed above can be nicely illustrated by considering the terms proportional to the large parameter $\omega = \ln(m_\mu/m_e) \approx 5.332$ in the corrections to the spectrum (cf. Eqs.(\ref{eqtena}, \ref{eqtenb}, \ref{eqtend})). They are proportional to
\be 
\frac{\alpha}{2\pi} \omega dx\left\{ (6-4x) x^2 \left[\frac{3}{2} + 2\ln \left(\frac{1-x}{x}\right)\right] + \frac{(1-x)}{3} [5+17x-34x^2]\right\} \, , \label{eqtwelve}
\ee
 and contain the electron ``mass singularity'' since $\omega$ diverges in the $m_e\to0$ limit. When integrated over the full spectrum, {\it i.e.}\ in the range $1 \ge x \ge0$,  Eq.(\ref{eqtwelve}) vanishes, leading to the cancellation mentioned above. Furthermore, the expression between curly brackets is negative in the upper part of the spectrum $(x\gsim0.68)$ and positive for $x\lsim0.68$. Using Eq.(\ref{eqtenc}), one readily verifies that the terms proportional to $(\alpha/2\pi)\omega$ in the $\cos\theta$ term of Eq.(\ref{eqtena}) also cancel when integrated over the full range $1\ge x\ge 0$. The cancellation of mass singularities has been also verified in the $\caloa$ contributions to $1/\tau_\mu$ proportional to $g^2_S$, $g^2_T$, and $g^2_P$ in the general Fermi theory, as well as in the corrections to the $\beta$-decay lifetime in the framework of the $V$-$A$ theory (Section~\ref{sec:2.5}). Furthermore, it has provided one of the main motivations for the Kinoshita-Lee-Nauenberg (KLN) Theorem \cite{Kinoshita:1962ur, Lee:1964is}.

An observable for which the $\caloa$ corrections become extremely large is the asymmetry of low energy $e^\mp$ (Kinoshita and Sirlin, 1957c). Their effect on the asymmetry parameter $\xi$ is also discussed in  \cite{Kinoshita:1958ru}. Another important result of the two-component neutrino theory was the prediction of the photon spectrum and rate in radiative muon decay $\mu\to e+\nu+\bar\nu+\gamma$ before its detection \cite{Kinoshita:1959bb}. As an example, for photons of energy $\ge20m_e$, the branching ratio was predicted to be 1.2\%.

As emphasized before, the two-component theory of the neutrino leads to the definite predictions $\rho=\delta=3/4$ (cf. Eq.(\ref{eqninea})). In order to measure with high precision these basic parameters (as well as $\xi$, $\eta$ and $A$) in the four-component neutrino framework of the general Fermi theory (cf. Eq.(\ref{eqfive})), one approach has been  to employ the fractional radiative corrections of the two-component neutrino theory which, as discussed before, are finite and well defined. Specifically \cite{Sherwood:1967zz}, the expression between curly brackets not involving $\cos\theta$ in Eq.(\ref{eqfive}), is multiplied by
$$ 
1+[(\alpha/2\pi)f(x)]/[3-2x-m^2_e/E^2_0x+6\eta (m_e/E_0)(1-x)/x] \, , 
$$
\noi while the expression proportional to $\cos\theta$ is multiplied by
$$
1+[(\alpha/2\pi)g(x)]/[1-2x+m^2_e/m_\mu E_0] \, . \nonumber
$$
\noi Comparison with Eq.(\ref{eqtena}) shows that these factors are indeed the corresponding fractional corrections in the two-component neutrino theory. The justification for this procedure is that, to a high degree of precision, the current experimental information is consistent with pure $V$, $A$, $V^\prime$ and $A^\prime$ interactions. Possible deviations, which in the four-component neutrino framework involve quadratic expressions in $g_i$, $g^\prime_i (i=S,T,P)$ are expected to be very small and can therefore be treated at the tree level. The products of these small deviations with $(\alpha/2\pi)f(x)$ and $(\alpha/2\pi)g(x)$ are of second order in the small quantities and, therefore, are not considered significant.

At present, very precise measurements of $\rho$, $\delta$, $\xi$ and $\eta$ are carried out in the TWIST experiment at TRIUMPH \cite{Bayes:2011T}, and a very accurate determination of $\tau_\mu$ has been made by the Mulan collaboration at PSI \cite{Webber:2010zf}.

\boldmath
\subsection{The $V$-$A$ Theory}
\label{sec:2.4}
\unboldmath

 The discovery of parity non-conservation led to another very important development: by greatly increasing the number of observables available for experimental and theoretical study, it opened the way for the determination of the basic phenomenological interaction. This led \textcite{Sudarshan:1957xx, Sudarshan:1958vf} and 
\textcite{Feynman:1958ty} to propose a universal $V$-$A$ Fermi Interaction for charged current processes, such as muon decay, $\beta$ decay and the semileptonic decays of hyperons.

In the case of muon decay, this theory implies the validity of Eqs.(\ref{eqeighta},\ref{eqeightb}) and furthermore states that 
\be
g_A = -g_V. \label{eqthirteen}
\ee
Using the Fierz transformations \cite{Fierz:1937xx}, Eqs.(\ref{eqeighta},\ref{eqeightb},\ref{eqthirteen}) lead to the following coupling constants $\tilde g_i$, $\tilde g^\prime_i$ in the charge-exchange order:
%
%\alpheqn
\bea
\tilde g_S & = & \tilde g^\prime_S = \tilde g_T = \tilde g^\prime_T = \tilde g_P = \tilde g^\prime_P = 0,  \label{eqfourteena} \\
\tilde g_V & = & -\tilde g_A = g_V = -\tilde g_V^\prime = \tilde g^\prime_A \, .  \label{eqfourteenb}
\eea
%\reseteqn
%
\noi Defining $G_\mu \equiv \sqrt{2} g_V$,  Eqs.(\ref{eqeighta},\ref{eqeightb},\ref{eqthirteen},\ref{eqfourteena},\ref{eqfourteenb}) lead to
%
%\alpheqn
%
\bea
{\cal{L}} & = & -\frac{G_\mu}{\sqrt{2}} [\bar\psi_{\nu_\mu} \gamma^\mu (1-\gamma_5) \psi_\mu][\bar\psi_e\gamma_\mu (1-\gamma_5)\psi_{\nu_e}] + {\rm h. c.} \, ,\label{eqfifteena} \\
& = & -\frac{G_\mu}{\sqrt{2}} [\bar\psi_e \gamma^\mu(1-\gamma_5)\psi_\mu] [ \bar\psi_{\nu_\mu} \gamma_\mu(1-\gamma_5)\psi_{\nu_e}] +  {\rm h. c.} \, .\label{eqfifteenb}
\eea
%
%\reseteqn
%
Thus, the interaction Lagrangian for muon decay in the $V$-$A$ theory has a very simple and elegant form that involves a single coupling constant and is preserved in passing from the charge-retention to the charge-exchange order. Eqs.(\ref{eqeighta},\ref{eqeightb},\ref{eqthirteen}) lead also to the sharp predictions:
%
%\alpheqn
\bea
\rho &=& \delta = 3/4 \,  ,  \label{eqsixteena} \\
\eta &=&  0 \, ,  \label{eqsixteenb} \\
\xi &=& 1 \, ,  \label{eqsixteenc} 
\eea
%\reseteqn
%
as can be readily verified by inserting Eq.(\ref{eqthirteen}) into Eqs.(\ref{eqnineb},\ref{eqninec}).

With the neglect of strong interaction effects, in the original version of the $V$-$A$ theory other four-fermion interaction processes were described by Lagrangian densities of the same form as Eq.(\ref{eqfifteena}). For example, for $n\to p+e^-+\bar\nu_e$, the basic process for $\beta$ decay, the Lagrangian density was postulated to be of the form.
\be
{\cal{L}}_{\beta{\rm -decay}}= -\frac{G_V}{\sqrt{2}} [\bar\psi_p\gamma^\mu(1-\gamma_5) \psi_n] [\bar\psi_e\gamma_\mu(1-\gamma_5)\psi_{\nu_e}]  + {\rm h. c.}\, , \label{eqseventeen}
\ee
where $G_V$ is the vector coupling constant in $\beta$-decay.

\boldmath
\subsection{Radiative Corrections to Muon Decay in the $V$-$A$ Theory and the Fermi Constant}
\label{sec:2.5}
\unboldmath

Taking into account Eqs.(\ref{eqthirteen},\ref{eqsixteena},\ref{eqsixteenb},\ref{eqsixteenc}), we see that in the $V$-$A$ theory, the energy-angle distributions of $e^-(e^+)$ in muon decay are simply obtained by setting $|g_A| = g_V = G_\mu/\sqrt{2}$, $\eta=0$, $\xi=1$ in the two-component theory expression (Eq.(\ref{eqtena})). In particular, the $\caloa$ corrections are still governed by the functions $f(x)$ and $g(x)$. Furthermore, using the transformation $\psi_e\to\psi^\prime_e=\gamma_5\psi_e$, $m_e\to-m_e$ discussed in Section~\ref{sec:2.1}, it can be shown that in the $V$-$A$ theory there are no contributions to the differential decay rate (Eq.(\ref{eqtena})) that involve odd powers of $m_e$ \cite{Roos:1971mj}. This implies that corrections of ${\cal{O}}((\alpha/\pi)m_e/m_\mu)$ are absent and that the leading mass-dependent corrections to the differential decay rate are of ${\cal{O}}((\alpha/\pi)m^2_e/m^2_\mu \ln(m^2_\mu/m^2_e))$. On the other hand, in the calculation of integrated observables such as the total decay rate, the integration over the electron or positron momentum does give rise to corrections of $\caloa$ proportional to $(m_e/m_\mu)^3$, as well as even powers of $m_e/m_\mu$ \cite{vanRitbergen:1998yd}.

Radiative corrections of ${\mathcal O}(\alpha^2)$ to the electron spectrum were evaluated by \textcite{Arbuzov:2002A,  Arbuzov:2002B, Arbuzov:2002C, Anastasiou:2005A}.

Recently, the TWIST collaboration \cite{Bayes:2011T} reported very accurate measurements of the parameters $\rho$, $\delta$ and  ${\mathcal P} ^\pi_\mu\xi$ in the four-component neutrino framework of the general Fermi theory (${\mathcal P}_\mu^\pi$ is the initial degree of polarization of the muon from $\pi$ decay):
\begin{align}
\rho &= 0.74977 \pm 0.00012~({\rm stat.}) \pm 0.00023~({\rm syst.}) \, ; \\
\delta &=0.75049 \pm  0.00021~({\rm stat.}) \pm 0.00027~({\rm syst.}) \, ; \\
{\mathcal P}_\mu^\pi \xi &= 1.00084 \pm 0.00029~({\rm stat.}) ^{+ 0.00165}_{-0.00063} ~({\rm syst.}) \, .
\end{align}
These results are in very good agreement with the predictions of the $V$-$A$ theory, Eqs.(\ref{eqsixteena}, \ref{eqsixteenc}) and ${\mathcal P}_\mu^\pi =1$, at a high level of precision. As mentioned before, the radiative corrections (RC) play a crucial role in the analysis. The authors also use these results to derive interesting bounds for the combinations $|(g_R/g_L) \zeta|$ and $(g_L/g_R) m_2$ in the generalized left-right symmetry model ($g_L$ and $g_R$ are the gauge couplings of $W_L$ and $W_R$, $\zeta$ the mixing angle when $W_L$ and $W_R$ are expressed in terms of the mass eigenstates $W_1$ and $W_2$, and $m_2$ the mass of $W_2$).

The radiative corrections to the muon lifetime $\tau_\mu$ have been the subject of great interest and detailed studies. In fact, the argument given at the end of Section~\ref{sec:2.1} can be generalized: it has been shown that to leading order in $G_\mu$, but all orders in $\alpha$, the radiative corrections to muon decay in the $V$-$A$ theory are finite after mass and charge renormalization \cite{Berman:1962xx}. The detailed calculations reach now the two-loop level and lead to:
%
%\alpheqn
\be
\frac{1}{\tau_\mu} = \frac{G^2_\mu m^5_\mu}{192\pi^3} F(x)[ 1+\delta_\mu],  \label{eqeighteena} 
\ee
where $x=m^2_e/m^2_\mu$, $F(x) = 1-8x-12x^2\ln x+8x^3-x^4$ is a   tree-level phase-space factor and $\delta_\mu$ is the radiative correction.

Neglecting very small terms proportional to powers of $m_e/m_\mu$, we have
\be
\delta_\mu = \frac{\alpha}{2\pi} \left(\frac{25}{4} - \pi^2\right) \left[ 1+\frac{2\alpha}{3\pi} \ln \left(\frac{m_\mu}{m_e}\right)\right] + 6.700 \left(\frac{\alpha}{\pi}\right)^2 + \cdots \, .\label{eqeighteenb} 
\ee
The $\caloa$ term has been known since the end of the 1950's \cite{Kinoshita:1958ru, Berman:1958ti}, the logarithmic term of $\caloasq$ was derived in 1971 \cite{Roos:1971mj}, and the last term in 1999 
\cite{vanRitbergen:1998yd, vanRitbergen:1999fi, Steinhauser:1999bx}, about 40 years after the one-loop correction! The two terms of $\caloasq$ nearly cancel each other. Including very small one and two-loop contributions proportional to powers of $m_e/m_\mu$ \cite{vanRitbergen:1998yd, Pak:2008qt}, we have
\be
\delta_\mu = -4.19948\times10^{-3} + 1.06\times10^{-6}\, ,\label{eqeighteenc}
\ee
where the first and second terms stand for the one and two-loop contributions, respectively. This reveals that when the corrections are expressed in terms of $\alpha$, as in Eq.(\ref{eqeighteenb}), the $\caloasq$ effects are very small, and the original $\caloa$ calculation turns out to be very accurate. Alternatively, $\delta_\mu$ is frequently written in the form  \cite{vanRitbergen:1998yd, vanRitbergen:1999fi, Steinhauser:1999bx}
\be
\delta_\mu = \frac{\alpha(m_\mu)}{2\pi} \left(\frac{25}{4} - \pi^2\right) + 6.700 \left(\frac{\alpha(m_\mu)}{\pi}\right)^2 + C(x)+ \cdots \, , \label{eqeighteend} 
\ee
%\reseteqn
%\setcounter{equation}{18}
%
\noi where $\alpha(m_\mu)=1/135.9026283\dots$ is the running $\alpha(\mu)$ parameter at the $m_\mu$ scale. In this second form the logarithmic term of $\caloasq$ has been absorbed in the ${\cal{O}}(\alpha(m_\mu))$ contribution, and the ${\cal{O}}(\alpha^2(m_\mu))$ effects are $\approx3.6\times10^{-5}$, considerably larger than in 
Eq.(\ref{eqeighteenc}). The correction $\delta_\mu$ has been also studied using optimization methods that select  the optimal scale in $\alpha(\mu)$, permit to analyze the scheme dependence of the calculations and estimate the unknown terms of ${\cal{O}}(\alpha^3(m_\mu))$ \cite{Ferroglia:1999tg}. This analysis leads to an estimated error of $\approx 2.6\times10^{-7}$ in $\delta_\mu$ due to the truncation of the perturbative series.

$C(x)$ in Eq.(\ref{eqeighteend}) denotes very small RC proportional to powers of $x$. Specifically,
\be
C(x) = \frac{\alpha(m_\mu)}{\pi} \left[x (-12 \ln{x} -9 - 4 \pi^2 +16 \pi^2 x^{1/2}) + {\mathcal O}(x^2)\right] - 
\left(\frac{\alpha(m_\mu)}{\pi}\right)^2 0.0784 + \cdots \, . \label{eq:1.35}
\ee
The terms of ${\mathcal O}(\alpha(m_\mu) x^l/ \pi)$ ($l=1,3/2$) were derived by \textcite{vanRitbergen:1998yd}. Their expression differs from that in Eq.(\ref{eq:1.35})  because of the factorization of $F(x)$ in our Eq.(\ref{eqeighteenc}), which was not employed by those authors. For clarity,  we point out that  to the stated level of accuracy, our result for $1/\tau_\mu$ based on Eqs.(\ref{eqeighteena}, \ref{eqeighteend}, \ref{eq:1.35}) through the terms of ${\mathcal O}(\alpha(m_\mu) x^l/ \pi)$ ,  is equivalent to that obtained in their 1999 paper. The contribution of ${\mathcal O}((\alpha(m_\mu)/ \pi)^2)$ was derived years later  \cite{Pak:2008qt} and amounts to $-4.3\times 10^{-7}$. An interesting feature is that its leading contribution is linear in $m_e/m_\mu$:
$-\left(\alpha(m_\mu)/\pi\right)^2 (5/4) \pi^2 x^{1/2} = -3.27 \times 10^{-7}$.

Because of the high precision of the $\tau_\mu$ measurement \cite{Webber:2010zf} and the theoretical clarity of Eqs.(\ref{eqeighteena},\ref{eqeighteenb}, \ref{eqeighteend},\ref{eq:1.35}), $G_F$, the universal Fermi constant of the weak interactions, is identified with $G_\mu$. Inserting the experimental value $\tau_\mu = 2196980.3(2.2)~{\rm ps}$, Eqs.(\ref{eqeighteena}, \ref{eqeighteend}, \ref{eq:1.35}) lead to $\delta_\mu = - 4.19818 \times 10^{-3}$ and 
\be
G_F = G_\mu = 1.1663788(7)\times 10^{-5}{\rm~GeV}^{-2} \, , \label{eqnineteen}
\ee
 an important 0.6 ppm determination \cite{Webber:2010zf}.

We note that the evaluation of $\delta_\mu$ in the $\alpha$ and $\alpha(m_\mu)$ schemes, namely $\delta_\mu= -4.19842 \times 10^{-3}$ (Eq.(\ref{eqeighteenc})) and $\delta_\mu = -4.19818 \times 10^{-3}$, respectively, differ by $-2.4 \times 10^{-7}$. This difference is consistent with the estimate of the third order coefficient in the $\alpha(m_\mu)$ expansion on the basis of the optimization methods, namely $(c_3)_{\rm est.} \approx -20$
\cite{Ferroglia:1999tg}. The effect of this difference on the determination of $G_F$ (Eq.(\ref{eqnineteen})), is also small in comparison with the current experimental error.

We also note that, in some theoretical discussions of $1/\tau_\mu$, a factor $(1+ 3 m_\mu^2/M_W^2)$ that represents the tree level correction from the $W$-boson propagator, is applied in the r.~h.~s. of  Eq.(\ref{eqeighteena}). Since this factor does not arise in the Fermi theory framework, it is not included in our  Eq.(\ref{eqeighteena}). It has been pointed out by \textcite{vanRitbergen:1998yd} that, in ST calculations, it can be more naturally included in the electroweak correction $\Delta r$ (cf.  Eq.(\ref{eq:3_1})). More generally, it can be included in the expressions of the form $G_F(1- {\rm \, EWC})$ where EWC denotes a generic electroweak correction such as $\Delta \hat{r}, \Delta \hat{r}_W$, and $\Delta r_{\rm eff}$ (cf.  Eqs.(\ref{eq:3_4}, \ref{eq:3_5}, \ref{eq:3_12})). On the other hand, it is useful to observe that this factor would amount to an addition of only $\approx 5 \times 10^{-7}$ to such electroweak correction, which is negligible at the current level of accuracy.

\subsection{The Universality of the Weak Interactions and the Conserved Vector Current Hypothesis}

The principle of universality of the weak interactions is a concept of enduring significance. In fact, it has motivated, at least in part, several important developments in particle physics.

The origin of the idea can be traced to 1947--49, when several authors 
\cite{Pontecorvo:1947vp, Puppi:1948qy, Puppi:1949xx, Klein:1948zz, Tiomno:1949zz, Lee:1949qk} noted that the  basic processes $\mu^-\to e^-+\nu_\mu +\bar\nu_e$, $n\to p+e^-+\bar\nu_e$, and $\mu^-+p\to n+\nu_\mu$ are characterized approximately by the same coupling constant, of magnitude $\approx 10^{-5}$ GeV$^{-2}$. On this basis they proposed a universal weak interaction among the doublets $(\nu_e,e)$, $(\nu_\mu,\mu)$ and $(p,n)$. In 1951, Enrico Fermi stated that this similarity is probably not accidental and has a deep meaning not understood at the time \cite{Fermi:1951xx}. He also suggested a possible analogy with the universality of electric-charge.

In their 1958 paper, \textcite{Feynman:1958ty} compared $G_\mu$ with $G_V$, the vector coupling in $\beta$-decay extracted from ${}^{14}{\rm O}$ decay, a superallowed $(0^+\to0^+)$ Fermi transition in which only the vector current contributes to zeroth order in $\alpha$. They found $G_V=G_\mu$ within roughly 1\%. The result was very surprising, since even if one assumed $G_V=G_\mu$ at the Lagrangian level as a manifestation of universality, a close equality was not expected because nucleons in $\beta$-decay are affected by strong interactions, while this is not the case for the leptons in muon decay. This prompted \textcite{Feynman:1958ty} to invoke the conserved vector current (CVC) hypothesis, previously discussed by \textcite{Gershtein:1955fb}. Specifically, the hadronic vector current in $\beta$ decay is assumed to be conserved in the presence of the strong interactions. Since conservation laws are generally associated with symmetries of the theory, they further identified it with the $\Delta I_3=1$ isospin current. The near equality $G_V\approx G_\mu$ could then be understood on the basis of two concepts: the principle of universality that states $G_V=G_\mu$ at the Lagrangian level, and CVC that implies that the strong interactions do not renormalize $G_V$ at $q^2=0$ in the limit of isospin invariance.

CVC, in turn, had another important consequence. If the strangeness conserving $(\Delta S=0)$ vector current is conserved, it would be natural to assume that the strangeness non-conserving $(\Delta S=1)$ vector current in semileptonic decays is also conserved in some suitable limit. This was one of the main motivations for the search for higher partial symmetries of the Strong Interactions. A number of possibilities were considered \cite{Behrends:1962xx}, culminating with the phenomenologically successful SU$(3)_{\rm flavor}$ symmetry 
\cite{GellMann:1962xb, GellMann:1964xy}. Gell-Mann also noted that a normalization of the hadronic currents is necessary in order to define precisely the concept of universality. This was an important motivation for Current Algebra \cite{GellMann:1964tf}. In fact, the non-linearity of the basic Current Algebra relation
\be
[J^a_0 (x), J^b_0(y)]_{x_0=y_0} = i\, f^{abc}J^c_0 (x) \delta^3 (\vec x - \vec y)\, , \label{eqtwenty}
\ee
where $f^{abc}(a,b,c=1\dots8)$ are the SU(3) structure constants, determines the normalization of the hadronic currents. SU$(3)_{\rm flavor}$ also led to the fundamental concept of quarks \cite{GellMann:1964Q, Zweig:1964Q} and the quark model of hadrons.

\boldmath
\subsection{Radiative Corrections to $\beta$ Decay in the $V$-$A$ Theory}
\unboldmath
\label{sec:2.7}

When the CVC hypothesis was formulated, it was natural to suspect that the $\approx 1\%$ difference between $G_V$ and $G_\mu$ was due to electromagnetic corrections. Here, we have in mind electromagnetic corrections not contained in Fermi's Coulomb-function which is automatically included in the theory of $\beta$-decay. However, when the $\caloa$ corrections to the decay probability of neutron $\beta$-decay were calculated by \textcite{Berman:1958ti} and \textcite{Kinoshita:1958ru} in the $V$-$A$ theory (cf. Eq.(\ref{eqseventeen})), a striking result was found: contrary to the case of muon decay, the $\caloa$ corrections to $\beta$-decay were logarithmically divergent! In particular, the detailed expression found by 
\textcite{Kinoshita:1958ru}  for the $\caloa$ corrections to the electron or positron spectrum is given by
%
%\alpheqn
\begin{align}
\Delta Pd^3p  & =  \frac{\alpha}{2\pi} P^0d^3p\left\{6\ln\left(\frac{\Lambda}{m_p}\right)+g(E,E_m) +\frac{9}{4}\right\}, \label{eqtwentyonea} \\
g(E,E_m)  & =  3\ln \left(\frac{m_p}{m_e}\right)-\frac{3}{4}-\frac{4}{\beta} \Li_2 \left(\frac{2\beta}{1+\beta}\right)   +4\left[\frac{\tanh^{-1}\beta}{\beta} -1 \right] \left[ \frac{(E_m-E)}{3E} - \frac{3}{2} +\ln \left\{\frac{2(E_m-E)}{m_e}\right\}\right] \nonumber \\
 &  +\frac{\tanh^{-1}\beta}{\beta} \left[ 2(1+\beta^2) + \frac{(E_m-E)^2}{6E^2} - 4\tanh^{-1}\beta\right] \, , \label{eqtwentyoneb}
\end{align}
%\reseteqn
%
\noi where $p$ and $E$  are the momentum and energy of the electron or positron, $E_m$ is the end-point energy, $\beta=p/E$, $m_p$ the proton mass, $\Lambda$ the ultraviolet cutoff, and
\be
P^0d^3p = \frac{8G^2_V}{(2\pi)^4} (E_m-E)^2 d^3p 
\ee
%\reseteqn
%\setcounter{equation}{21}
%
 is the uncorrected spectrum.
In deriving  Eq.(\ref{eqtwentyonea}), strong interactions have been neglected, so these results represent the corrections to the $\beta$-decay of ``bare nucleons'' devoid of hadronic structure. Very small contributions of ${\cal{O}}(E/m_p)$ have been also neglected.

The reason why the corrections to $\beta$ decay are divergent in the $V$-$A$ theory while those for muon decay are finite, can be understood in two ways:
\begin{itemize}
\item[i)] In contrast to the muon decay case, starting with the interaction Lagrangian of  Eq.(\ref{eqseventeen}) appropriate to $\beta$-decay, it is not possible to bring the two charged particles into the same covariant while retaining only $V$ and $A$ interactions. Thus, the analogy with QED  discussed in Section~\ref{sec:2.1}  is lost in the case of $\beta$-decay and the corrections are divergent.

\item[ii)] Using a current algebra formulation, it can be shown that in the $V$-$A$ theory the divergent part of the corrections to Fermi transitions is of the form
\be
\frac{\alpha}{2\pi} P^0d^3p~ 3[1+2\bar Q] \ln (\Lambda/M) \, , \label{eqtwentytwo}
\ee
where $\bar Q$ is the average charge of the underlying hadronic fields in the process and $M$ a relevant mass. In the case of  Eq.(\ref{eqseventeen}), the underlying fields are the neutron and proton so that $\bar Q=1/2$ and the divergent part is $(\alpha/2\pi) P^0d^3p\; 6\ln(\Lambda/M)$, in agreement with  Eq.(\ref{eqtwentyonea}). In the case of muon decay, the roles of $p$ and $n$ are played by $\nu_\mu$ and $\mu^-$, so that $\bar Q=-1/2$ and  Eq.(\ref{eqtwentytwo}) vanishes, consistent with the fact that the corrections to muon decay are finite in the $V$-$A$ theory. It is interesting to note that
in the corrections proportional to $|M_F|^2$, where $M_F$ is the Fermi matrix element,  the terms $3\ln(\Lambda/M)$ and $6\bar Q\ln(\Lambda/M)$ in  Eq.(\ref{eqtwentytwo}) arise from the vector and axial vector currents, respectively. Similarly, in  Eq.(\ref{eqtwentyonea}) $3\ln(\Lambda/m_p)+g(E,E_m)$ is the contribution from the vector current while the remaining $3\ln(\Lambda/m_p)+9/4$ emerges from the axial vector current. Thus, although the axial vector current does not contribute to the Fermi matrix element at the tree-level, it plays a very important role in ${\cal{O}}(\alpha)$.
\end{itemize}

The finding that the radiative corrections to $\beta$-decay in the $V$-$A$ theory are divergent, while those to muon-decay are convergent, created a serious theoretical problem since both processes are fundamental observables. Originally, Feynman, Berman, Kinoshita and Sirlin thought that this conundrum was due to the fact that strong interactions had been ignored in the calculations of the $\beta$-decay corrections. In fact, it was easy to imagine that strong interactions could give rise to form factors that would cut off the high-energy contributions of the virtual photons. If so, $\Lambda$ in  Eqs.(\ref{eqtwentyonea},\ref{eqtwentytwo}) was expected to be of the order of magnitude of the nucleon mass $M_N\approx 1$ GeV\null. The same point of view was strongly advocated by \cite{Kallen:1967xx}. A further complication at the time was that for $\Lambda\gsim 1$ GeV the radiative corrections increased the difference between $G_V$ and $G_\mu$! The situation, as it existed in 1960, was summarized by \cite{Feynman:1960xx}.

The statement of universality was significantly changed when \textcite{Cabibbo:1963yz} proposed his theory of semileptonic decays, constructed on the basis of SU(3)$_{\rm flavor}$ currents. Rather than stating $G_V(\Delta S=0)=G_\mu$, the principle of universality was expressed as
%\alpheqn
\be
G_V(\Delta S=0) = G_\mu \cos\theta_c;~ G_V(\Delta S=1) = G_\mu\sin\theta_c\, , \label{eq:23a}
\ee
\noi where $\theta_c$ is the Cabibbo angle. Thus, in the new framework we had
\be
G^2_\mu = G^2_V(\Delta S=0) + G^2_V(\Delta S=1)\, . \label{eq:23b}
\ee
%\reseteqn
%
 Eq.(\ref{eq:23a}) had two important consequences: by adjusting appropriately $\sin \theta_c$, it successfully described the fact that $\Delta S=1$ semileptonic decays are significantly suppressed relative to $\Delta S=0$ processes and, furthermore, the radiative corrections with $\Lambda\gsim 1$ GeV had an effect that was at least in the right direction to comply with  Eqs.(\ref{eq:23a},\ref{eq:23b}).

In the 1960's there were other developments that also contributed significantly to the analysis of universality.
\textcite{Behrends:1960nf} showed that if the conservation of SU$(2)$ vector currents (such as the isospin currents) is broken by mass splittings, their matrix elements at zero momentum transfer are not renormalized to first order in the symmetry-breaking parameters. They also conjectured the generalization of this theorem to higher symmetries. The results were confirmed by \textcite{Terentev:1963xx} on the basis of a different argument.  
\textcite{Ademollo:1964sr} independently derived the analogous theorem for SU$(3)$ vector currents. This non-renormalization theorem plays an important role in the analysis of universality: in the SU$(2)$ case it applies to $\beta$ decay, while in the SU$(3)$ context it is relevant for the $\Delta S=1$ semileptonic decays.

In 1966, there was another very important and surprising development. 
\textcite{Bjorken:1966zz}, using current algebra methods, reached the conclusion that the strong interactions do not tame the logarithmic divergence of the radiative corrections to the Fermi transitions in $\beta$ decay. Thus, according to this approach, the cutoff $\Lambda$ did not arise from the strong interactions! The analysis was extended by 
\textcite{Abers:1967xx}, who studied the divergent part of the corrections to the Fermi amplitude arising from the axial vector current. In their work, they applied the Bjorken-Johnson-Low limit 
\cite{Bjorken:1966zz, Johnson:1966xx}
 with a simplified, canonical evaluation of the relevant commutators.  \textcite{Sirlin:1967zza}, using a very different approach, showed that the function $g(E,E_m)$, which describes the corrections to the electron or positron spectrum in $\beta$ decay (cf.  Eq.(\ref{eqtwentyoneb})), is valid in the presence of the strong interactions, provided one neglects small contributions of ${\cal{O}}(\alpha E/M)$, where $M$ is a relevant hadronic mass. The approach employed in 
\textcite{Sirlin:1967zza}, 
the so-called $1/k$ method, consists in separating out, in a gauge-invariant manner, the contributions that behave as $1/k$ as $k\to0$ in the hadronic parts of the Feynman integrals, where $k$ is the virtual photon four-momentum. Such contributions are not affected by the strong interactions and lead to the function $g(E, E_m)$. The remaining contributions are shown to fall into two classes: constant amplitudes, independent of $E$ and $E_m$, which are affected by the strong interactions, but can be absorbed by suitable redefinitions of the vector and axial vector coupling constants $g_V$ and $g_A$, and very small terms of $\caloaem$ which are neglected. This method was extended to treat other observables such as the longitudinal polarization of electrons or positrons \cite{Sirlin:1967zza} and the asymmetry from polarized nuclei in $\beta$ decay 
\cite{Shann:1971fz, Yokoo:1973mz, Garcia:1978bq, Gluck:1992qy}. The current algebra formulation and the $1/k$ method finally overlapped when, in a subsequent paper, \textcite{Abers:1968xx} were able to obtain not only the divergent parts, but also the corrections to the energy spectrum described by the function $g(E,E_m)$. In fact, the current algebra formulation led to the important conclusion that, neglecting very small contributions of $\caloaem$, the $\caloa$ corrections to the Fermi amplitude arising from the vector current are not affected by the strong interactions, and it appeared that the divergent contributions involving the axial vector current were also known. Although other methods to evaluate the radiative corrections to $\beta$-decay were pursued, most notably by 
\textcite{Kallen:1967xx}, the current algebra formulation became the prevalent approach. 

Thus, in 1967 the situation regarding the radiative corrections to $\beta$-decay was both interesting and perplexing. On the one-hand, the current algebra approach had been the basis of great technical progress. On the other hand, there was the great difficulty that in the $V$-$A$ local Fermi theory the corrections are divergent! At the time, two different solutions to this serious problem were suggested: {\it i}) 
\textcite{Cabibbo:1967axx, Cabibbo:1967bxx} and 
\textcite{Johnson:1967xx} proposed to modify the space-space commutators of the current algebra of hadronic currents in such a way that the radiative corrections to $\beta$-decay become convergent {\it ii}) 
\textcite{Sirlin:1967zzb} proposed that the solution to the dilemma lies instead in an extension of the Fermi theory involving charged intermediate bosons $W^\pm$. The argument was that in this framework the leading divergent contributions to muon and $\beta$-decay are the same, so that they can be absorbed in a universal renormalization of $G_\mu$ and $G_V$, as discussed by
\textcite{Sirlin:1967zzb} and  \textcite{Abers:1968xx}. This approach, however, was not complete since the intermediate boson theory employed was not renormalizable and, as a consequence, logarithmic divergences with very small coefficients were not canceled. An additional limitation was that in this theory the effective cutoff was $\Lambda \approx m_W$, and its magnitude was unknown at the time.

Analogous results had been previously obtained by 
\textcite{Lee:1962zz, Shaffer:1962xx, Shaffer:1963xx, Dorman:1964zz}
 and \textcite{Bailin:1964xx, Bailin:1965xx}, who had studied the radiative corrections in the intermediate boson framework in the case of ``bare'' nucleons, devoid of strong interactions. The situation in 1968 was summarized by 
 \textcite{Sirlin:1968xx}.

As explained in Section~\ref{sec:SM}, the solution of the serious problem affecting the radiative corrections to $\beta$-decay had to wait until the emergence of the Standard Theory, a renormalizable theory of electroweak interactions!

Recently, a close, analytic correction for the ${\mathcal O}(\alpha)$  radiative correction to the $\bar{\nu}_e$ ($\nu_e$)
spectrum in allowed $\beta$ decay was derived \cite{Sirlin:2011wg}. The motivation of this calculation is that knowledge of the  $\bar{\nu}_e$ ($\nu_e$) spectrum is currently important for reactor  studies of neutrino oscillations. One finds:
\be
d P_\nu = d P_\nu^0 \left[ 1+\left(\frac{\alpha}{2 \pi} \right) h\left(\hat{E}, E_m\right)\right] \, ,
\label{eq:2_43}
\ee
where
\be
d P_\nu^0 =  A\, \hat{p}\, \hat{E}\, F(Z,\hat{E})\, K^2 dK \, ,
\label{eq:2_44}
\ee 
is the zeroth order spectrum, 
\begin{align}
h(\hat{E},E_m) &= 3 \ln{\left(\frac{m_p}{m_e}\right)} +\frac{23}{4} -\frac{8}{\hat{\beta}} \mbox{Li}_2 \left(\frac{2 \hat{\beta}}{1+\hat{\beta}} \right)+
8 \left(\frac{\mbox{tanh}^{-1} \hat{\beta}}{\hat{\beta}} -1 \right) \ln\left(
\frac{2 \hat{E} \hat{\beta}}{m_e} \right) 
%\nonumber \\ & 
+ \,4\, \frac{\mbox{tanh}^{-1} \hat{\beta}}{\hat{\beta}} \left[ \frac{7 +3 \hat{\beta}^2}{8} - 2 \,\mbox{tanh}^{-1} \hat{\beta} \right]\, , \label{eq:2_45}
\end{align}
$m_p$ is the proton mass, $K$ is the $\bar{\nu_e}$ energy, $\hat{E} = E_m - K$, $E_m$ the end-point energy of the electron in the $\beta$-decay, $\hat{p} = \sqrt{\hat{E}^2 - m_e^2}$, $\hat{\beta} = \hat{p}/\hat{E}$, $F(Z,E)$ the Fermi Coulomb function, $A$ a constant independent of $K$  and $\Li_2(z)$ the dilogarithm function defined in  Eq.(\ref{eq:2_19}).
As in the case of the ${\mathcal O}(\alpha)$ correction to the $e^-$ spectrum (cf.  Eq.(\ref{eqtwentyoneb})), the function $h(\hat{E}, E_m)$ is valid in the presence of the strong interactions, provided small contributions of ${\mathcal O}(\alpha \hat{E}/M)$ are neglected.

Including the ${\mathcal O}(\alpha)$ radiative corrections, the theoretical expressions for the $e^-$ and $\bar{\nu}_e$ spectra in allowed $\beta$-decay can be written in the form
\be
\frac{d P_e}{d E} = f_e(E, E_m) \,  ; \qquad \frac{d P_\nu}{d K} = f_\nu(K,E_m) \, ,
\ee
where 
\begin{align}
f_e(E, E_m) &=  A\, p\, E\, (E_m-E)^2\, F(Z, E)\, \left[1+ \frac{\alpha}{2 \pi} g(E, E_m) \right]\, , \label{eq:2_47}\\
f_\nu(K, E_m) &=  A\, \hat{p}\, \hat{E}\, K^2\, F(Z, \hat{E})\, \left[1+ \frac{\alpha}{2 \pi} h(\hat{E}, E_m) \right]\, , \label{eq:2_48}
\end{align}
$h(\hat{E}, E_m)$ is defined in  Eq.(\ref{eq:2_45}) and $g(E,E_m)$, the function that describes the ${\mathcal O}(\alpha)$ radiative correction to the $e^-$ spectrum, is shown in  Eq.(\ref{eqtwentyoneb}).

Comparing  Eqs.(\ref{eq:2_47}) and (\ref{eq:2_48}), neglecting contributions of ${\mathcal O}(\alpha^2)$, and recalling $\hat{E} = E_m - K$, one finds \cite{Sirlin:2011wg}:
\be
f_\nu\left(K,E_m\right)= f_e (\hat{E}, E_m) \left[ 1+ \frac{\alpha}{2 \pi} \left( h(\hat{E}, E_m) 
- g(\hat{E}, E_m )\right)\right] \, .
\label{eq:2_49} 
\ee
 Eq.(\ref{eq:2_49}) describes the conversion form the $e^-$ spectrum in a specific decay into the corresponding $\bar{\nu}_e$ spectrum when the ${\mathcal O}(\alpha)$ radiative corrections are included. This conversion procedure is the method currently employed to determine the $\bar{\nu}_e$ spectrum from the measured electron spectrum. In turn, as mentioned before, knowledge of the $\bar{\nu}_e$ spectrum is currently important for reactor studies of neutrino oscillations.

An interesting theoretical property of $h(\hat{E}, E_m)$ is  that its $m_e \to 0$ limit converges and leads to a very simple expression ($m_e$ is the electron mass). This is in sharp contrast with the behavior of $g(E, E_m)$ that diverges as $m_e \to 0$. This important difference can be explained in the following way  \cite{Sirlin:2011wg}. For given $K$, as $m_e \to 0$ all collinear $e-\gamma$ configurations become energy degenerate and generally give rise to mass singularities. An elementary, but powerful theorem in quantum mechanics on degenerate systems and mass singularities, due to \textcite{Lee:1964is}, leads to the conclusion that these singularities are canceled in the power series expansions of transition probabilities if the  latter are summed over an appropriate ensemble of such degenerate states. In the derivation of the radiative corrections to the $\bar{\nu}_e$ ($\nu_e$) spectrum, one performs the $d^3 p$ and $d^3 k$ integrations, where $p$ and $k$ are the electron and photon momenta, so  indeed one sums over the set of collinear $e-\gamma$ configurations that become energy degenerate in the $m_e \to 0$ limit. Therefore, according to this theorem,  $h(\hat{E}, E_0)$ should be free of $\ln m_e$ singularities, as found in the explicit calculation.  
In contrast, this is not the case in the derivation of the radiative corrections to the $e^-$ spectrum, since the $d^3p$ integration is not carried out. As a consequence, the Lee-Nauenberg theorem is not applicable to $g(E, E_m)$ and, as is well known, this function  diverges in the $m_e \to 0$ limit. Analogous examples of mass singularities in the ${\mathcal O} (\alpha)$ radiative corrections to the differential spectra, and their cancellation in the lifetimes, integrated asymmetries and some partial decay rates in muon and $\beta$ decays were extensively discussed in \cite{Kinoshita:1958ru}.

Pion $\beta$-decay, $\pi^+ \to \pi^0 +e^+ +\nu_e$ and its charge conjugate, $\pi^- \to \pi^0 + e^- + \bar{\nu}_e$, are processes of special interest, since their interpretation is devoid of the complications of nuclear structure that affect nuclear $\beta$-decays. In this sense, they may be regarded as the simplest examples of super-allowed $0 \to 0$ Fermi transitions. On the other hand, their branching ratio, $(1.036 \pm 0.006) \times 10^{-8}$ \cite{Nakamura:2010zzi, POC},
is very small and, consequently, the measurement of their decay rate is much less precise  than in the nuclear transitions.

Recently, \textcite{PPS} compared the radiative corrections involving the weak hadronic vector current in  pion $\beta$-decay, as evaluated in the $V-A$ theory in two different frameworks:
\begin{itemize}
\item[i)] the current algebra formulation, in which quarks are the fundamental underlying fields, and
\item[ii)] the elementary approach in which pions are regarded as the fundamental fields.
\end{itemize}
The comparison of the two calculations revealed a small difference that was shown to arise from a specific short-distance contribution that depends on the algebra satisfied by the weak and electromagnetic currents\footnote{The fact that this particular contribution is model-dependent was already pointed out by \textcite{Abers:1968xx}.}. In fact, the space-space components of the algebra are different in i) and ii) and this was shown to explain the discrepancy
 discussed above. The results were also compared with a recent calculation based on chiral perturbation theory ($\chi$PT) \cite{CKNP}. Taking into account its theoretical error, the $\chi$PT
 calculation was found to be consistent with those based on either i) or ii). \textcite{PPS} also discussed the important differences between the radiative corrections to pion $\beta$-decay as evaluated in the $V$-$A$ and Standard Theories.

\section{RADIATIVE CORRECTIONS IN THE STANDARD THEORY OF PARTICLE PHYSICS}
\label{sec:SM}

%\subsection{Introduction}

The Standard Theory of Particle Physics (ST) originally proposed by \textcite{Weinberg:1967tq}, \textcite{Salam:1968},
and \textcite{Glashow:1961tr}, emerged, with very important contributions from other physicists, in the period 1967-1974. At present, it is a gauge theory of the Electromagnetic, Weak and Strong Interactions based on the 
$\mbox{SU}(2)_L \times \mbox{U}(1) \times \mbox{SU}(3)_C$ symmetry group. Here $\mbox{SU}(2)_L \times \mbox{U}(1)$ is the symmetry group of the electroweak (EW) sector and $\mbox{SU}(3)_C$ that of Quantum Chromodynamics
(QCD), the current theory of the Strong Interactions.

As shown by \textcite{'tHooft:1971rn}, \textcite{'tHooft:1972fi, 'tHooft:1972xx}, \textcite{Lee:1971kj},
\textcite{Lee:1974zg, Lee:1973fn}, \textcite{Zinn-Justin:1975xx}, \textcite{Becchi:1974xx},
and others, it is a renormalizable theory. This implies that the electroweak corrections (EWC) in this theory can be evaluated by perturbative field theoretic methods, since the ultraviolet divergences found in the calculations can be absorbed as unobservable contributions to the masses and couplings  of the theory. In the domain in which the strong interaction running coupling $\alpha_s(\mu)$ is small, the same is true of the QCD corrections.

In 1972, dimensional regularization, a very ingenious method to regularize ultraviolet divergences, was proposed by \textcite{'tHooft:1972fi}, \textcite{Bollini:1972ui}, and \textcite{Ashmore:1972uj}. It is particularly useful in the context of gauge theories such as the ST. Dimensional regularization of  infrared divergences was proposed by  \textcite{Gastmans:1973uv}, and \textcite{Marciano:1974tv}, and that of mass singularities by \textcite{Marciano:1975de}. Dimensional regularization of infrared and mass singularities is widely used at present, particularly in QCD calculations.

Once the renormalizability of the ST was recognized, it was natural to study the EW and QCD corrections of the theory. The aim of these studies are:
\begin{itemize}
\item[i)] To verify the ST at the level of its quantum corrections.
\item[ii)] To search for discrepancies that may signal the presence of new physics beyond the ST.
\end{itemize}  
In the EW sector, these are essentially the objectives of what is now called Precision Electroweak Physics.

\subsection{Early Developments}
\label{sec:3.2}

Already in the 1970's there were a number of important developments:
\begin{itemize}
\item[i)] The evaluation of one-loop EWC to $g_{\mu}-2$ \cite{Jackiw:1972jz, Bars:1972pe, Fujikawa:1972fe}.
\item[ii)] \textcite{Weinberg:1973ua} showed that there are no violations of ${\mathcal O}(\alpha)$ to parity and strangeness conservation in strong interaction amplitudes.
\item[iii)] \textcite{Gaillard:1974hs} studied processes which are forbidden at the tree level, but occur via loop effects, and showed that the GIM mechanism \cite{Glashow:1970gm} generally suppresses neutral current amplitudes of 
${\mathcal O}(G_F \alpha)$.
\item[iv)] \textcite{Bollini:1973qg}  studied the cancellation of ultraviolet divergences in  fundamental natural relations 
of the ST.
\item[v)] Using a simplified version of the ST involving integer-charged quarks, and neglecting the effect of the Strong Interactions,  \textcite{Sirlin:1974ni} showed explicitly that the one-loop EWC to $\beta$-decay are indeed finite in the ST
and that the ``cutoff'' is given by $M_Z$. This leads to large EWC of ${\mathcal O} (4 \%)$, a result that has important consequences in the test of the universality of the Weak Interactions. Indeed, this result was one of the early ``smoking guns''  of the EW sector of the ST at the level of its quantum corrections. On the other hand, as discussed in Section~\ref{sec:sub3_11}, an evaluation of the EWC in the ``real'' ST, based on fractionally charged quarks, and taking into account the effect of the Strong Interactions, had to wait until the development of the  Current Algebra formulation of radiative corrections in gauge theories \cite{Sirlin:1977sv}.   
\item[vi)] \textcite{Veltman:1977kh}, and \textcite{Chanowitz:1978uj} discovered that heavy particles do not generally 
decouple in the EWC of the ST, and that a heavy top quark gives contributions of ${\mathcal O}(G_F M_t^2)$ to the $\rho$ parameter, defined as the ratio of the neutral and charged current coupling constants at zero momentum transfer.
\end{itemize}

\subsection{Input Parameters}
\label{sec:3.3}

Three very precisely measured constants play a particularly important role as input parameters in Electroweak Physics:
\begin{itemize}
\item[i)] The fine structure constant $\alpha = 1/137.035999679(94)$ \cite{Nakamura:2010zzi}, with a relative error
$\pm 6.9 \times 10^{-4}$~parts per million (ppm), obtained most precisely from $g_{(e)}-2$.
\item[ii)] The Fermi constant $G_F = G_\mu = 1.1663788(7) \times 10^{-5} \mbox{GeV}^{-2}$, with a relative error of $0.6$~ppm
(see Section~\ref{sec:2.5}).
\item[iii)] $M_Z = 91.1876 \pm 0.0021$~GeV \cite{Nakamura:2010zzi}, with a relative error of $23$~ppm.

This very precise determination of $M_Z$ required sophisticated experimental techniques and a very accurate study of the $Z$ Line Shape, in which QED and EW corrections play an important role (see for example \textcite{Berends:xxxx}) 
\end{itemize}

\subsection{The On-Shell Scheme of Renormalization}
\label{sec:3.4}

Towards the end of the 1970's it seemed likely that experimental physicists would search for the $W$ and $Z$ intermediate vector bosons of the ST and hopefully measure their masses. This motivated the idea of studying at the loop level the relation between $M_W$, $M_Z$, $G_F$, $\alpha$, and the EW mixing parameter $\sin^2 \theta_W$, as well as other fundamental parameters of the theory, such as the quark masses and the Higgs boson mass $M_H$.
The hope was that this analysis would lead to more accurate predictions for $M_W$ and $M_Z$. At the time, $G_F$ and $\alpha$ were accurately known, and $\sin^2 \theta_W$  was determined with less precision from $\nu-N$ deep inelastic scattering via the neutral and charged currents. Thus, it became clear that  it was necessary to evaluate the EWC
to the last two processes to extract $\sin^2 \theta_W$, and to muon decay to obtain the relation with $G_F$ and 
$\alpha$.

Since this required the analysis of a number of processes involving neutral and charged currents, in order to facilitate the evaluation of the corresponding EWC, \textcite{Sirlin:1980nh} proposed a simple, physically motivated framework
to renormalize the EW sector of the ST. This approach, with important contributions from other physicists\footnote{See, for example, \textcite{Aoki:1982R, Boehm:1986R, Consoli:1989fg, Hollik:1988ii}.} is currently known as the On-Shell scheme (OS). In the same 1980 paper, the OS scheme was applied to 
evaluate the one-loop EWC to muon decay in the ST. The analysis leads to the basic OS relations \cite{Sirlin:1980nh, Sirlin:1983ys}
\bea
s^2 c^2 &=& \frac{\pi \alpha}{\sqrt{2} G_F M_Z^2 (1- \Delta r)} \, ,\label{eq:3_1}\\
s^2 &=& \sin^2 \theta_W = 1 -\frac{M_W^2}{M_Z^2} \, , \label{eq:3_2}
\eea
where $G_F = G_\mu$ is the Fermi constant discussed in Sections~\ref{sec:2.5} and \ref{sec:3.3}, and $\Delta r$ is the EWC to muon decay. From  Eq.(\ref{eq:3_2}) we see that in the OS scheme the EW mixing parameter $\sin^2 \theta_W$ is simply defined in terms of the physical masses $M_W$ and $M_Z$, to all orders in perturbation theory. In two subsequent papers, the OS scheme was applied to the study of the EWC to $\nu-N$ deep inelastic scattering via the neutral current \cite{Marciano:1980pb} and via the charged current \cite{Sirlin:1981yz}. 
This trilogy of papers achieved the aim of establishing contact, at the level of the EWC, between the theory and the expected measurements of $M_W$ and $M_Z$. In fact, using  Eqs.(\ref{eq:3_1},\ref{eq:3_2}) and the information from $\nu-N$ scattering, they led to more accurate predictions of $M_W$ and $M_Z$ before the actual measurements. As the experiments on $\nu-N$ scattering improved, the role of the EWC became more important. A detailed analysis
\cite{Amaldi:1987fu} led to the estimates $M_W = 80.2 \pm 1.1$~GeV, $M_Z = 91.6 \pm0.9$~GeV, with central values that differ from the current ones by about $0.2$~GeV and $0.4$~GeV, respectively. It should be pointed out that this closeness is rather accidental (for example, the top quark mass used in calculations at the time was much smaller than its present value). Nonetheless, the early predictions were very useful because they provided what turned out to be realistic mass ranges for the experimental searches of the $W$ and $Z$ bosons. Furthermore, as shown in \cite{Amaldi:1987fu}, they also turned out to be in good agreement with the early measurements of the $W$ and $Z$ masses.

The EWC $\Delta r$ in  Eq.(\ref{eq:3_1}) depends on various physical parameters of the ST such as $\alpha$, $M_W$,
$M_Z$, $M_H$, $M_f$, $\alpha_s(M_Z)$, \ldots, where $M_H$ is the Higgs boson mass, $M_f$ a generic fermion mass and $\alpha_s(M_Z)$ the QCD running coupling evaluated at the scale $\mu =  M_Z$. It follows from 
 Eqs.(\ref{eq:3_1},\ref{eq:3_2}) that $\Delta r$ is a physical observable.  Eqs.(\ref{eq:3_1},\ref{eq:3_2}) can be viewed as the relation between the physical parameters of the Fermi Theory (low-energy effective theory), namely $G_F$ and $\alpha$, and those of the ST (underlying theory), namely $\alpha$, $M_W$, $M_Z$, $M_H$, $M_f$, \ldots, at the level of the quantum corrections. These relations are currently used to calculate $M_W = M_W(M_H)$, leading to very sharp constraints on $M_H$.

The On-Shell Scheme is also used in the ZFITTER program \cite{Arbuzov:2002Z, Bardin:2001Z} and the Gfitter project \cite{Flacher:2009Z}, extensively employed in the analysis of the electroweak precision observables.

\subsection{The $\ms$ Scheme of Renormalization}
\label{sec:3.5}

Another important and very useful approach is the $\ms$ renormalization framework, in which the electroweak mixing parameter is identified with the running coupling $\sin^2\theta_W(\mu) = e^2(\mu)/g^2(\mu)$ evaluated at the
$\mu =M_Z$ scale. (Here $g$ is the $\mbox{SU}(2)_L$ gauge coupling).

In this scheme, the renormalization of $\sin^2\theta_W(\mu)$ and the various couplings is implemented by modified minimal subtraction ($\ms$) \cite{Bardeen:1978yd, Buras:1979yt}. At the one-loop level, this involves subtracting 
\be
\delta = \frac{1}{n-4} + \frac{1}{2} \left[ \gamma_E - \ln(4 \pi) \right] \, 
\ee
from the EWC, where the first term is the characteristic pole in dimensional regularization and $\gamma_E = 0.5772\ldots$
is Euler's constant. Since at the one-loop level, $\delta$ always appears in combination with $\ln(1/\mu)$ where $\mu$
is the 't~Hooft mass scale, an equivalent procedure is to rescale $\mu$ according to
$\mu = \mu' e^{\gamma/2}/(4 \pi)^{1/2}$, subtract only the $(n-4)^{-1}$ pole term and then set $\mu'$, rather $\mu$, at the relevant mass scale. This second formulation can be conveniently generalized to higher order EWC and one can define the $\ms$ renormalization procedure as the subtraction of the pole terms  $(n-4)^{-m}$ ($m \ge 1$), and the identification of the rescaled parameter $\mu'$ with the relevant mass scale.

Although  masses can also be defined as running parameters, a hybrid scheme in which couplings and 
$\sin^2\theta_W(\mu)$ are renormalized by $\ms$ subtractions, but masses are still the physical ones, has proved to be very useful and is frequently employed.

An early application of the $\ms$ scheme \cite{Marciano:1980be}  was the derivation of  precise $\mbox{SU(5)}$
predictions for the neutral current amplitude $\sin^2\theta_W^{\mbox{{\tiny exp}}}(q^2)$ (defined in  Eq.(\ref{eq:3_7})), and 
$M_W$ and $M_Z$.

It was also employed in the early papers of \textcite{LSW, WLS} on the EWC  to deep inelastic neutrino and electron scattering.

Two important relations in this scheme were derived by \textcite{Sirlin:1989uf} and  \textcite{Fanchiotti:1990wv}:
\bea
\hat{s}^2 \hat{c}^2 &=& \frac{\pi \alpha}{\sqrt{2} G_F M_Z^2 (1-\Delta \hat{r})} \, , \label{eq:3_4} \\
\hat{s}^2 &=& \frac{\pi \alpha}{\sqrt{2} G_F M_W^2 (1-\Delta \hat{r}_W)} \, , \label{eq:3_5}
\eea
where $\hat{s}^2 \equiv \sin^2 \hat{\theta}_W (M_Z)$ is the $\ms$ electroweak mixing parameter evaluated at the scale $\mu= M_Z$, $\hat{c}^2 =1-\hat{s}^2$, and $\Delta \hat{r}$ and $\Delta \hat{r}_W$ are the corresponding EWC. In 1989,
 Eq.(\ref{eq:3_4}) and the early $M_Z$ measurements at the Large Electron-Positron collider (LEP) were applied to improve significantly the determination of $\sin^2 \hat{\theta}_W (M_Z)$ \cite{Sirlin:1989uf}. In fact, $\hat{\alpha}(M_Z)$, 
$\sin^2 \hat{\theta}_W (M_Z)$ and $\alpha_s(M_Z)$ provide the initial values for the renormalization group equations
(RGE)  satisfied by the running  $\mbox{SU}(2)_L \times \mbox{U}(1) \times \mbox{SU}_C (3)$ gauge couplings,
which play a crucial role in the study of  Grand Unified Theories (GUTs) and in the discovery of supersymmetric Grand Unification (see, for example, \textcite{Langacker:1993rq, Langacker:1995fk}). 
In particular, the 1989 analysis \cite{Sirlin:1989uf} found that the improved value of $\sin^2 \hat{\theta}_W (M_Z)$
was indeed consistent with supersymmetric Grand Unification.

A modification of the renormalization prescription for $\sin^2\hat{\theta}_W (M_Z)$ was proposed by \textcite{Marciano:1990dp} and \textcite{Marciano:1991ix, Marciano:1993jd, Marciano:1993ep}. According to this prescription, aside from the $1/(n-4)$ pole terms, the contributions from particles of mass $m > M_Z$ that do not decouple in the $m \to \infty$ limit, are also subtracted from the amplitude  multiplying 
$\sin^2 \hat{\theta}_W (M_Z)$, and are therefore absorbed in this renormalized parameter. The aim of this prescription
is to obtain values of $\sin^2 \hat{\theta}_W (M_Z)$ from the on-resonance observables which are very insensitive to heavy particles of mass $m > M_Z$, a property that facilitates the analysis of the evolution of $\sin^2 \hat{\theta}_W(M_Z)$ to the GUT scale.

The neutral current vertex of the $Z$ boson into a fermion-antifermion pair ($f \bar{f}$) has the form
\be \label{eq:3_6}
\langle f \bar{f} | J_Z^\mu | 0\rangle = V_f(q^2) \bar{u}_f \gamma_\mu 
\left[ \frac{I_{3f} \left( 1- \gamma_5\right)}{2} - \hat{k}_f(q^2) \hat{s}^2 Q_f\right] v_f \, ,
\ee
where $V_f(q^2)$, $\hat{k}_f(q^2)$, and its OS counterpart $k_f(q^2)$ are electroweak form factors. $I_{3f}$ and 
$Q_f$ denote the third component of the weak isospin and the charge of fermion $f$.

In terms of the $\hat{k}_f$ and $k_f$ form factors, the neutral current amplitude $\sin^2\theta_W^{\mbox{{\tiny exp}}}(q^2)$ 
discussed by \textcite{Marciano:1980be}, is
\be \label{eq:3_7}
\sin^2\theta_W^{\mbox{{\tiny exp}}}(q^2) \equiv \hat{k}_f (q^2) \hat{s}^2 = k_f (q^2) s^2 \, .
\ee

The $\ms$ and OS definitions of the electroweak mixing angle are related by \cite{Degrassi:1990tu}:
\bea
\hat{s}^2 &=& s^2 \left(1+\frac{c^2}{s^2} \Delta \hat{\rho} \right) \, , \label{eq:3_8} \\
\Delta \hat{\rho} &=& \mbox{Re} \left[ \frac{A_{WW}(M_Z^2)}{M_W^2} - \frac{A_{ZZ}(M_Z^2)}{M_Z^2 \hat{\rho}}\right]_{\ms}  \label{eq:3_9} \, ,
\eea
where $A_{WW}(q^2)$ and $A_{ZZ}(q^2)$ are the $W-W$ and $Z-Z$ transverse self-energies, $\hat{\rho} = (1-\Delta \hat{\rho})^{-1}$, and $\ms$ denotes the $\ms$ renormalization and the choice $\mu=M_Z$.
 
The $\ms$ scheme is also used in the radiative correction program GAPP \cite{ErlerGAPP}, extensively employed by J.~Erler and P.~Langacker in their by-annual reviews of the Electroweak Model and Constraints on New Physics (see, for example, \textcite{Nakamura:2010zzi}).

Early studies of the QCD contributions to EWC include \textcite{DV87, D88, K90, HK91}.
The incorporation of QCD effects in the basic EWC $\Delta \hat{r}$, $\Delta \hat{r}_W$ and $\Delta r$ was implemented by \textcite{Fanchiotti:1992tu} (see
also references cited therein). They include perturbative ${\mathcal O}(\alpha \alpha_s)$ contributions and $t\bar{t}$
threshold effects. Here $\alpha_s$ is evaluated at a relevant mass scale such as $M_Z$ or $M_t$.

\subsection{The Effective Electroweak Mixing Parameter}
\label{sec:3.6}

Another very useful version of the electroweak mixing parameter is $\sef \equiv \sefl$, extensively employed by the Electroweak Working Group (EWWG) to analyze the data at the $Z$ resonance. Here eff. and lept. are abbreviations for effective and leptonic, respectively. It is defined by \cite{Rolandi:1992aj}:
%Martinez:XXXX
\be
1  - 4 \sefl = \frac{g_V^l}{g_A^l} \, ,
\ee
where $g_V^l$ and $g_A^l$ are the effective vector and axial vector couplings of the $Z \to l\bar{l}$ amplitude at resonance ($q^2 = M_Z^2$) and $l$ stands for a charged lepton.

The relations between $\sef$ and $\hat{s}^2$, $s^2$ were obtained by \textcite{Gambino:1993dd}:
\be \label{eq:3_11}
\sef = \mbox{Re}\,\hat{k}_l(M_Z^2) \hat{s}^2 = \mbox{Re}\,k_l(M_Z^2) s^2 \, ,
\ee
where $\hat{k}_l(q^2)$  and $k_l(q^2)$ are the electroweak form factors introduced in  Eq.(\ref{eq:3_6}) and following lines. Because of a fortuitous cancellation of EWC, $\mbox{Re}\,\hat{k}_l(M_Z^2)$ is very close to $1$.
Applying the Marciano-Rosner renormalization prescription (cf. Section~\ref{sec:3.5}), \textcite{Gambino:1993dd} found
\be
\Delta = \sef -\hat{s}^2 \approx 3 \times 10^{-4} \, . \label{eq:3_11p}
\ee
They also pointed out that, if this prescription is not applied, so that the complete  top-quark contribution is included in the calculation of $\hat{k}_l(M_Z^2)$, the difference becomes even smaller, namely $\Delta \approx 1 \times 10^{-4}$ for $M_t = 173.2$~GeV.

Combining  Eq.(\ref{eq:3_4})  and  Eq.(\ref{eq:3_11}), and writing $\mbox{Re} \,\hat{k}_l(M_Z^2) = 1+(\hat{e}^2/\hat{s}^2) \Delta \hat{k}(M_Z^2)$, one finds \cite{Ferroglia:2001cr}:
\bea
\sef \cef &=& \frac{\pi \alpha}{\sqrt{2} G_F M_Z^2 \left(1 - \Dref \right)} \, , \label{eq:3_12} \\
\Dref &=& \Delta \hat{r} + \frac{e^2}{\sef} \Delta \hat{k} \left( 1- \frac{\sef}{\cef}\right) \left( 1+ x_t \right) + 
\cdots \, ,  \label{eq:3_13}
\eea
where $x_t = 3 G_F M_t^2 /(\sqrt{2} \, 8 \pi^2)$ is the leading contribution to $\Delta \hat{\rho}$.  Eq.(\ref{eq:3_13})
includes the complete one-loop EWC, as well as the two-loop contributions enhanced by factors $\left( M_t^2 /M_Z^2\right)^n  \, (n = 1,2)$. We note that  Eq.(\ref{eq:3_12}) has a form analogous to  Eq.(\ref{eq:3_1}) and 
 Eq.(\ref{eq:3_4}). The one-loop approximation to  Eq.(\ref{eq:3_13}) had been previously applied to discuss
the mass scale of new physics in the Higgs-less scenario \cite{Kniehl:1999an}. 

The asymptotic behaviors for large $M_t$, $M_H$, of the basic corrections $\Delta r$, $\Delta \hat{r}$, and 
$\Dref$ are very instructive. At the one-loop level, we have 
\bea
\Delta r &\sim& -\frac{3 \alpha}{16 \pi s^4} \frac{M_t^2}{M_Z^2} + \frac{11 \alpha}{24 \pi s^2} \ln\left( \frac{M_H}{M_Z}\right) + \cdots \, , \label{eq:3_14} \\
\Dref \approx \Delta \hat{r} &\sim& - \frac{3 \alpha}{16 \pi \hat{s}^2 \hat{c}^2} \frac{M_t^2}{M_Z^2} + 
\frac{\alpha}{2 \pi \hat{s}^2 \hat{c}^2} \left( \frac{5}{6} - \frac{3}{4} \hat{c}^2\right)\ln\left( \frac{M_H}{M_Z}\right) + \cdots \, . \label{eq:3_15}
\eea 
 Eqs.(\ref{eq:3_14},\ref{eq:3_15}) reveal a quadratic dependence on $M_t$, a logarithmic dependence on $M_H$. The asymptotic behaviors in $M_t$ and $M_H$ have opposite signs, a fact that helps to explain a well known $M_t-M_H$ correlation, namely increasing (decreasing)  values of $M_t$ favor increasing (decreasing) values of $M_H$.
The cofactor of $M_t^2/ M_Z^2$ in $\Delta r$ is approximately larger by $c^2/s^2 \approx 3.5$ than in $\Delta \hat{r}$,
$\Dref$. This implies that $\Delta r$ is significantly more sensitive to $M_t$ than $\Delta \hat{r}$  and $\Dref$.

The asymptotic behavior for large $M_t$ of the neutral current amplitude is
\be
\mbox{N.~C. ampl.} \sim \frac{G_F}{1-x_t} \, , \label{eq:3_16}
\ee
where $x_t$ is defined after  Eq.(\ref{eq:3_13}).

Additional contributions to $\Delta r$ and $\Dref$ lead to shifts $\delta M_W /M_W \approx -0.205~ \delta(\Delta r)$,
$\delta \sef/\sef \approx 1.52~\delta (\Dref)$.

Current values for the three versions of the electroweak mixing parameter discussed above, are
\be
\sin^2 \theta_W = 0.22290(29)\, ;  \qquad
\sefl = 0.23153(16) \, ; \qquad
\sin^2 \hat{\theta}^2_W (M_Z) = 0.23123(16) \,. \label{eq:3_17}
\ee
The value of $\sin^2 \theta_W$ was obtained using  Eq.(\ref{eq:3_2}) and the experimental values $M_Z = 91.1876(21)$~GeV (\ref{sec:3.3}) and $M_W = 80.385(15)$~GeV (the average of the Tevatron and LEP2 measurements).
The $\sefl$ value is the average of the values obtained from all the asymmetries measured at LEP and at the Stanford Linear Collider (SLC), and dates back to 2005. The value of $\sin^2 \hat{\theta}_W (M_Z)$ has been derived from that of $\sefl$ taking into account  Eq.(\ref{eq:3_11p}).

The values of  the QCD coupling and the $\ms$ fine structure constant  at the scale $\mu= M_Z$ are given by \cite{ErlerPDG, DissertoriPDG} 
\be
\alpha_s(M_Z) = 0.1184(7)\, , \qquad \hat{\alpha}(M_Z) = (127.916(15))^{-1}\, .
\ee

%For completeness, we note that the global fit to all the data in the ST framework \cite{Nakamura:2010zzi} leads to the values $M_W = 80.384(14)$~GeV, %$\sin^2 \theta_W = 0.22292(28)$, $\sin^2 \hat{\theta}_W(M_Z)  = 0.23116(13)$,
%$\sefl = 0.23146(12)$. These values are consistent  with those in  Eq.(\ref{eq:3_17}). On the other hand,
%the errors in the global fit are smaller (as expected), particularly in the case of $M_W$ and $\sin^2 \theta_W$. 

\subsection{Renormalization Schemes: General Observations}
\label{sec:3.7}

As discussed in Sections~\ref{sec:3.4}-\ref{sec:3.6}, the EWC have been evaluated in specific renormalization schemes.
An interesting  feature is that each scheme is associated with a specific definition of the renormalized electroweak mixing parameter.

Two of the most frequently employed schemes are:
\begin{itemize}
\item[i)] The On-Shell (OS) Scheme, discussed in Section~\ref{sec:3.4}. It is very ``physical'',
 since it identifies renormalized couplings and masses with physical, scale-independent observables, such as $G_F$,
 $\alpha$, $M_W$, $M_Z$, $M_H$, $M_f$ \dots It has also provided the framework for very accurate calculations
such as the complete two-loop evaluation of $\Delta r$ and $\sefl$ (cf. Sections~\ref{sec:3.15}). As mentioned in Section~\ref{sec:3.4}, it is used in the ZFITTER and Gfitter programs, extensively employed by the LEP EW and Gfitter Groups in the analysis of the precision electroweak observables.
  
\item[ii)] The $\ms$ Scheme, discussed in Section~\ref{sec:3.5}.   It has very good convergence properties. In fact, in this scheme one essentially subtracts the pole terms, and therefore the calculations follow closely the structure of the unrenormalized theory. In this way it avoids large finite corrections that are often induced by renormalization. It employs scale dependent couplings such as $\alpha(\mu)$, $\hat{s}^2(\mu)$, which play a crucial role in the study of Grand Unification. On the other hand, the use of such couplings generally leads to a residual scale dependence in the evaluation of observables, due to the truncation of the perturbative series.  As explained in Section~\ref{sec:3.5}, it is used in the GAPP program, extensively employed by J.~Erler and P.~Langacker in their by-annual contributions to the Review of Particle Physics.
\end{itemize}

More recently, a novel approach, called the Effective Scheme, was proposed by \textcite{Ferroglia:2001cr}. It employs scale-independent parameters such as $\sef$, $G_F$, $M_W$, $M_Z$ \dots Consequently, the calculation of observables is strictly scale-independent in finite orders of perturbation theory. Furthermore, it shares the good convergence properties of the $\ms$ scheme, a fact that is related to the numerical closeness of $\sef$ and $\hat{s}^2(M_Z)$ (cf.   Eq.(\ref{eq:3_11p})).

The comparative evaluation of the EWC using different renormalization schemes is often very useful, because it provides an estimate of the theoretical error due to the truncation of the perturbative series.

\boldmath
\subsection{The Running of $\alpha(\mu)$ and $\sin^2 \theta_W (\mu)$}
\unboldmath
\label{sec:3.8}

A very important contribution to the EWC is associated with the running of $\alpha$ to the $M_Z$ scale via vacuum polarization contributions, an effect usually parametrized as
\be
\frac{\alpha(M_Z)}{\alpha} = \frac{1}{1-\Delta \alpha} \, . \label{eq:3_18}
\ee
The leptonic contribution is
\be
\Delta \alpha_l = 314.97686 \times 10^{-4} \simeq 0.03150 \, . \label{eq:3_19}
\ee
This result includes three-loop contributions evaluated  by \textcite{Steinhauser:1998rq}. 
The contribution of the five lightest quarks ($u-b$) is evaluated using dispersion relations involving the experimental cross-section for $e^+ e^- \rightarrow$~hadrons at low $\sqrt{s}$ and perturbative QCD (PQCD) at large $\sqrt{s}$.
Important studies of these effects were carried out by  \textcite{EJ95} and \textcite{Jegerlehner:2001wq} (and references cited therein). Recent, accurate values include 
$\Delta \alpha_h^{(5)} = 0.02750(33)$ \cite{Burkhardt:2011ur}, $\Delta \alpha_h^{(5)} = 0.027626(138)$ \cite{Hagiwara:2011af}, and $\Delta \alpha_h^{(5)} = 0.02757(10)$ \cite{Davier:2010nc}. The smaller error in the last reference is partly due to the use of PQCD in the $\sqrt{s}$ range  between $1.8$~GeV and $3.7$~GeV.
Combining the result obtained in that reference with $\Delta \alpha_l$ (cf.  Eq.(\ref{eq:3_19})),  one finds the accurate value
\be
\Delta \alpha = \Delta \alpha_l + \Delta \alpha_h^{(5)} = 0.05907(10) \, . \label{eq:3_20}
\ee 
 Eq.(\ref{eq:3_20}) does not include the top-quark contribution,  which is evaluated perturbatively, amounts to 
\be
\Delta \alpha_{\rm top} = -0.72 \times 10^{-4} \, ,
\ee
 and is usually taken into account together with other $M_t$-dependent EWC.

Running versions of the electroweak mixing parameter were proposed by \textcite{Czarnecki:2000ic} and by 
\textcite{Ferroglia:2003wa}. For $q^2 < 0$, the first authors define
\be
\sin^2 \theta_W (Q^2) = \kappa(Q^2) \sin^2 \hat{\theta}_W (M_Z) \, , \label{eq:3_21}
\ee 
where $Q^2 = -q^2$ and $\kappa(Q^2)$ is identified with the $\ms$ form factor $\hat{k}_e(-Q^2)$ (cf.  Eq.(\ref{eq:3_6}) in the case $f=e$). \textcite{Czarnecki:2000ic} found  that the EWC lead to $\kappa(0) = 1.0301 \pm 0.0025$,
and pointed out that this $+ 3 \%$ increase in the value of the electroweak mixing parameter, 
appropriate for low $Q^2$, gives rise to a $38 \%$ reduction in the left-right polarization asymmetry  $A_{\rm LR}$ in M\o ller scattering! The reason is that $A_{\rm LR}$ is proportional to $1 -  4 \sin^2 \theta_W(Q^2)$, a factor close to zero, and a small shift in the value of the electroweak mixing parameter has a very pronounced effect. In the same work, $\sin^2 \theta_W(Q^2)$ was evaluated and displayed over a large range $0 \le Q \le 1$~TeV, where $Q \equiv \sqrt{Q^2}$. 

%The analysis includes a comparison with existing experiments, as well as predictions of $\sin^2 \theta_W(Q^2)$ that can be probed by potential future 
%M\o ller and $e^+ e^-$ asymmetry measurements at high energies. 
%In particular, at $Q \approx 100$~GeV, $\sin^2 \theta_W(Q^2)$ is in good agreement with the value of the electroweak mixing parameter derived  from the %$Z$-resonance observables.

\textcite{Ferroglia:2003wa} proposed an alternative ``running'' version of the electroweak mixing parameter. Specifically, they define 
\be
\sin^2 \hat{\theta}_W (q^2) = \left(1- \frac{\hat{c}}{\hat{s}} \frac{a_{\gamma Z} (q^2, M_Z)}{q^2} \right) \hat{s}^2 \, ,
\label{eq:3_22}
\ee
where $\hat{s}^2 = \sin^2 \hat{\theta}_W (M_Z)$ (cf. Section~\ref{sec:3.5}) and $a_{\gamma Z}(q^2,M_Z)$ is the``Pinch Technique'' (PT) $\gamma Z$ self-energy evaluated at $\mu = M_Z$.

We recall that the Pinch Technique \cite{Cornwall:1981xx, Cornwall:1981zr, Cornwall:1989fg, Papavassiliou:1990xx} is a prescription that combines the conventional self-energies with ``pinch parts'' from vertex and box diagrams in such a manner that the modified self-energies are gauge-independent and are endowed with very desirable theoretical properties. The PT self-energies in the electroweak sector of the ST were derived by
\textcite{Degrassi:1992ue}. In the same paper it was shown that the ``pinch parts'' can be identified with amplitudes involving appropriate equal-time commutators of currents, which explains the fact that they are process independent and are not affected by the strong interactions. \textcite{Ferroglia:2003wa} evaluated and displayed $\sin^2 \hat{\theta}_W (q^2)$ both in the space-like ($q^2 < 0$) and time-like ($q^2 >0$) domains, appropriate to $e^--e^-$ and 
$e^+-e^-$ colliders, respectively. In the second case $a_{\gamma Z}(q^2, M_Z)$ is generally complex and $\sin^2 \hat{\theta}_W (q^2)$ is defined by the real part of the r.h.s. of  Eq.(\ref{eq:3_22}).

Interestingly, the authors obtained 
\be
1 - 4 \sin^2 \hat{\theta}_W (0) = 0.0452 \pm 0.023 \, , \label{eq:3_23}
\ee
which is very close to $0.0450 \pm 0.023 \pm 0.0010$, the result previously found by \textcite{Czarnecki:1996fw}
for the complete one-loop EWC to $A_{\rm LR}$ at $Q^2 = 0.025~{\rm GeV}^2$ and $y\equiv Q^2/s = 1/2$, appropriate to the SLAC experiment E158 \cite{Kumar:1995ym}. Setting $q^2 = M_Z^2$ in  Eq.(\ref{eq:3_22}), one finds $\sin^2 \hat{\theta}_W (M_Z^2) = 0.23048$, which is lower than $\sefl = 0.23153$ by $0.45 \%$; although not in precise agreement, the two parameters are rather close.

It is also interesting to note that both the running of $\alpha$ and of the weak mixing angle have been derived directly in the $\ms$ scheme (\textcite{E99, ERM05}, respectively).

\boldmath
\subsection{The $M_t$ Prediction}
\unboldmath

An important example of the successful interplay between theory and experiment was the prediction of the top quark mass $M_t$ and its subsequent measurement.

Before 1995, the top quark could not be produced directly, but it was possible to estimate its mass because of its virtual contributions to the EWC. In 1994, a global analysis by the EWWG led to the indirect determination \cite{Pietrzyk:1994A} 
\be
M_t = 177 \pm 11^{+18}_{-19}~{\rm GeV}\, , \label{eq:3_24}
\ee
where the central value corresponds to $M_H = 300$~GeV , the first error is experimental, and the second represents the shift in the central value assuming
$M_H = 65~{\rm GeV} \, (-19~{\rm GeV})$, or $M_H = 1~{\rm TeV} \,  (+18~{\rm GeV}) $.
%\end{displaymath}
%
This can be compared with the current experimental value $M_t = 173.2 \pm 0.9$~GeV \cite{TEWWGMT}.

This successful prediction was possible because of the very sensitive $M_t$-dependence of the basic EWC (cf., for example,   Eqs.(\ref{eq:3_14},\ref{eq:3_15})).

\subsection{Evidence for Electroweak Corrections}
\label{sec:sub3_10}
\begin{itemize}
\item[a)] \emph{Evidence for EWC beyond the running of $\alpha$} \cite{Sirlin:1993pu}. Using the experimental values of $\alpha$, $G_F$, $M_Z$, $M_W = 80.385 \pm 0.015$~GeV \cite{TEWWGMW, EWWG}, and  Eqs.(\ref{eq:3_1},\ref{eq:3_2}), one finds
\be
\left( \Delta r\right)_{\rm exp} = 0.03506 \pm 0.00090 \, . \label{eq:3_25}
\ee
The contribution to $\Delta r$ from the running of $\alpha$ is $\Delta \alpha = 0.05907 \pm 0.00010$ (cf.  Eq.\ref{eq:3_20})). Thus, the contribution to $\Delta r$ beyond the running of $\alpha$ is
\be
\left( \Delta r\right)_{\rm exp} - \Delta \alpha = -0.02401 \pm 0.00091 \, , \label{eq:3_26}
\ee
which differs from $0$ by $26~\sigma$!

An alternative argument is to compare the values of $\sefl = 0.23153 \pm 0.00016$ and $\sin^2 \theta_W = 0.22290 \pm 0.00029$ (cf.  Eq.(\ref{eq:3_17}) and following discussion). The difference is $0.00863 \pm 0.00033$, also $26~\sigma$, and it arises from EWC not including $\Delta \alpha$. Indeed, the difference is mainly due to the EWC $c^2 \Delta \hat{\rho}$ in  Eq.(\ref{eq:3_8}).

\item[b)] \emph{Evidence for Electroweak Bosonic Corrections (EWBC)} \cite{Gambino:1994vk}. They include loops involving the bosonic sector: $W$, $Z$, $H$ and unphysical scalars. They are subleading numerically relative to the fermionic contributions, but very important conceptually. Strong evidence for the EWBC can be found by measuring
$\Dref$. Using the experimental values of $\alpha$, $G_F$, $M_Z$, $\sefl$ and   Eq.(\ref{eq:3_12}) one finds:
\be
\left( \Dref \right)_{\rm exp}  = 0.06059 \pm 0.00045 \, . \label{eq:3_27}
\ee

Subtracting the contribution of the EWBC from the theoretical expression for $\Dref$ given in Eq.~(\ref{eq:3_13}), but retaining the fermionic EWC, the theoretical value is $\left( \Dref \right)_{\rm theor.} ^{(f)} = 0.05045 \pm 0.00056$. The difference $\left( \Dref \right)_{\rm exp} - \left( \Dref \right)_{\rm theor.} ^{(f)} = 0.01014 \pm 0.00072$ provides an estimate of the EWBC to $\Dref$. Thus, they differ from $0$ by $14~\sigma$! 

\end{itemize}

\subsection{Precise Test of CKM Unitarity}
\label{sec:sub3_11}

Since the Cabibbo-Kobayashi-Maskawa (CKM) matrix $V_{ij}$ is unitary, a fundamental prediction is that
\be
\sum_j |V_{ij}|^2 = 1 \, , \qquad  \sum_i |V_{ij}|^2 = 1 \, . \label{eq:3_28}
\ee

In particular, in the three-generation case, the elements of the first row must satisfy the equality
\be
|V_{ud}|^2 + |V_{us}|^2 +|V_{ub}|^2 = 1 \, . \label{eq:3_29}
\ee
$|V_{ud}|^2$, the dominant term in  Eq.(\ref{eq:3_29}), is obtained most precisely from the $0^+ \to 0^+$ Superallowed Fermi Transitions in $\beta$-decay. Using the Current Algebra formulation to evaluate the ${\mathcal O}(\alpha)$ EWC in the Standard Theory, one finds the following expression for the probability of these important transitions
\cite{Sirlin:1977sv}:
\bea
P d^3p &=& P^0 d^3 p \left\{ 1+\frac{\alpha}{2 \pi} \left[3 \ln\left(\frac{M_Z}{m_p} \right) +g \left(E,E_m \right)
+6 \bar{Q}  \ln\left(\frac{M_Z}{M} \right) +2 C +{\mathcal A}_{\bar{g}} \right] \right\}\, , \label{eq:3_30} \\
P^0 d^3 p &=& \frac{G_F^2 (V_{ud})^2}{8 \pi^4} |M_F|^2 F\left(Z,E\right) (E_m-E)^2 d^3 p \, ,
\label{eq:3_31}
\eea
where $p$, $E$, and $E_m$ are the momentum,  energy and end-point energy of the electron or positron in the decay, $F\left(Z,E\right)$ is the Fermi Coulomb function, $g \left(E,E_m \right)$ is defined in  Eq.(\ref{eqtwentyoneb}) in Section~\ref{sec:2.7}, and $M_F$ is the matrix element of the weak hadronic vector current between the initial and final nuclei. For iso-triplet transitions $|M_F|^2 = 2$.  

The terms between square brackets in  Eq.(\ref{eq:3_30}) represent the ${\mathcal O}(\alpha)$ corrections not contained in $F(Z,E)$ in the approximation of neglecting contributions of ${\mathcal O} ((\alpha/\pi) E/m_p)$. The first two terms in that expression arise from the weak hadronic vector current and are not affected by the strong interactions
(S.I.). In particular, the proton mass $m_p$ cancels in the sum. We recall that the function $g(E,E_m)$ describes the ${\mathcal O}(\alpha)$ radiative corrections to the electron or positron spectrum in $\beta$ decay in the presence of the 
S.I. (cf. Section~\ref{sec:2.7}). The third term is a short-distance contribution to the Fermi amplitude arising from the weak hadronic axial vector current and $\bar{Q}$ is the average charge of the fundamental doublet involved in the transition. In the ST this is the $u-d$ doublet and $\bar{Q} = (2/3-1/3)/2 = 1/6$. 
$M$ is a hadronic mass of ${\mathcal O}(1~{\rm GeV})$.
The $2 C$ term is the corresponding non-asymptotic part and ${\mathcal A}_{\bar{g}} \approx -0.34$ is a very small asymptotic QCD contribution proportional to $\alpha_s(M_Z)$. Although the axial vector current does not contribute to the Fermi amplitude at the 
tree level, we see that it gives rise to an important EWC in  Eq.(\ref{eq:3_30}). 

The EWC to $\beta$ decay are dominated by a large logarithmic term, $(3 \alpha/2 \pi) \ln (M_Z/2 E_m)$. For example, in the superallowed ${}^{14} {\rm O}$ decay, $E_m = 2.3$~MeV, and this contribution amounts to $3.4 \%$. As we will see, such large correction is phenomenologically crucial to verify  Eq.(\ref{eq:3_29}). As mentioned in Section~\ref{sec:3.2}, this result was one of the early ``smoking guns'' of   the EW sector of the ST at the level of its quantum corrections.

Contributions of ${\mathcal O}(Z \alpha^2)$ and ${\mathcal O}(Z^2 \alpha^3)$ are denoted by $\delta_2$  and $\delta_3$. In particular, in the mid-eighties a re-evaluation of $\delta_2$ played an important role  in the test of the conserved vector current (CVC) hypothesis. In fact, at the time the analysis of eight accurately measured Superallowed Fermi transactions showed a significant departure from CVC expectations. Simple theoretical arguments strongly suggested that the problem arose in the evaluation of the two-loop $\delta_2$ that had been done numerically long before. The correction was then evaluated analytically by \textcite{Sirlin:1986cc} and \textcite{Sirlin:1987sy} and, when applied to the eight transitions, led to very good agreement with CVC, a result confirmed by a new numerical evaluation \cite{Jaus:1986te}.
One finds that $\delta_2$ varies from $0.22 \%$ for ${}^{14}{\rm O}$ decay to $0.50 \%$ for the ${}^{54}{\rm Co}$
transition, while $\delta_3$ is much smaller \cite{Jaus:1986te}.

There is also a correction $\delta_c$ that describes the lack of perfect overlap between the wavefunctions of the parent and daughter nuclei due to Coulomb forces and configuration mixing effects in the shell-model wavefunctions, as well as a nuclear-structure-dependent correction $\delta_{\rm NS}$. They have been extensively discussed in the literature (see \textcite{Towner:2007np} and references cited therein).

Over the years, a number of refinements have been incorporated in the evaluation of the EWC. For example leading logarithmic contributions ${\mathcal O}(\alpha^n \ln^n(M_Z/m_p))$ and ${\mathcal O}(\alpha^n \ln^n(m_p/2 E_m))$
have been summed via a renormalization group analysis by \textcite{Marciano:1986pd} and  \textcite{Czarnecki:2004cw}. They lead to the replacements
\bea
1+ \left(\frac{2 \alpha}{\pi}\right) \ln\left(\frac{M_Z}{m_p} \right) &\rightarrow& S\left(m_p, M_Z\right) = 1.02248 \, ,
\label{eq:3_32} \\
1+ \left(\frac{3 \alpha}{2 \pi} \right) \ln \left(\frac{m_p}{2 E_m} \right) &\rightarrow& L\left( 2 E_m, m_p\right) \, ,
\label{eq:3_33}
\eea
where $(3 \alpha/2 \pi) \ln(m_p/ 2 E_m)$ is a leading contribution to $g(E, E_m)$. In the case of  neutron $\beta$ decay, for example, $L\left( 2 E_m, m_p\right) = 1.02094$. \textcite{Sirlin:1981ie} showed that all semileptonic processes mediated by the $W$ boson are enhanced by a short-distance EWC analogous to  Eq.(\ref{eq:3_32}), namely of the form $1+ (2 \alpha/\pi) \ln(M_Z/M) \rightarrow S(M,M_Z)$, where $M$ is a relevant hadronic mass.
Interesting examples include the hadronic decays of the $\tau$ \cite{MS88}, $\pi_{l2}$ decays \cite{MS93}, and muon capture \cite{Czarnecki:2007th}, where short-distance effects of this type play a very important role in the EWC. More recently, 
 \textcite{Marciano:2005ec}
developed a new method to compute hadronic effects on EWC to low-energy weak interaction semileptonic processes.
It employs high order perturbative QCD results originally derived for the Bjorken sum rule for polarized electroproduction, as well as a large $N$ QCD -motivated interpolating function that matches long and short distance EWC. When applied to the Superallowed Fermi transitions, it improves the evaluation of the axial vector current contribution in  Eq.(\ref{eq:3_30}) and reduces by a factor of $2$ the theoretical loop uncertainty in the extraction of $V_{ud}$.

A critical survey \cite{Hardy:2008gy} examines 20 Superallowed $0^+ \rightarrow 0^+$ $\beta$ decays. The analysis leads to the evaluation of the ${\mathcal F}t$ values for the 20 transitions, where ${\mathcal F}$ is a phase space factor that includes the Fermi Coulomb function, the electroweak corrections and the nuclear corrections $\delta_c$
and $\delta_{\rm NS}$, and $t$ is the partial half-life.

The CVC hypothesis predicts that the ${\mathcal F}t$ values should be the same for all these transitions, a demanding test that is very well satisfied by the results. For the weighted average of the 13 most accurate ${\mathcal F}t$ values
(those with errors less than $\pm 0.4 \%$) the authors obtain $\overline{{\mathcal F}t} = 3071.87 \pm 0.83\, s$, a result that leads to the important determination
\be
|V_{ud}| = 0.97425(22) \, . \label{eq:3_34} 
\ee
The value of $|V_{us}|$ can be determined from $K_{l3}$ decays and that of $|V_{us}|/|V_{ud}|$ from the ratio
of $K^+ \to \mu^+ \nu$ and $\pi^+ \to \mu^+ \nu$ decay rates. Combining the two inputs the authors find
\be
|V_{us}| = 0.22534(93) \, . \label{eq:3_35} 
\ee

Inserting  Eqs.(\ref{eq:3_34},\ref{eq:3_35}) and $|V_{ub}| =(3.93\pm0.35)\times 10^{-3}$ \cite{Amsler:2008zzb},
\textcite{Hardy:2008gy} obtain
\be
|V_{ud}|^2 +|V_{us}|^2+ |V_{ub}|^2 = 0.99995(61) \, , \label{eq:3_36}
\ee
an impressive $0.06 \%$ test of the three generation ST at the level of its quantum corrections. It is interesting to note that the overall EWC to  Eq.(\ref{eq:3_36}) are of ${\mathcal O}(4 \%)$, i.~e. 66 times larger than the $0.061 \%$ error!

EWC of ${\mathcal O}(\alpha)$ to neutron $\beta$ decay were included in the classic work of \textcite{Wilkinson:1982hu}. More recently, a number of refinements were introduced by \textcite{Czarnecki:2004cw}. Since the axial vector current is not conserved, in the case of the Gamow-Teller amplitude the Current Algebra analysis of the  EWC does not lead to a simple expression, independent of the S.I., in contrast with the corrections involving the vector current (cf.  Eq.(\ref{eq:3_30}) and the discussion following that equation). The strategy followed by the authors was to define $g_A = G_A/G_V$ ($G_V \equiv G_F |V_{ud}|$) in terms of the neutron lifetime $\tau_n$ by means of the expression
\be
\frac{1}{\tau_n} = \frac{G_F^2 |V_{ud}|^2 }{2 \pi^3} m_e^5 \left(1+3 g_A^2 \right) f\left(1+ {\rm RC} \right) \,,
\label{eq:3_37}
\ee
where $f = 1.6887$ is a phase space factor that includes the Coulomb Fermi function contribution, as well as smaller corrections, and $(1+{\rm RC})$ is identified with the well known EWC involving the vector current. This implies that some EWC  are absorbed in this definition of $g_A$ and, therefore, $G_A$. An interesting point is that the correction $1+(\alpha/2 \pi) g(E,E_m)$ \cite{Sirlin:1967zza} and the short distance contribution $1+2 (\alpha/\pi)
\ln(M_Z/m_p)$ \cite{Sirlin:1981ie} affect both the Fermi and Gamow-Teller transitions, so they are well described by the factorization of the EWC in  Eq.(\ref{eq:3_37}). It follows that the same is true of the large logarithmic term
$(3 \alpha/2 \pi) \ln(m_p/2 E_m)$ contained in $(\alpha/2 \pi) g(E,E_m)$.

The authors proceeded then to evaluate a number of higher order EWC to  Eq.(\ref{eq:3_37}): the sum of the corrections 
of ${\mathcal O}(\alpha^n \ln^n(M_Z/m_p))$ and ${\mathcal O}(\alpha^n \ln^n(m_p/2 E_m))$ according to  Eqs.(\ref{eq:3_32},\ref{eq:3_33}), the contribution of $\delta_2$, and next-to leading log  corrections of ${\mathcal O}(\alpha^2 \ln{(M_Z/m_p)})$ and ${\mathcal O}(\alpha^2 \ln{(m_p/m_f)})$ arising from fermion vacuum polarization insertions in loops with photon propagators. The analysis led to the useful relation
\be
|V_{ud}|^2 (1 +3 g^2_A) \tau_n = 4908 \pm 4~{\rm s} \, . \label{eq:3_38}
\ee
Using the experimental averages $\tau_n = 885.7(7)$~s and $g_A = 1.2720(18)$ (from the polarized neutron decay asymmetry),  Eq.(\ref{eq:3_38}) leads to
\be
|V_{ud}| = 0.9729(12) \qquad (\mbox{neutron decay}) \, , \label{eq:3_39}
\ee
which is consistent with  Eq.(\ref{eq:3_34}), but much less precise.  Eq.(\ref{eq:3_38}) can also be applied to calculate $g_A$ using the accurate $|V_{ud}|$ value from the Superallowed Fermi transitions and the experimental value of 
$\tau_n$. In this way the authors obtained the precise prediction
\be
g_A = 1.2703(8) \, , \label{eq:3_40}
\ee 
which was compared with the experimental values derived from the asymmetry.

\vspace*{2mm}
Over the years, the test of unitarity of the CKM matrix  shown in  Eq.(\ref{eq:3_29}) has been used to set bounds on certain forms of new physics.
The strategy is to attribute to the new physics the deviation from unity  of the experimental value of $\sum_{i=1}^3 |V_{u i}|^2$, so that exact CKM unitarity is satisfied.
See, for example, \textcite{Sirlin95L}.
\begin{itemize}
\item[i)] \emph{4th generation}. For a long time, the determination of $\sum_{i=1}^3 |V_{u i}|^2$ led to values smaller than unity by about $2 \sigma$. This suggested the possibility of a fourth generation \cite{Marciano:1986pd} and the derivation of an upper bound for $V_{u b'}$, where $b'$  denotes the additional down quark. Since the current result (Eq.(\ref{eq:3_36})) is in excellent agreement with 3-generation unitarity, at present this test does not provide a signal for a fourth generation. Nonetheless, if a fourth generation exists, from  Eq.(\ref{eq:3_36}) one finds $|V_{u b'}|\leq 0.03~(90 \% {\rm CL})$,
which is not very restrictive since $|V_{ub}| \simeq 4 \times 10^{-3}$.
\item[ii)] \emph{$Z'$ bosons}. In some models with additional $U(1)$ factors, the new $Z'$ bosons have different couplings to quarks and leptons and, consequently,  give rise to EWC involving box diagrams that distinguish $\mu$ and semileptonic decays \cite{MS1987}. As a consequence, the experimental value of $\sum_{i=1}^3 |V_{u i}|^2$  is modified by a contribution that depends on the ratio $M_{Z'}/M_W$, where $M_{Z'}$ is the $Z'$ mass. The analysis leads to lower bounds for $M_{Z'}$. Typically, they are of the order of a few hundreds GeV and are not competitive with the bounds from direct searches, precision electroweak data and Atomic Parity Violation \cite{ErlerPDG, ELMR, dAdBPV}, which are of ${\mathcal O}(1~{\rm TeV})$.
\item[iii)] \emph{Compositness}. It is frequently discussed in terms of a residual four-fermion interaction with a coupling $1/\Lambda^2$, where $\Lambda$ represents the composite mass scale. If we assume that the new interaction involves only particles of the same generation, it would affect $\beta$ transitions but not muon decay. If we further assume that it is of the form of  Eq.(\ref{eqseventeen}) with $G_V/\sqrt{2}$ replaced by $1/\Lambda^2$,
$G_V^2/G_{\mu}^2 = V_{u d}^2$  is modified to $ V_{ud}^2 (1+ 2 \sqrt{2}/ (G_V \Lambda^2))$. Using  Eq.(\ref{eq:3_36}) one then obtains the bound
$2 \sqrt{2} V_{ud}/(G_\mu \Lambda^2) < 9.7 \times 10^{-4}$ or $\Lambda > 16~ {\rm TeV} ~(90 \% {\rm CL})$.
\item[iv)] \emph{Left-Right Symmetry}. In the``manifest'' left-right symmetry models \cite{BBMS}, there are two small parameters: the mixing angle $\zeta$  that 
relates the $W_1$ and $W_2$ mass eigenstates to the left and right handed fields $W_L$ and $W_R$, and $\delta = (m_1/m_2)^2$, where
$m_i$ ($i=1,2$) are the corresponding masses. 
Corrections linear in the small parameter $\delta$ contribute to $G_V$ and $G_{\mu}$, but cancel in their ratio. 
This can be shown using the results of \textcite{BBMS}. 
In particular, if terms of second and higher order in the small parameters $\zeta$ and $\delta$ are neglected, one finds
$G_V/G_{\mu} = (1-\zeta) V_{ud}$, with analogous shifts for the other semileptonic decays. (For other predictions in the manifest left-right symmetric model, see also \textcite{HT}.) As a consequence,  Eq.(\ref{eq:3_36}) becomes
\be
\sum_{j=1}^3 |V_{uj}|^2 = 0.99995 \pm 0.00061 + 2 \zeta (V_{u d})^2 \,.
\ee
Thus, CKM unitarity (Eq.(\ref{eq:3_29})) leads to 
\be
\zeta = (0.3 \pm 3.2) \times 10^{-4}\,. 
\ee

\end{itemize}

\subsection{Electroweak Corrections to Muon Capture}
\label{sec:3.12}

The study of muon capture by nuclei, $\mu^-  N \to \nu_\mu N'$,  has played an important role in the development of Weak Interaction Physics. See, for example, \textcite{Primakoff:1959fs}, \textcite{Mukhopadhyay:1976hu}, and \textcite{Gorringe:2002xx}.

In 2007, the MuCap collaboration \cite{Andreev:2007wg} reported a precise measurement of the $1S$ singlet capture rate in hydrogen:
\be
\Gamma\left( \mu^- p \to \nu_\mu n \right)^{\rm singlet}_{1S} = 725.0 \pm 13.7 \pm 10.7 \, s^{-1} \, . 
\label{eq:3_41} 
\ee
A major aim of the experiment is an accurate determination of the induced pseudoscalar coupling $g_{{\rm P}}(q^2)$ in the matrix element of the axial vector current between nucleon states:
\be
\langle n| A_\alpha | p \rangle = \bar{u}_n (p_2) \left[g_A (q^2) \gamma_\alpha \gamma_5 + g_{P}(q^2) \frac{q_\alpha}{m_\mu} \gamma_5 \right] u_p(p_1) \, ,
\label{eq:3_42} 
\ee
where $q = p_2-p_1$. On the theoretical side, PCAC (partially conserved axial current) and chiral perturbation theory predict (\textcite{Kaiser:2003dr} and references cited therein):
\be
g_{P}(q_0^2) = 8.2 \pm 0.2 \, ,
\label{eq:3_43} 
\ee
where $q_0^2 = - 0.88 m_\mu^2$, as appropriate for $\mu^-$ capture in H.
Comparing  Eq.(\ref{eq:3_41}) with the theoretical expression used at the time for the capture rate (which did not take into account the EWC), it was found that $g_{P}^{{\rm exp}}(q_0^2) = 6.0 \pm 1.2$, which is about $2 \sigma$ below the prediction in  Eq.(\ref{eq:3_43}).

In order to advance the theory of muon capture to a higher level of precision, \textcite{Czarnecki:2007th} incorporated the EWC in the theoretical expression for the capture rates. They found that they enhance  the capture rates for ${\rm H}$ and ${}^3{\rm He}$ by $2.8 \%$ and $3.0 \%$, respectively. It turns out that the $g_{P}$ values extracted by comparing the theoretical and experimental results are very sensitive to the effect of the EWC. In fact, in the case of ${\rm H}$, when the EWC are included, the authors found
\be
g_{P}^{{\rm exp}}(q_0^2) = 7.3 \pm 1.2 \qquad ({\rm H})\, , \label{eq:3_44}
\ee 
an increase of $g_{P}^{{\rm exp}}(q_0^2)$ by about $+ 22\%$! Furthermore,  Eq.(\ref{eq:3_44}) agrees, within the error, with the theoretical prediction of  Eq.(\ref{eq:3_43}). The implications of the EWC in the case of ${}^3{\rm He}$
capture, $\mu^- \,{}^3{\rm He} \to \nu \,  {}^3{\rm H}$, were also analyzed.

\subsection{Electroweak Corrections to Neutrino-Lepton Scattering}
\label{sec:3.13}

Before the advent of the ST, the QED corrections to the process $\nu_e + e \to \nu_e +e$ were studied by \textcite{Lee:1964jq} and \textcite{Ram:1967zz}. After the emergence of the ST, neutrino-lepton scattering became a subject of special interest. Aside from the fact that they are fundamental processes, they provide instructive and interesting examples of scattering reactions in the Weak Interactions. In particular, their theory is relatively simple: at the tree level, they are not affected by the strong interactions and, at the one-loop EW level, they are less sensitive to strong interactions than $\nu N$ and $e N$ scattering, and $e^+ + e^- \to f + \bar{f}$ annihilation.

Including the EW and QED corrections of ${\mathcal O}(\alpha)$, and using the $\ms$ scheme of renormalization, the differential cross-section for $\nu_\mu + e \to \nu_\mu + e$ is given by \cite{Sarantakos:1983bp}:
\begin{align}
\frac{d \sigma}{d z} &= \frac{2 G_F^2\left( \rho_{\rm N.C.}^{(\nu;l)} \right)^2 (p_1\cdot p_2)}{\pi \left(1-\frac{q^2}{M_Z^2} \right)^2} \Biggl\{ 
\varepsilon_-^2(q^2) \left[ 1 + \frac{\alpha}{\pi} f_-(z)\right] + \varepsilon_+^2 (q^2) (1-z)^2 \left[ 1 + \frac{\alpha}{\pi} 
f_+(z) \right] 
\nonumber \\
& 
- \varepsilon_+(q^2) \varepsilon_-(q^2) \frac{m_e^2}{(p_1\cdot p_2)} z \left[ 1 + \frac{\alpha}{\pi}f_{+-}(z)\right]
\Biggr\} \, , \label{eq:3_45}
\end{align}
where $p_1$ is the four-momentum of the incident neutrino, $p_2$ and $p_2'$ the four-momenta of the initial and final electrons, $q^2 =  (p_2 - p_2')^2$, 
\bea
z &=& -\frac{q^2}{2 (p_1 \cdot p_2)} = \frac{E_e'-m_e}{E_\nu}  \, ,\label{eq:3_46}\\
\rho_{\rm N.C.}^{(\nu;l)}  &=& 1 + \frac{\hat{\alpha}}{4 \pi \hat{s}^2} \Biggl\{\frac{3}{4 \hat{s}^2} \ln c^2 - \frac{7}{4}
+\frac{2\hat{c}_Z}{\hat{c}^2}  +\frac{3}{4} \xi  
\left[ \frac{\ln(c^2/\xi)}{c^2 -\xi} + \frac{1}{c^2} \frac{\ln\xi}{1-\xi} \right] +\frac{3}{4} \frac{M_t^2}{M_W^2}
\Biggr\} \, , \label{eq:3_47} \\
\varepsilon_-(q^2) &=& \frac{1}{2} \left(1- 2 \hat{\kappa}^{(\nu_\mu;l)}(q^2) \hat{s}^2 \right) \, , \label{eq:3_48} \\
\varepsilon_+(q^2) &=& - \hat{\kappa}^{(\nu_\mu;l)}(q^2) \hat{s}^2 \, , \label{eq:3_49} \\
\hat{\kappa}^{(\nu_\mu;l)}(q^2) &=& 1 -\frac{\alpha}{2 \pi \hat{s}^2} \left[ \sum_i \left( C_{3i} Q_i - 4 \hat{s}^2 Q_i^2\right)J_i(q^2) - 2 J_{\mu}(q^2) +\ln c \left(\frac{1}{2} - 7 \hat{c}^2 \right) + \frac{\hat{c}^2}{3} + \frac{1}{2} +\frac{\hat{c}_\gamma}{\hat{c}^2}\right] \, ,\label{eq:3_50}\\
J_i(q^2) &=& \int_0^1 dx\, x (1-x) \ln\left( \frac{m_i^2 -q^2 x (1-x)}{M_Z^2}\right) \, , \label{eq:3_51}\\
\hat{c}_Z &=& \frac{19}{8} -\frac{7}{2} \hat{s}^2 + 3 \hat{s}^4 \, , \label{eq:3_52}\\
\hat{c}_\gamma &=& \frac{19}{8} -\frac{17}{4} \hat{s}^2 + 3 \hat{s}^4 \, . \label{eq:3_53}
\eea
In these expressions terms of ${\mathcal O} (\alpha q^2/M_Z^2)$ have been neglected. In this approximation, we see from  Eq.(\ref{eq:3_47}) that $\rho_{\rm N.C.}^{(\nu;l)}$ is independent of $q^2$. It is also independent of the $\nu$ and charged lepton flavors. In contrast, $\hat{\kappa}^{(\nu_\mu;l)}(q^2)$ depends on $q^2$. It also depends on the incident neutrino flavor via the term $- 2 J_{\mu}(q^2)$ in  Eq.(\ref{eq:3_50}) (which arises from the ``$\nu_\mu$ charge radius'' diagrams). As in previous Sections, $s^2 =1-c^2 = \sin^2 \theta_W$ (cf.  Eq.(\ref{eq:3_2})) and $\hat{s}^2 = 1- \hat{c}^2 = \sin^2 \hat{\theta}_W (M_Z)$ (cf. Section~\ref{sec:3.5}). In  Eq.(\ref{eq:3_47}), $\hat{\alpha} = \hat{\alpha}(M_Z) \simeq 1/127.9$ is the $\ms$ QED coupling at scale $\mu = M_Z$ and $\xi = M_H^2/M_Z^2$. In  Eq.(\ref{eq:3_50}), the sum is over the charged leptons and quarks (in the quark sector $\sum_i = 3 \sum_f$ where $f$ denotes the flavors and the factor 3 represents the color degrees of freedom), and $m_i$, $Q_i$ and $C_{3i}$ are the mass, charge (in units of the proton charge $e_p$) and twice the third component of the weak isospin  of the $i$th fermion, respectively.
In  Eq.(\ref{eq:3_46}), $E'_e$ and $E_\nu$ are the energies of the outgoing electron and the incident neutrino in the rest frame of the incoming electron. Thus, in that frame, $z = T/E_\nu$, where $T$ is the kinetic energy of the scattered electron.

The expressions for $\hat{\kappa}^{(\nu_\mu;l)}(q^2)$ in  Eq.(\ref{eq:3_50}) and $J_i(q^2)$ in  Eq.(\ref{eq:3_51}) have been updated from the paper by \textcite{Sarantakos:1983bp} to take into account the use of $\sin^2 \hat{\theta}_W (M_Z)$
in  Eqs.(\ref{eq:3_48},\ref{eq:3_49}), while the early work employed $\sin^2 \hat{\theta}_W(M_W)$.

The functions $f_-(z)$, $f_+(z)$ and $f_{+-}(z)$ in  Eq.(\ref{eq:3_45}) describe QED corrections. The two first functions have been evaluated analytically in the relativistic approximation \cite{Sarantakos:1983bp}, assuming $m_e/E_e$, $m_e/E_\nu$ and $m_e/(E_{{\rm max}} -E_e) \ll 1$. Exact expressions for $f_-(z)$ and $f_+(z)$ can be obtained from \textcite{Ram:1967zz}; $f_{+-}(z)$ was evaluated exactly by \textcite{Passera:2000ug}. However, these expressions are long and complicated, and are best treated using numerical tabulations.

The differential cross-sections for $\bar{\nu}_\mu + e \to \bar{\nu}_\mu + e$,
 $\nu_e + e \to \nu_e + e$ and   $\bar{\nu}_e + e \to \bar{\nu}_e + e$ are obtained from the $\nu_\mu + e \to \nu_\mu + e$ case by making simple changes explained in \cite{Sarantakos:1983bp}.
In particular, in  $\nu_e + e \to \nu_e + e$ there are two distinct classes of contributions, one involving the neutral currents as in $\nu_\mu + e \to \nu_\mu + e$, the other mediated by the W boson.

If the tree-level propagator factors $(1-q^2/M_W^2)^{-2}$ and $(1-q^2/M_Z^2)^{-2}$ are ignored (i.~e. if $q^2 /M_W^2 \ll 1$), in passing from $\nu_\mu + e \to \nu_\mu +e$ to $\nu_e + e \to \nu_e +e$ the only changes are 
\begin{itemize}
\item[i)] $\varepsilon_-(q^2)$ in  Eq.(\ref{eq:3_48}) is changed to 
\be
\varepsilon_-(q^2) = \frac{1}{2}\left(1 - \hat{\kappa}^{(\nu_e; l)}(q^2) \hat{s}^2 \right) - \frac{1}{\rho_{\rm N.C.}^{(\nu;l)}} \, ,
\label{eq:3_54}
\ee
where $\rho_{\rm N.C.}^{(\nu;l)}$ is defined in  Eq.(\ref{eq:3_47}),
\item[ii)] $\varepsilon_+(q^2)$ in  Eq.(\ref{eq:3_49}) is changed to
\be
\varepsilon_+(q^2) = - \hat{\kappa}^{(\nu_e; l)}(q^2) \hat{s}^2 \, ,\label{eq:3_55}
\ee
\item[iii)] $\hat{\kappa}^{(\nu_e; l)}(q^2)$ is obtained from $\hat{\kappa}^{(\nu_\mu; l)}(q^2)$ by replacing $-2 J_\mu(q^2) \to 
-2 J_e(q^2)$ in  Eq.(\ref{eq:3_50}).
\end{itemize}
 
We note that the additional $(-\rho_{\rm N.C.}^{(\nu;l)})^{-1}$ term in   Eq.(\ref{eq:3_54}) reflects the
tree-level contribution of the $W$ mediated amplitude, and the change in (iii) arises from the ``charge radius'' diagrams that depend on the neutrino flavor.

The results discussed in this Section have been applied to the study of the electron recoil-energy spectra and the total cross-sections for neutrino-electron scattering by solar neutrinos \cite{Bahcall:1995mm}.
This paper also presents simple modifications of the relativistic expressions for the QED functions $f_-(z)$ and $(1-z)^2 f_+(z)$ so that they can be applied approximately in the non relativistic domain. An approximate expression for $f_{+-}(z)$ (a function that had not been calculated previously) is also included.

As mentioned before, the expressions in  Eqs.(\ref{eq:3_45}-\ref{eq:3_55}) have been derived in the $\ms$ scheme of renormalization. If the analysis is carried out, instead, in the OS scheme, the expression for $\rho_{\rm N.C.}^{(\nu;l)}$
 is essentially the same as  Eq.(\ref{eq:3_47}), except that $\hat{s}^2$, $\hat{c}^2$,and $\hat{\alpha}$ are changed to
 $s^2$, $c^2$ and $\alpha$.
On the other hand, the OS form factor $\kappa^{(\nu_\mu;l)}(q^2)$ \cite{Marciano:1980pb, Sarantakos:1983bp} that multiplies $\sin^2 \theta_W$ in the EWC, has a considerably more complex structure than  the $\ms$
form factor $\hat{\kappa}^{(\nu_\mu;l)}(q^2)$ given in  Eq.(\ref{eq:3_50}).
In particular, in ${\mathcal O}(\alpha)$, $\kappa^{(\nu_\mu;l)}(q^2)$ depends on $M_H$, while $\hat{\kappa}^{(\nu_\mu;l)}(q^2)$ does not. This more complex structure can be traced to the contributions of the counterterm
$(c^2/s^2) {\rm Re}[A_{ZZ}(M_Z^2)/M_Z^2 - A_{WW}(M_W^2)/M_W^2 ]$ present in $\kappa^{(\nu_\mu;l)}(q^2)$
(we recall that $A_{ZZ}(q^2)$ and $A_{WW}(q^2)$ are the $Z$  and $W$ transverse self-energies).

\subsection{Electron-Positron Annihilation}
\label{sec:3.14}

Since LEP was an $e^--e^+$ collider, the study of the annihilation process into fermion-antifermion pairs,
$e^-+e^+\to f+ \bar{f}$, became a subject of great interest.

An early paper by \textcite{Passarino:1978jh} examined the EW and QED corrections to $e^- +e^+ \to \mu^- +\mu^+$.
This paper also introduced a method to reduce one-loop tensor integrals to scalar ones, which has been frequently employed in the calculation of the EWC to
several important processes.
Since that time, detailed studies of EW, QED, and QCD corrections to $e^- + e^+ \to f + \bar{f} $ were carried out by several authors. See, for example,  \textcite{Ellis:1986jba} and references cited therein; \textcite{Alexander:1988mw} and references cited therein; \textcite{Altarelli:1989R}; \textcite{Bardin:1995R};
\textcite{Kuhn:1989fd} and references cited therein. A paper by \textcite{Degrassi:1990ec} analyzed the EWC to cross-sections, asymmetries and $Z$ partial widths using both the On-Shell and the $\ms$ renormalization frameworks.
The results of the partial widths and asymmetries for some final-state modes were then compared numerically with those obtained in the formulation of \textcite{Consoli:1989fg} and \textcite{Hollik:1988ii}. The corrections to the $Z b \bar{b}$ vertex involve a significant $M_t^2$ dependence, and played an important role in the indirect determination of the top quark mass before the discovery of this fundamental particle \cite{ABR86, BH88, BPS91}.

The asymmetries measured at LEP and SLC are of special interest because they provide the most precise determination of $\sefl$ (cf. Section~\ref{sec:3.6}). They include: a) the measurement at LEP of the forward-backward 
asymmetries $A_{\rm FB}^{0,f}$ (for $f=e,\mu,\tau,s,c,b$), the $\tau^-$ polarization asymmetry $P_\tau$
in $e^-+e^+ \to \tau^- +\tau^+$ and the forward-backward asymmetry $Q_{\rm FB}^{\rm had}$ between positive and negative charge in hadronic $Z$-events; b) the measurements at SLC of the left-right $e^-$-polarization asymmetry $A_{LR}^0$ and the combined forward-backward $e^-$-polarization asymmetries $A_{\rm LR}^{0 {\rm FB}}$, separately analyzed for hadronic and leptonic final states.
For a recent discussion, see \textcite{ErlerPDG}, particularly Section~10.4.

For a long time, there has been an intriguing difference, at the $3~\sigma$ level, between the values of $\sefl$ derived from the leptonic and hadronic asymmetries.
In fact, one finds $(\sef)_l = 0.23113(21)$ from the leptonic asymmetries ($A_{\rm FB}^{0,l}$, $A^0(P_\tau)$, 
$A^0_{\rm LR}$, $A^{0,{\rm FB}}_{\rm LR}$) ($l=e,\mu,\tau$) and $(\sef)_h = 0.23222(27)$ from the hadronic asymmetries ($A^{0,q}_{\rm FB}$, $Q_{\rm FB}^{\rm had}$) ($q = s,c,b$). Furthermore, the results within each group
are in good agreement with each other. The intriguing question remains of whether the difference between $(\sef)_l$
and $(\sef)_h$ is due to a statistical fluctuation or arises from new physics involving perhaps the third generation of quarks. The second scenario is difficult to implement because of  the constraints imposed by  the $Z \to b \bar{b}$ branching ratio.
In the first case, a possible approach to take into account the difference is to enlarge the error, as discussed by
\textcite{Gurtu:1996nm}; \textcite{Degrassi:1997iy} and \textcite{Ferroglia:2002rg}. For example, if the $\sefl$
error is increased by a factor $[\chi^2/{\rm D.O.F.}]^{1/2}$ following the Particle Data Group prescription \cite{Barnett:1996yz},
one obtains the value $\tilde{s}^2_{{\rm eff}} = 0.23153(25)$.
 The discrepancy discussed above is of   particular significance for the indirect estimate of $M_H$, which is very sensitive to the precise value of $\sefl$. Since this issue has not been resolved, the usual procedure is to employ the average value obtained from all the asymmetries.

\subsection{Estimates of the Higgs Boson Mass}
\label{sec:3.15}

The Higgs boson is the fundamental missing piece of the ST. Thus, an important question is to what extent can $M_H$
be estimated using the precision electroweak data and the theoretical expressions for the relevant observables, which
depend on $M_H$ via EWC. In fact, such estimates may provide very useful information for explorations at the 
Large Hadron Collider (LHC), since one of its main objectives is the search for this fundamental particle.

At the one-loop level, for large $M_H$, the dependence of the EWC on $M_H$ is proportional to $\ln(M_H/M_Z)$
(cf.  Eqs.(\ref{eq:3_14},\ref{eq:3_15})), a slowly-varying function\footnote{It is interesting to note that the evaluation of higher order corrections to the $\rho$ parameter has a long history, starting with the paper of \textcite{vdBV}, where the contributions of ${\mathcal O}(\alpha^2 M_H^2)$ were obtained.
The important two-loop QCD and EW contributions to the $\rho$ parameter were evaluated by \textcite{DV87, FAT94, CKS95}.
Later developments include calculations, at the three- and four-loop levels, of pure EW and mixed EW and QCD corrections in the large $M_H$ or $M_t$ limits \cite{BTB, FKSV, BCFJS, Cetal, BoCz}.}. Thus, precise calculations are needed! Theorists distinguish two classes of errors: a) parametric, such as $\delta M_W$, $\delta \sef$, $\delta M_t$, $\delta \Delta \alpha_h^{(5)}$ \dots; b) uncertainties due to the truncation of the perturbative series (i.~e. uncalculated higher order effects). As mentioned at the end of Section~\ref{sec:3.7}, estimates of the second class of errors are often obtained by comparing the evaluation of the EWC using different renormalization schemes. In the case when the expansion parameters are scale-dependent, as in the $\ms$ scheme of renormalization, errors of the second class are frequently
estimated by examining the scale-dependence of the calculations.

The comparison of the accurate experimental values of $M_W$ and $\sefl$ with their theoretical calculations have been subjects of particular interest, since they provide important information about $M_H$.

Over the years, a number of higher order EWC were incorporated in the theoretical calculations. Contributions of ${\mathcal O}(\alpha)$, ${\mathcal O}(\alpha^n \ln^n M_Z/M_W)$, and ${\mathcal O}(\alpha^2 \ln M_Z/M_f)$
(where $f$ is a generic quark or lepton) were analyzed in the period 1979-84. Those of 
${\mathcal O}(\alpha^2 (M_t/M_W)^4)$, ${\mathcal O}(\alpha \alpha_s)$, and ${\mathcal O}(\alpha \alpha_s^2 (M_t/M_W)^2)$ were studied from the late 80's to the middle 90's. EWC of ${\mathcal O}(\alpha^2(M_t/M_W)^2)$ were evaluated by  \textcite{Degrassi:1996mg}; \textcite{Degrassi:1997ps}; \textcite{Degrassi:1997iy}; \textcite{Degrassi:1999jd}; (see also the references cited in those papers).

Very simple analytic formulas for the theoretical calculation of $\sefl$, $M_W$, and the leptonic partial widths $\Gamma_l$ of the $Z$ boson were presented by \textcite{Ferroglia:2002rg}. They reproduced accurately the results of the detailed calculations in the On-Shell, $\ms$, and Effective Schemes (cf. Section~\ref{sec:3.7}) as functions of $M_H$, $M_t$, $\Delta \alpha_h^{(5)}$,
and $\alpha_s(M_Z)$, over the range $20\, {\rm GeV} \le M_H \le 300\, {\rm GeV}$. In particular, they incorporated the complete one-loop EWC, as well as the two-loop contributions enhanced by factors $(M_t^2/M_Z^2)^n$ $(n=1,2)$
that had been studied previously. These simple formulas were applied to estimate $M_H$ and its $95 \%$ C.L. upper 
bound $M_H^{95}$ using either $(\sef)_{\rm exp}$, $(M_W)_{\exp}$, or, simultaneously, $(\sef)_{\rm exp}$, $(M_W)_{\exp}$ and $(\Gamma_l)_{\rm exp}$ as input parameters.

An important advance has been the calculation of the complete two-loop contribution to $\Delta r$ in the OS scheme of renormalization. It includes the fermionic contribution, which involves diagrams with one or two closed fermion loops 
\cite{Freitas:2000gg,  Freitas:2002ja} and the purely bosonic two-loop contribution\footnote{For clarity, we point out that in the recent higher order calculations
  $\Delta r$ is introduced by the relation $s^2c^2 = (\pi \alpha/\sqrt{2} G_F M_Z^2)
   (1 + \Delta r)$, with $\Delta r$ in the numerator, which coincides
  with the expression originally derived by \textcite{Sirlin:1980nh}. Of course, at
  the one-loop level, this expression and Eq.(\ref{eq:3_1})  are equivalent.} \cite{Awramik:2002wn, 
Awramik:2002vu, Onishchenko:2002ve}. Since $\Delta r$ is the quantum correction in the relation of $M_W$ with $\alpha$, $G_F$, and $M_Z$, this result provides directly the two-loop EWC in the theoretical calculation of $M_W$.

Another important achievement has been the calculation, also in the OS scheme, of  the complete two-loop EWC in the theoretical evaluation of $\sefl$  \cite{Awramik:2004ge, Awramik:2006ar, Awramik:2006uz, Hollik:2005ns, Hollik:2006ma, Hollik:2005va}.

Simple analytic formulas that incorporate accurately the contribution of the one and  two-loop EWC in the theoretical calculations of $M_W$ and $\sefl$, as functions of $M_H$, $M_t$, $\Delta \alpha$, $\alpha_s(M_Z)$ and $M_Z$, were given, respectively, by \textcite{Awramik:2004ge} and \textcite{Awramik:2006uz}.

Next, we illustrate the application of these accurate formulas to the estimate of the Higgs boson mass $M_H$ and its $95 \%$ C.L. upper bound $M_H^{95}$. We use as inputs $M_W = 80.385(15)$~GeV \cite{TEWWGMW, EWWG}, $M_Z = 91.1876(21)$~GeV (Section~\ref{sec:3.3}),
$M_t = 173.2(0.9)$~GeV \cite{TEWWGMT}, $\sefl=0.23153(16)$ (cf.  Eq.(\ref{eq:3_17})), $\Delta \alpha = 0.05907(10)$ (cf. Eq.(\ref{eq:3_20})), $\alpha_s(M_Z)=0.1184(7)$ \cite{DissertoriPDG}. On this basis, we obtain the following estimates:
\bea
M_H &=& 98^{+25}_{-21} \, {\rm GeV} \, ; \quad M_H^{95} = 142 \, {\rm GeV} \quad \left(M_W + \sef \right)\, ,
\label{eq:3_56} \\
M_H &=& 81^{+28}_{-24} \, {\rm GeV} \, ; \quad M_H^{95} = 131 \, {\rm GeV} \quad \left(M_W \right)\, ,
\label{eq:3_57} \\
M_H &=& 129^{+53}_{-38} \, {\rm GeV} \, ; \quad M_H^{95} = 226 \, {\rm GeV} \quad \left(\sef \right)\, ,
\label{eq:3_58} 
\eea
 Eq.(\ref{eq:3_56}) was obtained by means of a $\chi^2$ analysis based on the theoretical expressions for both $M_W$ and $\sefl$, while  Eq.(\ref{eq:3_57}) and  Eq.(\ref{eq:3_58}) were derived from the separate application of the $M_W$
and $\sefl$ formulas, respectively. 

As a comparison, a recent standard global fit to the EW data \cite{Baak:2011ze} employs the inputs 
$M_W = 80.399(23)$~GeV,
$M_Z = 91.1875(21)$~GeV,
$M_t = 173.3(1.1)$~GeV, 
$\sefl=0.23153(16)$,
$\Delta \alpha = 0.05899(10)$, $\alpha_s(M_Z) = 0.1193(28)$, and derives the estimate
$M_H = 96^{+31}_{-24}$~GeV, $M^{95}_H = 169$~GeV (this last value includes the effect of the estimated theoretical error).

Since the inputs in our calculations are somewhat different (particularly in the case of $M_W$ for which we use a more recent and precise value), for comparison purposes we repeat our calculation of  Eq.(\ref{eq:3_56}) employing the same inputs as in the global fit. This leads to $M_H = 103^{+32}_{-26}$~GeV, $M_H^{95} =  160$~GeV,
which can be compared with the values $M_H = 96^{+31}_{-24}$~GeV, $M_H^{95} = 169$~GeV obtained in the global fit with the same input parameters.
Thus, we see that the estimates obtained by combining the theoretical expressions for $M_W$ and $\sefl$
are rather close to those obtained in the global fit, an observation   that illustrates the importance and sensitivity of these two observables in the prediction of $M_H$ and $M_H^{95}$.

We note
that the central values of $M_H$ in both  Eq.(\ref{eq:3_56}) and the global fit are well below the $95 \%$ C.~L. lower
bound 
\be
\left( M_H\right)_{L.B.} = 114.4~{\rm GeV} \,  , \label{eq:3_59}
\ee
inferred from the direct experimental searches of the Higgs boson at LEP and the Tevatron.
On the other hand, the two $M_H$ estimates are compatible with  Eq.(\ref{eq:3_59}) when their errors are taken into account.

In Section~\ref{sec:3.14}, we pointed out that, for a long time, there has been an intriguing difference, at the $3 \sigma$ level, between the values of $\sefl$ derived from the leptonic and hadronic asymmetries, namely $(\sef)_l = 0.23113(21)$ and $(\sef)_h = 0.23222(27)$. In order to illustrate the potential effect of this dichotomy, we give below the $M_H$ and $M_H^{95}$ estimates obtained by using separately these values, as well as their combinations with the 
theoretical expression for $M_W$:
\begin{align}
M_H &= 54^{+33}_{-21} \, {\rm GeV} \, ; &M_H^{95} = 117\,{\rm GeV} \qquad &\left((\sef)_l\right) \, , 
\label{eq:3_60} \\
M_H &= 71^{+23}_{-18} \, {\rm GeV} \, ;  &M_H^{95} = 111\,{\rm GeV} \qquad &\left(M_W,(\sef)_l\right) \, , 
\label{eq:3_61} \\
M_H &= 513^{+387}_{-212} \, {\rm GeV} \, ;  &  &\left((\sef)_h\right) \, , 
\label{eq:3_62} \\
M_H &= 117^{+32}_{-27} \, {\rm GeV} \, ;  &M_H^{95} = 173\,{\rm GeV} \qquad &\left(M_W,(\sef)_h\right) \, , 
\label{eq:3_63} 
\end{align}
We see that the estimates based on $(\sef)_l$, either by itself (Eq.(\ref{eq:3_60})), or in combination with $M_W$
 (Eq.(\ref{eq:3_61})), are very low. In fact, at the $1 \sigma$ level, they disagree with $\left( M_H\right)_{L.B.}$  
 (Eq.(\ref{eq:3_59})). We also note that $M_H^{95}$ in  Eq.(\ref{eq:3_60}) is barely compatible with $\left( M_H\right)_{L.B.}$, while its value in  Eq.(\ref{eq:3_61}) is lower.  Thus, in an hypothetical scenario in which the $(\sef)_l$-$(\sef)_h$ discrepancy were to settle on the leptonic side, for example by bringing the value of $(\sef)_h$ close to the present determination of $(\sef)_l$, a serious discrepancy would arise
 between the $M_H$, $M_H^{95}$ estimates and $(M_H)_{L.B.}$.
 
 In contrast, $(\sef)_h$ leads to considerably larger estimates of $M_H$ and $M_H^{95}$ (cf.  Eqs.(\ref{eq:3_62},\ref{eq:3_63})). We note that in  Eq.(\ref{eq:3_62}) we have not included the value of $M_H^{95}$. The reason is that the 
 range of validity of the simple analytic formulas is $10~{\rm GeV} \le M_H \le 1~{\rm TeV}$, while their application
 to  Eq.(\ref{eq:3_62}) leads to a value of $M_H^{95}$ considerably larger than $1~{\rm TeV}$.
 
 Finally, we consider the estimates of $M_H$, $M_H^{95}$ based on $\tilde{s}^2_{\rm eff} = 0.23153(25)$, the value
 obtained from the weighted average of $(\sef)_l$ and $(\sef)_h$ by enlarging the error according
 to the Particle Data Group prescription (cf. the discussion toward the end of Section~\ref{sec:3.14}):
 \begin{align}
 M_H &= 129^{+89}_{-54} \, {\rm GeV} \, ; &M_H^{95} = 302 \, {\rm GeV} \qquad&(\tilde{s}^2_{\rm eff}) \, ,
 \label{eq:3_64} \\
 M_H &= 90^{+27}_{-22} \, {\rm GeV} \, ; &M_H^{95} = 137 \, {\rm GeV}\qquad &(M_W,\tilde{s}^2_{\rm eff}) \, .
 \label{eq:3_65} 
 \end{align} 
As expected, the central value in  Eq.(\ref{eq:3_64}) is the same as in  Eq.(\ref{eq:3_58}), but the errors and 
$M_H^{95}$ are larger. On the other hand, the central value and $M_H^{95}$ in  Eq.(\ref{eq:3_65}) are smaller than in  Eq.(\ref{eq:3_56}). The reason is that the increased error in $\tilde{s}^2_{\rm eff}$ gives greater weight to the 
$M_W$ contribution, which favors smaller values of $M_H$ and $M_H^{95}$.

In February 2012, the ATLAS Collaboration at LHC \cite{atlas} reported that their combined search for the ST Higgs boson excludes the $M_H$ ranges $112.9$ -- $115.5$~GeV, $131$ -- $238$~GeV and $251$ -- $466$~GeV at 
$95 \, \%$ C.L.. Thus, subject to that exclusion, the still-allowed domains are 
$115.5$ -- $131$~GeV, $238$ -- $251$~GeV, $\ge 466$~GeV. On the same day, the CMS Collaboration at LHC
\cite{cms} reported that their combined search excludes the $M_H$ range $127$ -- $600$~GeV at $95 \%$ C.L. and 
$129$ -- $525$~GeV at $99 \%$~C.L.. Thus, subject to the $95 \%$ C.L. exclusion, the still-allowed regions are 
$114.4$ -- $127$~GeV, $\ge 600$~GeV. At the same time, the ATLAS Collaboration reported an excess
of events above the expected ST background around $M_H \sim 126$~GeV with  a local significance
of $3.5~\sigma$, while the CMS Collaboration found an excess at $M_H = 124$~GeV with a local significance of 
$3.1~\sigma$. Both collaborations expect to collect a considerable amount of additional data in 2012 in order to ascertain whether the observed excesses represent real signals of the Higgs boson or they simply reflect statistical fluctuations of the ST background. For the moment, we observe that, when the $1~\sigma$ errors are taken into account, the estimates in both  Eq.(\ref{eq:3_56}) and the global fit are compatible with a Higgs boson in the neighborhood of $M_H =125$~GeV.

There are also very interesting theoretical upper and lower bounds for $M_H$, $M_{{\rm max}}(\Lambda)$ and $M_{{\rm min}}(\Lambda)$, where $\Lambda$
is the scale up to which the ST is assumed to be valid.  $M_{{\rm max}}(\Lambda)$ is obtained from the requirement that the Higgs self-coupling does not exhibit a Landau pole below $\Lambda$. $M_{{\rm min}}(\Lambda)$ is obtained from considerations of vacuum stability. If $\Lambda = M_P$, $M_{\rm max}(M_P) \approx 175$~GeV (cf. \textcite{BKKS} and references cited therein). Since the recent Higgs boson searches at LHC exclude the range 
$(129 - 525)$~GeV at $99 \%$ CL, this result indicates that, in the absence of new physics, the ST is a weakly coupled theory up to $M_P$. 
Recent analyses of $M_{{\rm min}}(M_P)$ include \textcite{BKKS} and \textcite{EMetal}.  The authors of the first paper find $M_{{\rm min}} = 129 \pm 6$~GeV,
which overlaps with the allowed region in the recent searches. The second paper derives both stability and metastability bounds. For their central values, they find $M_{{\rm min}}(M_P) = 130 \pm 3$~GeV, and $M^{{\rm metas}}_{{\rm min}}(M_P) = 111 \pm 3$~GeV. The metastability bound is derived by requiring that the lifetime of the electroweak vacuum is larger than the age of the universe. Combining the results explained above, and assuming that the Higgs boson is discovered in the range $115.5~{\rm GeV} \le M_H \le 127~{\rm GeV}$ currently allowed by the direct searches at the LHC, \textcite{BKKS} conclude:
\begin{itemize}
\item[a)] a new energy scale between the Fermi and Planck scales is not necessarily required,
\item[b)] in the absence of such scale, the EW theory remains weakly coupled up to $M_P$,
\item[c)] the EW vacuum has a lifetime larger than the age of the universe.
\end{itemize}

% Added on August 10 2012

On July 4, 2012, the ATLAS \cite{ATLASHiggs} and CMS \cite{CMSHiggs} collaborations at LHC announced the discovery at the $5 \sigma$ level of a boson in the mass interval $124-126$ GeV. There is a widespread belief in the Physics community that this is the long-sought Higgs boson. To ascertain whether this is the case, further analyses are in progress to determine whether the spin of the newly discovered particle is indeed $0$ as befits the Higgs boson, and whether  its production and decay rates  conform with the ST expectations.

\boldmath
\subsection{The Muon $g_\mu -2$ \label{sec:3.20}}
\unboldmath

The anomalous magnetic moment of the muon, $a_\mu = (g_\mu-2)/2$, is one of the most interesting and precisely measured observables in particle physics. In fact, since each sector of the ST contributes in a significant way to its theoretical prediction, the $a_\mu$ measurement by the E821 experiment at the Brookhaven National Laboratory \cite{gm2a, gm2b, gm2c, Roberts:2010cj}, with a remarkable precision of $0.5$ parts per million, permits to test  the entire ST and examine possible new physics effects \cite{Czarnecki:2001zz, Stockinger:2007zz}. It is important to note that even more precise measurements are planned at the Fermilab experiment P989 and J-PARC with anticipated errors that are smaller than the current one by factors of $4$ and $5.4$, respectively.

The ST prediction of $a_\mu$ includes QED, electroweak (EW) and  hadronic (leading- and higher-order) contributions: $a_\mu^{{\rm ST}} = a_\mu^{{\rm QED}} + a_\mu^{{\rm EW}}+ a_\mu^{{\rm HLO}} +a_\mu^{{\rm HHO}} $. The QED contribution, computed to four loops and estimated to five\footnote{After this paper was submitted for publication, the calculations of the five loop contributions to $a_e$ and $a_{\mu}$ were completed \cite{5L1, 5L2}, leading to  $a^{\mbox{QED}}_{\mu} = 116 584 718. 845 (37) \times 10^{-11}$ and $\Delta a_{\mu} =  260(80) \times 10^{-11}$.}, currently stands at $a_\mu^{{\rm QED}} = 116584718.08(15) \times 10^{-11}$ \cite{Kinoshita:2004zz, Kinoshita:2006zz, Kinoshita:2006zzb, Aoyama:2007dv, Aoyama:2007mn, Aoyama:2008gy, Aoyama:2008hz, Aoyama:2010yt, Aoyama:2010pk, Aoyama:2010zp, Aoyama:2011rm, Aoyama:2011zy,Aoyama:2011dy, Aoyama:2012fc, Laporta:1992pa, Laporta:1996mq, Passera:2006gc, Kataev:2006yh}, while the EW effects, suppressed by a factor $(m_\mu/M_W)^2$, amount to
$a_\mu^{{\rm EW}} = 154(2) \times 10^{-11}$ \cite{Czarnecki:1995wq, Czarnecki:1995sz, Degrassi:1998es, Czarnecki:2002nt}.

Recent calculations of the hadronic leading-order contribution, based on the hadronic $e^+ e^-$ annihilation data, include: $a_\mu^{{\rm HLO}} = 6949.1(42.7) \times 10^{-11}$ \cite{Hagiwara:2011af}, $a_\mu^{{\rm HLO}} = 6903(53) \times 10^{-11}$ \cite{Jegerlehner:2009ry}, $a_\mu^{{\rm HLO}} = 6923(42) \times 10^{-11}$ \cite{Davier:2010nc}. The three results agree within errors. A recent analysis by \textcite{JS11} finds good agreement between 
the calculations based on the $e^+ e^-$ annihilation and $\tau$ decays data leading to $a_\mu^{{\rm HLO}} = 6909.6(46.5) \times 10^{-11}$.

The higher-order hadronic contribution is divided into two parts: $a_\mu^{{\rm HHO}} = a_\mu^{{\rm HHO}}({\rm vp}) + a_\mu^{{\rm HHO}}({\rm lbl})$. The first one $a_\mu^{{\rm HHO}} ({\rm vp}) = -98(1) \times 10^{-11}$ \cite{Hagiwara:2006jt}, is the 
${\mathcal O}(\alpha^3)$ contribution of diagrams containing  hadronic vacuum polarization insertions.
The second one, also of ${\mathcal O}(\alpha^3)$, is the hadronic light-by-light contribution; since it cannot be derived from data, its evaluation is based on specific models. Two of the most recent determinations, $116(39) \times 10^{-11}$ \cite{Jegerlehner:2009ry, Nyffeler:2009tw} and 
$105(26) \times 10^{-11}$ \cite{Prades:2009tw} are in good agreement. If one adds the latter to $a_\mu^{{\rm HLO}} = 6949.1(42.7) \times 10^{-11}$ and the rest of the ST contributions, one obtains $a_\mu^{{\rm ST}} = 116591828(50) \times 10^{-11}$. The difference with the experimental value $a_\mu^{{\rm exp}} = 116592089(63) \times 10^{-11}$  \cite{Roberts:2010cj} is
$\Delta a_{\mu} =a_\mu^{{\rm exp}} - a_\mu^{{\rm ST}}=261(80)\times 10^{-11}$, i.~e. $+3.3\, \sigma$ (all errors have been added in quadrature).  A somewhat larger discrepancy, $3.6\, \sigma$, is obtained if one employs $a_\mu^{{\rm HLO}} = 6923(42) \times 10^{-11}$.

It has been pointed out that Supersymmetry (SUSY) may provide a natural explanation for the $3-4\, \sigma$ discrepancy between $a_\mu^{{\rm exp}}$ and $a_\mu^{{\rm ST}}$ (for a review, see \textcite{Stockinger:2007zz}).
Assuming, for simplicity, a single mass $M_{{\rm SUSY}}$ for the supersymmetric particles that contribute to $a_\mu^{{\rm SUSY}}$, one finds \cite{Kosower:1983zz, Yuan:1984zz, Moroi:1996zz, Ibrahim:2000zz, Heinemeyer:2000zz, Heinemeyer:2000zzb}
\be
a_\mu^{{\rm SUSY}} \simeq {\rm sgn}(\mu) \times 130 \times 10^{-11}\left( \frac{100\, {\rm GeV}}{M_{{\rm SUSY}}}\right)^2
\tan \beta \, , \label{eq:3_95}
\ee  
where $\tan \beta > 3-4$ is the ratio of the two scalar vacuum expectation values and ${\rm sgn}(\mu)$ the sign of the $\mu$ term  in SUSY models. Assuming that $a_\mu^{{\rm SUSY}}$ cancels the discrepancy, so that $a_\mu^{{\rm SUSY}} = \Delta a_\mu$, and using, for example, the value $\Delta a_\mu = 261(80) \times 10^{-11}$, one finds 
${\rm sgn}(\mu) = +$ and 
\be
M_{{\rm SUSY}} \simeq 71^{+14}_{-9} \sqrt{\tan \beta} \, {\rm GeV}  \, .
\label{eq:3_96}
\ee 
For $\tan \beta \sim 4 -50$,  Eq.(\ref{eq:3_96}) leads to the very rough estimate $124~{\rm GeV} \leq M_{{\rm SUSY}} \leq 601~{\rm GeV}$. On the other hand, signals of supersymmetric particles have not been uncovered so far. Other new physics explanations of the $a_\mu$ discrepancy have also been discussed \cite{Czarnecki:2001zz}. 

In an alternative approach, not involving new physics, \textcite{Passera:2008jk, Passera:2008hj, Passera:2010ev} considered whether an increase in the hadroproduction cross section $\sigma(s)$  in low-energy $e^+ e^-$ collisions, due to hypothetical experimental errors, could bridge the $a_\mu$ discrepancy. They found that this is unlikely in view of the current experimental error estimates. If, nonetheless, this turns out to be the explanation of the discrepancy, it has an interesting consequence: the increase in $\sigma(s)$ also increases $\Delta \alpha_{{\rm had}}^{(5)}(M_Z)$ which, in turn, affects the estimate of $M_H$. The authors found that, in this hypothetical scenario, the $95 \%$ CL upper bound on the Higgs boson mass is reduced to about $135$~GeV which, in conjunction with $(M_H)_{{\rm LB}} = 114.4$~GeV, leaves a narrow window for the mass of this fundamental particle. This window is slightly larger than  the range allowed by the very recent LHC direct searches (cf. previous to last pararaph in Section~\ref{sec:3.15}).

\boldmath
\subsection{Atomic Parity Violation \label{sec:3.21}}
\unboldmath

The interference of the electromagnetic and weak neutral current amplitudes leads to parity violating effects in atomic transitions that have been the subject of ingenious experiments and detailed theoretical studies.

The pseudoscalar component of the electron-quark interaction, arising from the $Z$ boson exchange at $q^2 = 0$, is usually expressed in the form
\begin{align}
{\mathcal H}_{{\rm PV}} &= \frac{G_\mu}{\sqrt{2}} \Biggl\{ \left[C_{1u} \bar{u} \gamma^\mu u + C_{1 d} \bar{d}  \gamma^\mu d \right]
\left[\bar{e} \gamma_\mu \gamma_5 e \right] + 
\left[C_{2u} \bar{u} \gamma^\mu  \gamma_5 u  + C_{2 d} \bar{d}  \gamma^\mu \gamma_5 d \right]
\left[\bar{e} \gamma_\mu  e \right] + \cdots  
\Biggr\} \, , \label{eq:a_1}
\end{align} 
where the ellipses represent heavy-quark contributions ($q = s,c,b,t$).

The $C_{2i}$ ($i=u,d$) are suppressed by a factor $1-4 \sin^2 \hat{\theta}_W (M_Z) \simeq 0.075$ that arises from the electron's vector coupling to the $Z$ boson. Also, the $C_{1i}$ ($i=u,d$) terms are of primary importance for heavy atoms because they add up coherently over all quarks in the nucleus. As a consequence, parity violating effects are dominated by contributions proportional to the weak charge
\be
Q_W  (Z,A)  \equiv 2 \left[ \left(A+Z\right) C_{1 u} + \left( 2 A -Z\right) C_{1 d}\right]\, ,  \label{eq:a_2}
\ee
where $Z $ and $A$ are the atomic and mass numbers of the atom.

The dominance of the $C_{1 i}$ ($i = u, d$) terms is also theoretically fortunate, because the corresponding hadronic currents are conserved and therefore are not affected by the strong interactions at $q = 0$.

As pointed out by \textcite{Bouchiat:1974zz}, parity violating effects in heavy atoms scale roughly as $Z^3$ (one $Z$ factor reflects the coherence effect in $Q_W$, while the others arise from the electron wave function and momentum near the nucleus).

Electroweak corrections of ${\mathcal O}(\hat{\alpha})$ to the $C_{1i}$ and $C_{2i}$ ($i=u,d$) coefficients were evaluated in the $\ms$ scheme by
\textcite{Marciano:1983APV, Marciano:1984APV} and \textcite{Lynn:APV}.

For the dominant coefficients $C_{1i}$ ($i=u,d$), the two first authors expressed their results in the form
\begin{align}
C_{1 u} &= \frac{\rho'_{{\rm PV}}}{2}  \left[1 - \frac{8}{3} \kappa'_{{\rm PV}}(0) \sin^2 \hat{\theta}(M_W)\right] \, , \label{eq:a_3} \\
C_{1 d} &= -\frac{\rho'_{{\rm PV}}}{2}  \left[1 - \frac{4}{3} \kappa'_{{\rm PV}}(0) \sin^2 \hat{\theta}(M_W)\right] \, . \label{eq:a_4}
\end{align}
The constants $\rho'_{{\rm PV}}$ and $\kappa'_{{\rm PV}}(0)$ contain the ${\mathcal O}(\hat{\alpha})$ EWC, which depend on $M_t, M_H, M_W$ and $M_Z$,
and are normalized so that $\rho'_{{\rm PV}} = \kappa'_{{\rm PV}}(0) = 1$ at the tree-level. The detailed expressions for $\rho'_{{\rm PV}}$ and $\kappa'_{{\rm PV}}(0)$
are given in \textcite{Marciano:1983APV, Marciano:1984APV}. A more recent version of these results, that employs $\sin^2 \hat{\theta}_W(M_Z)$ instead of 
$\sin^2 \hat{\theta}_W(M_W)$, was presented by \textcite{Marciano:1993ep}. 

Measurements of atomic parity violation have been made in bismuth, lead, thallium and cesium (for reviews see \textcite{Masterson:APV}; \textcite{Bouchiat:1997mj}; \textcite{Ginges:2004qt}). The most precise so far have been measurements of $Q_W$ in cesium, at the $0.4~\%$ level.
The analysis of the data requires detailed atomic physics calculations \cite{Blundell:1996qj, Porsev:2009pr} and QED corrections \cite{Ginges:2004qt}. 
A recent result \cite{Porsev:2009pr} is $Q_W({\rm Cs}) = -73.16(29)_{{\rm exp}}(20)_{{\rm th}}$, in impressive agreement with the ST expectation
$Q_W({\rm Cs})_{{\rm ST}} = -73.15(2)$ \cite{ErlerPDG}.

$Q_W$ is very insensitive to the $T$ parameter, and thus provides a direct probe of the $S$ parameter, as emphasized by \textcite{Marciano:1990dp} and  \textcite{Marciano:1991ix, Marciano:1993ep} (cf. Section~\ref{sec:3.16}).

Recently, sharp lower bounds on the mass of $Z'$ bosons associated with interesting models beyond the ST have been derived from atomic parity violation measurements \cite{DGT} The same paper also sets constraints on the $Z'$ couplings.

\boldmath
\subsection{Radiative Corrections in Flavor Physics: The $b \to s \gamma$ Case}
\unboldmath

Over the years, flavor physics played a crucial role in shaping our understanding of the interactions of elementary particles. The study of weak decays, including flavor and CP violating meson decays, led physicists to 
discover the GIM mechanism \cite{Glashow:1970gm} and the CKM matrix \cite{Cabibbo:1963yz, Kobayashi:1973vf},  both of which are essential elements in establishing 
the particle content of the ST. 

In recent years, flavor physics observables were measured with great accuracy at several experimental facilities. Currently, one of the experiments at LHC, named LHCb, is primarily devoted to the measurement of the properties of hadrons containing a bottom  quark. A second forthcoming experiment at CERN, called NA62, will measure very rare decays
of charged kaons. Two new super-B factories will be built in Frascati (Italy) and at KEK (Japan).
While experiments at high energy colliders allow physicists to search for new physics beyond the ST  by attempting to produce new particles, precise flavor physics experiments exploit the high luminosity of flavor factories in order to 
search for the effects of new physics in rare events. In this sense, the direct searches at high energy colliders are complementary to the indirect searches at flavor factories, which are sensitive to energy scales as high as  $\sim 10^4-10^5$~TeV.

An extended description of  all of the observables in weak decays  goes beyond the scope of the present review; the interested reader can find a comprehensive introduction to this topic in the classic Les Houches lectures by \textcite{Buras:1998raa}. 
Here, we focus on a single representative example: the inclusive radiative decay of the $B$ meson mediated by the partonic decay process $b \to s \gamma$.  There are three reasons for this choice:
\begin{itemize}
\item[i)] As all flavor-changing neutral current (FCNC) processes, the $b \to s \gamma$ decay is a loop-induced process in the ST. As such,
it is sensitive to new physics contributions, which can be of the same order in the coupling constants as the leading order contribution in the ST.
\item[ii)] As  will be shown below, inclusive decays are theoretically clean processes since they 
are not very sensitive to non perturbative effects and can be calculated with great accuracy within perturbation theory.
\item[iii)] The measurements of this process, which was carried out at CLEO (Cornell), BELLE (KEK Tsukuba), and BABAR (Stanford), are very precise; in order to match the current experimental accuracy it was necessary  to consider, in calculating the branching ratio, the effect of  NLO and NNLO QCD correction, as  well as  the effect  of NLO electroweak corrections.
\end{itemize}

At the hadron level, the processes of interest are the  inclusive radiative decays of B mesons into a photon and an arbitrary hadronic state of total strangeness $-1$, $\bar{B} \to X_s \gamma$, where $\bar{B}$ denotes a $\bar{B}^0$ or 
$B^-$ meson, while $X_s$ indicates an inclusive hadronic state not containing charmed particles.
At the parton level, these processes are induced by a FCNC decay of the $b$ quark contained in the $\bar{B}$ meson. The $b$
quark decays into a photon and a strange quark plus other partons, collectively indicated by the
symbol $X^{\rm parton}_s$. In the ST, such a decay first takes place at  one-loop order through  diagrams involving heavy particles; for example, through a triangle loop with two virtual top quarks and a virtual $W$ boson. 
 Such diagrams are now commonly referred to as ``penguin'' diagrams.

In contrast with the exclusive decay modes, inclusive decays of B mesons are
theoretically clean observables; in fact, it is possible to prove that the decay width $\Gamma(\bar{B} \to X_s \gamma)$
is well approximated by the partonic decay rate $\Gamma(b \to X_s^{\rm parton} \gamma)$:
\be
\Gamma(\bar{B} \to X_s \gamma) = \Gamma(b \to X_s^{\rm parton} \gamma) + \Delta^{\rm non-pert.} \, .
\label{eq:A_1}
\ee
The second term on the r.h.s. of  Eq.(\ref{eq:A_1}) represents non-perturbative corrections.
The latter are small, since they are suppressed at least by a factor $({\Lambda _{\rm QCD}}/m_b)^2$,
where $m_b$ is the $b$-quark mass and $\Lambda_{{\rm QCD}} \sim 200$~MeV. The relation in Eq. (\ref{eq:A_1}) is
known as {\em Heavy Quark Expansion}  (reviews of this topic and on Heavy-Quark Effective Theory can be found in \textcite{Neubert:1993mb} and in  \textcite{Manohar:2000dt}). 

The partonic process can be studied within the context of perturbative QCD.
However, the first-order QCD corrections to the partonic process are very large.
The large corrections originate from hard gluon exchanges between quark lines of
the one-loop electroweak graphs. In general, Feynman diagrams involving
different mass scales  depend on logarithms of the ratios of these scales. If there is a strong hierarchy among the mass scales, then the logarithms are numerically large. In the case of QCD
corrections to the partonic process $b \to  X^{{\rm parton}}_s \gamma$, the mass scales involved are the
$M_W$,  $M_t$, and $m_b$. $M_W$ and $M_t$ are of the same
order of magnitude $\mu_W \sim 100$~GeV, while the $b$-quark mass is considerably smaller:
$m_b \sim 5$~GeV. Consequently, one finds that $\alpha_s(m_b) \ln(\mu_W^2/m_b^2) \sim 1$; therefore, the perturbative expansion is spoiled, and the large logarithmic corrections must be resummed to all orders.

The easiest way to implement the resummation of the logarithms discussed
above is to work within the context of a renormalization-group-improved effective
theory with five active quarks. In such a theory, the heavy degrees of freedom
involved in the decay under study are integrated out. By means of an operator
product expansion, it is possible to factorize the contribution of the short-distance
and long-distance dynamics in the decay of the $B$ meson. In the ST, the short-distance
dynamics is characterized by mass scales of the order of the top-quark or
$W$-boson mass, while the long-distance dynamics is characterized by the $b$-quark mass.
The boundary between short-distance and long-distance is chosen at a low-energy
scale $\mu_b$, such that $m_b \sim \mu_b \ll M_W$. The scale $\mu_b$ is unphysical, and
therefore physical quantities cannot depend on it: This fact is employed in order to obtain 
renormalization group equations (RGE) satisfied by the various factors in the calculation. The large logarithmic corrections are resummed 
by solving these RGE.

The Lagrangian employed in calculating the $b \to X^{\rm parton}_s \gamma$ 
 decay rate can be written as 
\be
{\mathcal L} = {\mathcal L}_{{\rm QED} \otimes {\rm QCD}} (u,d,c,s,b) + \sum_{i=1}^8 \frac{ 4 G_F}{\sqrt{2}} V_{ts}^* V_{tb}
C_i(\mu,\mu_W ) Q_i(\mu) + {\mathcal O}\left( \frac{m_b}{M_W}\right) \, .
\label{eq:A_2}
\ee
In the equation above, ${\mathcal L}_{{\rm QED} \otimes {\rm QCD}} $  represents the usual QED and QCD
Lagrangians with five active quark flavors, while
 $Q_i$ are eight effective operators of dimensions five and six. Operators with
dimensions larger than six are suppressed by inverse powers of the $W$-boson mass
and are ignored. The short-distance dynamics is encoded in the``coupling
constants'' that multiply the effective operators, which are called {\em Wilson coefficients} and are
indicated by $C_i$ in  Eq.(\ref{eq:A_2}). The Wilson coefficients are the only elements of the
Lagrangian which depend on the heavy particles masses $M_W$ and $m_t$. 
The eight effective operators $Q_i$ appearing in the Lagrangian in Eq.(\ref{eq:A_2}) are listed 
for example in \textcite{Misiak:2006ab}. 

Any perturbative calculation of the $b \to X^{\rm parton}_s \gamma$ decay rate within the context of
the renormalization-group-improved perturbation theory applied to the Lagrangian
in  Eq.(\ref{eq:A_2}) involves three different steps:
\begin{enumerate}
\item The first step, called {\em matching}, consists in fixing the value
of the Wilson coefficients at the high-energy scale $\mu_W \sim M_W,m_t$. This is
achieved by requiring that Green functions calculated in the full ST and
in the effective theory provide the same result up to terms suppressed by
the ratio between the external momenta and $\mu_W$. At the scale $\mu_W$,
QCD corrections are free of large logarithmic corrections and  can therefore be evaluated in finite-order
perturbation theory.
\item Secondly, once the value of the Wilson coefficient at the electroweak scale
has been obtained from the matching step, it is then necessary to obtain the value of the Wilson coefficients at the low-energy scale $\mu_b \sim m_b$. This
can be achieved by solving the system of RGE satisfied by the Wilson coefficient. The RGE system  has the following form:
\be
\mu \frac{d}{d \mu} C_i(\mu) = \gamma_{ji} (\mu) C_j (\mu) \, ,
\label{eq:A_2b}
\ee
where the summation over $j$ is implied.
The matrix $\gamma$
 in the equation above is the anomalous dimension matrix
 of the effective operators. The elements of the matrix have perturbative expansions in powers of $\alpha_s$. Since the various operators mix under
renormalization, this step of the calculation is called {\em mixing}. By solving the
RGE, it is possible to resum  the large logarithms of the ratio
$\mu_W/\mu_b$ to all orders in $\alpha_s$ in the Wilson coefficients. 
\item Finally, it is necessary to calculate on-shell matrix elements of the partonic
process in the effective theory. QCD radiative corrections to the matrix
elements do not include large logarithms, since the dependence on the heavy
degrees of freedom is completely encoded within the Wilson coefficients.
\end{enumerate}

Radiative decays of the B meson were first experimentally observed in the
exclusive $B \to K^* \gamma$
 decay mode by the CLEO collaboration at Cornell in 1993.
Nowadays, the branching ratio of the inclusive decay $\bar{B} \to X_s \gamma$
 has been measured
by several collaborations. The current world average, obtained by averaging the
CLEO, BELLE, and BABAR measurements  \cite{Asner:2010qj} is
\be
{\mathcal B}\left( \bar{B} \to X_s \gamma\right)^{\rm exp}_{E_\gamma > E_0} = \left(3.55 \pm 0.24 \pm 0.09 \right) \times 10^{-4} \, .
\label{eq:A_3}
\ee
In  Eq.(\ref{eq:A_3}), the first error is due to statistical and systematic uncertainty, while the
second is due to the theoretical input on the $b$-quark Fermi motion. In order to eliminate
irreducible backgrounds, experimental collaborations impose a lower cut on the
photon energy. The value in  Eq.(\ref{eq:A_3}) refers to a lower cut $E_0 = 1.6$~GeV.

The measurement in  Eq.(\ref{eq:A_3}) has an experimental error of $7 \%$ and  must be compared with an equally accurate theoretical prediction within the ST.
In renormalization-group-improved perturbation theory, ${\rm N}^m{\rm LO}$ QCD calculations of this process involve the resummation of $\alpha_s^n \ln^{n-m}(m_b^2/\mu^2_W)$ logarithms, as well as the evaluation of ${\mathcal O}(\alpha_s^m)$ corrections to the Wilson coefficients at the scale $\mu_W$ and to the  matrix elements.
In order to obtain theoretically reliable predictions and to match the current experimental accuracy, it was necessary to evaluate both the NLO (i.~e. $m=1$) and NNLO (i.~e. $m=2$) QCD corrections, as well as the NLO electroweak corrections (i.~e.~${\mathcal O}(\alpha \alpha_s^n \ln^n(m_b^2/\mu_W^2))$).

The fascinating history of the calculation of the radiative corrections to the  $\bar{B} \to X_s \gamma$ process was recently reviewed in detail by \textcite{Buras:2011we}. 
The calculation of the LO QCD (i.e.~$m=0$) corrections in renormalization group improved perturbation theory was carried out in the period 1988-1993. An interesting technical feature of this calculation is that, in order to obtain the anomalous dimensions,  one needs to evaluate two-loop Feynman diagrams  already at LO QCD.   Once these LO QCD corrections became available \cite{Ciuchini:1993ks, Ciuchini:1993fk, Cella:1994np,Cella:1994px}, it was pointed out that their renormalization scale dependence  is very large \cite{Ali:1993cj}: by varying $\mu_b$ in the range $m_b/2 < \mu_b < 2 m_b$, the predicted branching ratio changed  by $\sim 60 \%$!
Consequently, the evaluation of the NLO QCD corrections was necessary \cite{Buras:1994yt}. 

The evaluation of the NLO QCD corrections was a challenging task involving several groups in the calculation
of the matching, mixing, and matrix elements; it was completed at the beginning of the last decade. Comprehensive 
reviews of this effort, along with complete lists of references to the contribution of the various groups, were written by
\textcite{Buras:2002er} and  \textcite{Hurth:2003vb}. It is worth emphasizing that NLO determinations of the branching ratio include electroweak effects of ${\mathcal O} (\alpha)$ \cite{Czarnecki:1998tn, Baranowski:1999tq, Gambino:2000fz}. 

While the calculation of the NLO QCD and electroweak corrections  considerably reduces the scale dependence of the $\bar{B} \to X_s \gamma$ branching ratio in the ST, \textcite{Gambino:2001ev} pointed out that this calculation is affected by a $\sim 10 \%$ theoretical uncertainty 
related to the choice of the charm quark mass in the two-loop matrix elements of the four-quark operators.
Consequently, in order to reduce this uncertainty, an evaluation of the NNLO QCD corrections became necessary.
A first estimate of the NNLO branching ratios, including all the numerically dominant effects, was completed 
by \textcite{Misiak:2006ab} and \textcite{Misiak:2006zs}. 
Reviews of the NNLO calculation, including references to the contributions of various groups, can be found for example in 
\textcite{Ferroglia:2008yq}, \textcite{Haisch:2008ar}, and   \textcite{Misiak:2011dz}.

%rearranged

One finds that the  NLO QCD, NNLO QCD, and NLO electroweak contributions amount to  approximately $30 \%$,
$10 \%$, and $4 \%$ of the LO QCD result, respectively.
The predicted value in the ST was found to be 
\be
{\mathcal B} \left( \bar{B} \to X_s \gamma \right)^{\rm ST}_{E_\gamma > 1.6~{\rm GeV}} = \left( 3.15 \pm 0.23 \right) \times 10^{-4} \, ,
\label{eq:A_4}
\ee
which agrees with the world average of the experimental measurements within $1.2~\sigma$. The error on the theoretical estimate is about
$7 \%$ and was obtained by combining four different uncertainties in quadrature: parametric
uncertainty ($3 \%$), uncertainty due to missing higher order corrections ($3 \%$),
uncertainty due to non-perturbative corrections ($5 \%$), and uncertainty due to the
$m_c$-interpolation ambiguity in the calculation of \textcite{Misiak:2006ab, Misiak:2010sk} ($3 \%$). The result in 
 Eq.(\ref{eq:A_4}) is affected  by a  theoretical uncertainty which is approximately of the same magnitude as the
experimental one. Additional perturbative NNLO corrections to the branching ratio were recently evaluated by 
\textcite{Asatrian:2010rq}, \textcite{Ewerth:2008nv},  \textcite{Ferroglia:2010xe}, and  \textcite{Misiak:2010tk}; although these corrections are not included in the calculation leading to 
 Eq.(\ref{eq:A_4}), their numerical impact is expected to be marginal. Additional work within perturbation theory is still required to eliminate the $m_c$-interpolation ambiguity mentioned above \cite{Boughezal:2007ny}.

The current theoretical error is dominated by the uncertainty associated to non-perturbative effects, estimated to be about $5 \%$ \cite{Misiak:2006zs}. The non-perturbative uncertainty primarily arises from corrections
of ${\mathcal O}(\alpha_s \Lambda_{\rm QCD}/m_b)$, which are very difficult to evaluate; they were analyzed by \textcite{Lee:2006wn} and \textcite{Benzke:2010js}.

New physics contributions to the partonic process can modify the matching conditions for the Wilson coefficients 
of the operators in the low-energy effective theory and can also induce new operators
besides those already present in the ST. Therefore, the good agreement between the ST prediction and the measured
value of the $\bar{B} \to X_s \gamma $ branching ratio sets strong constraints on the parameters of some new physics 
models. For example, an analysis of the decay within the type II Two-Higgs-Doublet-Model leads one to set a lower bound on the mass of the charged  Higgs boson: $M_{H^\pm} > 295$~GeV at $95 \%$ confidence level \cite{Misiak:2006zs}.

\boldmath
\subsection{Unstable Particles \label{sec:3.17}}
\unboldmath

In the early nineties, \textcite{Sirlin:1991fd} found that the conventional definitions of the mass and width of the 
$Z^0$ vector boson, namely
\begin{align}
M^2 &= M_0^2 + {\rm Re} \,A(M^2) \, , \label{eq:3_77} \\
M \Gamma &= - \frac{{\rm Im} \, A(M^2)}{(1- {\rm Re}\, A'(M^2))} \label{eq:3_78} \, ,
\end{align}
where $M_0$ is the bare mass, $M$ is the on-shell mass, and $A(s)$ the transverse self-energy, are gauge dependent in next-to-next-to-leading order (NNLO), i.~e. at the two and three-loop levels, respectively.
By extension, analogous conclusions hold true for other unstable particles. This led to a serious theoretical problem because, in the context of gauge theories, a fundamental requirement is that physical observables should be gauge independent.

The original argument was based on the observation that the complex-valued position $\bar{s}$ of the propagator's pole must be gauge independent, since it is a singularity of the analytically extended $S$  matrix. In the case of bosons, the inverse propagator is proportional to
\be
\Pi(s) = s - M_0^2 - A(s) \, , \label{eq:3_79}
\ee
where $s = q^2$ is the square of the four-momentum transfer. Thus, the pole position is
\be
\bar{s} = M_0^2 + A(\bar{s}) \, . \label{eq:3_80}
\ee 
 Writing $\bar{s} = m_2^2 - i m_2 \Gamma_2$, where $m_2$ and $\Gamma_2$ are real, gauge independent parameters, one has
\begin{align}
m_2^2 &= M_0^2 +{\rm Re}\,A(\bar{s}) \, , \label{eq:3_81} \\
m_2 \Gamma_2 &=  - {\rm Im}\, A(\bar{s})\, . \label{eq:3_82}
\end{align} 
If one expands $A(\bar{s})$ about $m_2^2$ and retains only leading terms in $\Gamma_2$,  Eqs.(\ref{eq:3_81},\ref{eq:3_82}) lead back to   Eqs.(\ref{eq:3_77},\ref{eq:3_78}). On the  other hand, if terms of higher order in $\Gamma_2$ are retained, the comparison of  Eqs.(\ref{eq:3_81},\ref{eq:3_82}) with  Eqs.(\ref{eq:3_77},\ref{eq:3_78})
show that indeed $M^2$ and $\Gamma$ are gauge dependent in higher orders. At the two-loop level, the analysis shows that the gauge dependence of $M^2$ occurs  only in a restricted range of the gauge parameter $\xi$ and, as a consequence, it is bounded. In fact, in a later letter, \textcite{Passera:1996nk} showed that the maximum shift in $M$
due to the gauge dependence at the two-loop level is about $2~$MeV. Although a small effect, it is of the same magnitude as the $2.1~$MeV experimental error. However, at the three-loop level and higher, the gauge dependence is unbounded, so that $M$  and $\Gamma$ (cf.  Eqs.(\ref{eq:3_77},\ref{eq:3_78})) are not only inconsistent with basic principles, but their numerical values depend in an arbitrary manner on the choice of $\xi$.

In fact, the comparison of the pole definitions of the mass and width $(m_2, \Gamma_2)$ with the conventional ones $(M, \Gamma)$ leads to the conclusion that the gauge dependencies of the latter are numerically very large, particularly in the case of a heavy Higgs boson \cite{Kniehl:1998fn, Kniehl:1998vy}. 

At this stage, it is instructive to point out the conceptual difference between the gauge independent parameter $m_2^2$  and the gauge dependent $M^2$. While $m_2^2$ is the real part of the zero of the inverse propagator, $M^2$  is the zero of the real part, an important difference.

In a second 1991  contribution, \textcite{Sirlin:1991rt} analyzed specific physical amplitudes and derived an independent proof of the need for additional higher order gauge dependent counterterms in  Eq.(\ref{eq:3_77}),
a result that gives additional support to the arguments and conclusions of the first paper.

It has also been emphasized that  Eq.(\ref{eq:3_78}) leads to serious unphysical singularities if $A(s)$ is not analytic in the neighborhood of $M^2$. This occurs when $M^2$ is very close to a physical threshold, as discussed by 
 \textcite{Fleischer:1980ub}; \textcite{Bardin:1991dp}; \textcite{Kniehl:1990xe, Kniehl:1991ze, Kniehl:1991hk, Kniehl:1993ay}: \textcite{Bhattacharya:1991gr}; \textcite{Kniehl:2000kk,
Kniehl:2002wn}, or, in the resonance region, when the unstable particle is coupled to massless quanta, as in the cases of the $W$ vector boson and unstable quarks. In particular, it has been pointed out that the on-shell mass of an unstable quark has an unbounded gauge dependence of ${\mathcal O} (\alpha_s (\xi_g-3) \Gamma)$, where $\xi_g$ is the gluon gauge parameter and $\Gamma$ the width 
\cite{Passera:1998uj, Passera:1998jv, Sirlin:1998dz}. 

In order to solve the serious problems raised by the gauge dependence of $M$ and $\Gamma$ (cf.  Eqs.(\ref{eq:3_77},\ref{eq:3_78})), \textcite{Sirlin:1991fd} proposed to define the mass and width of the $Z^0$ vector boson 
by means of the gauge independent parameters
\be
m_1 = \sqrt{m_2^2 + \Gamma_2^2} \, ; \qquad \Gamma_1 = \frac{m_1 \Gamma_2}{ m_2} \, .
\ee
As emphasized in the same work, the advantage of the $(m_1, \Gamma_1)$ definitions relative to the $(m_2, \Gamma_2)$ is that $m_1$ and $\Gamma_1$  can be identified with the $Z^0$ mass and width measured at LEP.

Formal proofs of the gauge independence of $\bar{s}$ and the gauge dependence of $M$ and $\Gamma$, based on the Nielsen identities that describe the gauge dependence of Green functions \cite{Nielsen:1975fs}, have been presented in the literature \cite{Gambino:1999ai, Grassi:2000dz, Grassi:2001bz}.

Applying the Nielsen identities to $\Pi(s, \xi_k)$ (cf.  Eq.(\ref{eq:3_79})), one finds 
\be
\frac{\partial}{\partial \xi_l} \Pi\left(s,\xi_k \right) = 2 \Lambda_l (s,\xi_k) \Pi(s,\xi_k)\, , \label{eq:3_84}
\ee
where we have indicated explicitly the dependence on the gauge parameters $\xi_k$ and $\Lambda_l(s, \xi_k)$ is a complex, amputated, 1PI, two point Green function of ${\mathcal O}(g^2)$ involving the gauge field, its BRST variation, and the gauge fermion.

As $\bar{s}$ is the zero of $\Pi(s,\xi_k)$, it follows that
\be
\Pi \left(\bar{s},\xi_k \right) = 0\, . \label{eq:3_85}
\ee
Differentiating  Eq.(\ref{eq:3_85}) with respect to $\xi_l$:
\be
\frac{\partial \bar{s}}{\partial \xi_l} \frac{\partial}{\partial \bar{s}} \Pi\left(\bar{s},\xi_k \right) +
\frac{\partial}{\partial \xi_l}  \Pi\left(\bar{s},\xi_k \right)  =0 \, .\label{eq:3_86}
\ee
 Eqs.(\ref{eq:3_84},\ref{eq:3_85}) imply that the second term on the l.h.s. of  Eq.(\ref{eq:3_86}) vanishes.
As $\left( \partial/ \partial \bar{s}\right) \Pi(\bar{s},\xi_k) = 1+ {\mathcal O}(g^2)$,  Eq.(\ref{eq:3_86}) leads to
\be
\frac{\partial \bar{s}}{\partial \xi_l} = 0 \, , \label{eq:3_87}
\ee
which expresses the gauge independence of $\bar{s}$. It is important to note that this conclusion is valid to all orders in perturbation theory.

Instead, taking the real part of  Eq.(\ref{eq:3_84}):
\be
\frac{\partial}{\partial \xi_l} {\rm Re}\, \Pi(s,\xi_k) = 2 \left[{\rm Re}\, \Lambda_l (s, \xi_k) {\rm Re} \,\Pi(s,\xi_k) - {\rm Im} \,\Lambda_l (s, \xi_k) {\rm Im}\, \Pi(s, \xi_k)\right] \, . \label{eq:3_88}
\ee
Recalling that the on-shell $M^2$ is the zero of the ${\rm Re}\, \Pi(s,\xi_k)$, it follows that
\be
{\rm Re}\, \Pi(M^2, \xi_k) = 0 \, . \label{eq:3_89}
\ee
Differentiating  Eq.(\ref{eq:3_89}) with respect to $\xi_l$ and using  Eqs.(\ref{eq:3_88},\ref{eq:3_89}), one obtains
\be
\frac{\partial M^2}{\partial \xi_l} {\rm Re}\, \Pi'(M^2, \xi_k) -  2 {\rm Im}\, \Lambda_l \left(M^2, \xi_k \right) {\rm Im}\,
\Pi(M^2, \xi_k) = 0 \, , \label{eq:3_90}
\ee
where the prime stands for derivative with respect to $M^2$. Noting that ${\rm Re}\, \Pi'(M^2,\xi_k) = {\mathcal O}(1)$ and that both ${\rm Im} \Lambda_l(M^2,\xi_k)$ and ${\rm Im}\,(M^2,\xi_k)$ are ${\mathcal O}(g^2)$,  Eq.(\ref{eq:3_90}) implies that $\partial M^2/\partial \xi_l  = {\mathcal O}(g^4)$. Thus, $M^2$ is gauge dependent at the two-loop 
level, i.~e. in NNLO, the same conclusion reached by \textcite{Sirlin:1991fd, Sirlin:1991rt}.

Similarly, for the conventional expression of the width (cf.  Eq.(\ref{eq:3_78})), a somewhat lengthier derivation leads in leading order to
\be
\frac{d}{d \xi_l} \frac{{\rm Im}\, \Pi(M^2, \xi_k)}{{\rm Re}\, \Pi'(M^2, \xi_k)} = 2 \left\{{\rm Im}\, \Lambda_l (M^2,\xi_k) 
\left[ {\rm Im}\, \Pi(M^2, \xi_k) \right]^2 \right\}' +{\mathcal O}(g^8) \, , \label{eq:3_91}
\ee
where $d/ d\xi_l$ stands for the total derivative with respect to $\xi_l$. Since ${\rm Im} \Lambda_l$ and ${\rm Im}\, \Pi$
are both of ${\mathcal O}(g^2)$,  Eq.(\ref{eq:3_91}) implies that  Eq.(\ref{eq:3_78}) is gauge dependent in ${\mathcal O}(g^6)$, i.~e. in NNLO, in agreement with the earlier conclusions \cite{Sirlin:1991fd, Sirlin:1991rt}.

\subsection{Renormalization of the Cabibbo-Kobayashi-Maskawa Matrix}
\label{sec:3.18}

An important problem associated with the Cabibbo-Kobayashi-Maskawa (CKM) matrix \cite{Cabibbo:1963yz,Kobayashi:1973vf} is its renormalization. An early analysis \cite{Marciano:1975cn} focused on the renormalization of ultraviolet (UV) divergences in the two-generation case. Since the CKM matrix  is one of the fundamental cornerstones of the weak interactions and, by extension, of the ST,  it is important to develop renormalization schemes that treat both the finite and divergent contributions with well-defined renormalization conditions. Over the last two decades several papers have addressed
 this basic problem at various levels of generality and complexity: \textcite{Denner:1990yz}; \textcite{Kniehl:1996bd};  \textcite{Gambino:1998ec};  \textcite{Kniehl:2000rb};  \textcite{Barroso:2000is};  \textcite{Yamada:2001px}; \textcite{Diener:2001qt}; \textcite{Pilaftsis:2002nc};  \textcite{Espriu:2002xv};  \textcite{Zhou:2003gb, Zhou:2003te};  \textcite{Liao:2003jy}; \textcite{Denner:2004bm}; \textcite{Kniehl:2006bs, Kniehl:2006rc, Kniehl:2009kk};  \textcite{Almasy:2008ep}.

The main difficulties in the CKM renormalization arise from external-leg mixing self-energy corrections. For instance, for an outgoing quark, these EWC are of the form
\be
\Delta {\mathcal M}_{i i'}^{{\rm leg}} = \bar{u}_i (p) \Sigma_{i i'} \left(p\!\!\!\slash \right)\frac{1}{p\!\!\!\slash -m_{i'}} \, ,
\label{eq:3_92}
\ee
where $i$ denotes the outgoing quark of momentum  $p$ and mass $m_i$, $i'$ the virtual quark of mass $m_{i'}$,
and $\Sigma_{i i'} \left(p\!\!\!\slash \right)$ the self-energy.

In the following, we outline the strategies followed in two of the most recently proposed on-shell schemes to renormalize the CKM matrix at the one-loop level.

\begin{itemize}

\item[A)] Using a simple procedure based on Dirac Algebra, \textcite{Kniehl:2006bs, Kniehl:2006rc} separated the contributions to $\Sigma_{i i'} \left(p\!\!\!\slash \right)/(p\!\!\!\slash -m_{i'})$ into two classes: 1) gauge independent self-mass (sm) contributions proportional to $(p\!\!\!\slash -m_{i'})^{-1}$ with a cofactor that involves the chiral projectors $a_\pm = (1 \pm \gamma_5)/2$, but not $p\!\!\!\slash$\,    2) gauge dependent wave-function renormalization (wfr) contributions in which the virtual quark propagator $(p\!\!\!\slash -m_{i'})^{-1}$ has been canceled. Furthermore, using the unitarity relation $V_{lm} V^\dagger_{mn} = \delta_{ln}$ satisfied by the CKM matrix elements $V_{lm}$, one finds that the wfr have an important property: all the gauge dependent and all the UV-divergent wfr contributions to the physical amplitude $W \to q_i \bar{q}_j$ depend only on an overall factor $V_{ij}$ and the external quark masses $m_i$ and $m_j$, a property shared by the one-loop proper vertex contributions. This leads to the cancellation of the gauge dependence and UV divergence of the wfr contributions to $W \to q_i + \bar{q}_j$ with those arising from the one-loop vertex corrections, exactly as in the unmixed, single generation case!

The renormalization of the sm contributions is implemented using the mass counterterms
\be
\overline{\psi}^Q \left( \delta m^{Q(+)} a_+ + \delta m^{Q (-)} a_-\right) \psi^Q \qquad (Q =U,D)\, ,
\label{eq:3_93}
\ee
where $U (D)$ stands for the up (down) quarks, and $\delta m^{Q(\pm)}$ are non-diagonal matrices subject  to the hermiticity condition $\delta m^{Q(+)} =  \delta m^{Q (-) \dagger}$.

The UV-divergent sm contributions obey the hermiticity condition, so they can be canceled by the $\delta m^{Q(\pm)}$
in all $i i'$ channels. However, this is not the case for some of the finite parts. For this reason, the authors used a specific renormalization prescription: the $\delta m^{Q(\pm)}$ were adjusted to cancel the full sm contributions in all diagonal $(i=i')$ channels, as well as the $u c$, $u t$, $c t$ channels for the $U$ quarks and the $s d$, $b d$,  $b s$
channels for the $D$ quarks. This implies that there are residual sm contributions in the reverse $c u$, $t u$, $t c$, $d s$,
$d b$, $s b$ channels, but they are finite, gauge independent and very small.
In fact, since these residual sm contributions converge in the limit $m_{i'} \to m_i$, they may be regarded as additional finite and gauge independent contributions to wfr that happen to be small.
An attractive feature of this renormalization prescription is that the external-leg sm contributions are fully canceled when the external particle is $u$, $d$, $\bar{u}$ or $\bar{d}$ quark, a very useful property since $V_{ud}$ is by far the most precisely determined CKM matrix element (cf.  Eq.(\ref{eq:3_34})).

The renormalization procedure outlined above presents interesting similarities with the approach followed by 
\textcite{Feynman:1949zx, FeynmanQED}  in Quantum Electrodynamics (QED). Thus, it may be regarded as a generalization of Feynman's approach to the case in which the self-energy $\Sigma_{i i'} \left(p\!\!\!\slash \right)$ contains non-diagonal, as well as diagonal components.

In the same work, \textcite{Kniehl:2006bs, Kniehl:2006rc} showed that an equivalent and interesting formulation of the same renormalization scheme is obtained by diagonalizing the complete mass matrix $m -\delta m^{Q(+)} a_+ - \delta m^{Q (-)} a_-$ ($m$ is the diagonal, renormalized mass matrix) by biunitarity transformations acting on the up and down quark spaces. This procedure generates an explicit CKM counterterm matrix $\delta V$, which automatically satisfies the following important properties: it is gauge independent, preserves unitarity in the sense that both the renormalized and bare CKM matrices $V$ and $V_0 = V -\delta V$ are unitary at the one-loop level, and leads to renormalized amplitudes that are non-singular in the limit in which any two quarks become mass degenerate. In this alternative formulation, the off-diagonal UV-divergent sm contributions are canceled by $\delta V$ while, as usual, the diagonal sm contributions are canceled by the mass counterterms that are also diagonal.

The renormalization scheme outlined above has been generalized to the case of an extended lepton sector  that includes Dirac and Majorana neutrinos in the framework of the seesaw mechanism \cite{Almasy:2008ep}.

\item[B)] A second on-shell renormalization scheme \cite{Kniehl:2009kk} is based on explicit mass counterterm matrices
\be
\delta m^Q_{i i'} = \delta m_{i i'}^{Q(+)} a_+ + \delta m_{i i'}^{Q(-)} a_- \qquad (Q=U,D) \, ,
\label{eq:3_94}
\ee
where $\delta m_{i i'}^{Q(+)}$ and $m_{i i'}^{Q(-)}$ are defined in terms of the Lorentz-invariant self-energy functions and obey two important properties: {\em (i)} they are gauge independent and {\em (ii)} they automatically satisfy the hermiticity condition $\delta m_{i i'}^{Q(+)} = m_{i i'}^{Q(-) \dagger}$ of the mass matrix. The second property implies that they can be applied directly to all diagonal and off-diagonal amplitudes and, in this sense, they are ``flavor-democratic'' since they do not single out particular flavor channels. As in the case of scheme A), diagonalization of the complete mass matrices leads to a gauge independent CKM counterterm   matrix $\delta V$ that preserves unitarity and now satisfies another highly desirable theoretical property, namely ``flavor-democracy''.

\end{itemize}

\boldmath
\subsection{$S$, $T$, and $U$ Parameters \label{sec:3.16}}
\unboldmath

``New physics'', i.~e., physics beyond the ST, may contribute to EWC. If the new physics is associated with a high-mass scale and contributes mainly to the self energies, the idea has been proposed to parametrize its contributions in terms of three amplitudes, $S$, $T$, and $U$, introduced by \textcite{Peskin:1990zt}. See also \textcite{Lynn:1985fg};  \textcite{Holdom:1990tc}; \textcite{Kennedy:1990ib, Kennedy:1991sn}
\textcite{ Altarelli:1990zd}; \textcite{Golden:1990ig}; \textcite{PT92}.

In the $\ms$ scheme we have \cite{Marciano:1990dp, Marciano:1993ep, Marciano:1991ix, Sirlin:1993cm, Sirlin:1993bu}:
\begin{align}
\Delta \hat{r} &= \left(\Delta \hat{r}  \right)_{{\rm ST}} +\frac{\hat{\alpha}}{4 \hat{s}^2 \hat{c}^2} S_Z -
\hat{\alpha} T \, , \label{eq:3_66} \\
\Delta \hat{r}_W &= \left(\Delta \hat{r}_W  \right)_{{\rm ST}}  + \frac{\hat{\alpha}}{4 \hat{s}^2} S_W \, , 
\label{eq:3_67} \\ 
\frac{\hat{\alpha}}{4 \hat{s}^2 \hat{c}^2} S_Z &= \left[ \frac{A_{ZZ}\left(M_Z^2 \right) - A_{ZZ}\left(0 \right)}{M_Z^2} \right]_{\ms}^{{\rm new}}
\, , \label{eq:3_68} \\
\frac{\hat{\alpha}}{4 \hat{s}^2 } S_W &= \left[ \frac{A_{WW}\left(M_W^2 \right) - A_{WW}\left(0 \right)}{M_W^2} \right]_{\ms}^{{\rm new}}
\, , \label{eq:3_69} \\
\hat{\alpha} T &=  \left[ \frac{A_{WW}(0)}{M_W^2} - \frac{A_{ZZ}(0)}{M_Z^2} \right]_{\ms}^{{\rm new}}
\, . \label{eq:3_70}
\end{align}
In  Eqs.(\ref{eq:3_68}-\ref{eq:3_70}), the $A$-functions are the unrenormalized self energies defined according to the conventions of \textcite{Marciano:1980pb}, $\ms$ means that the $\ms$ renormalization has been implemented and $\mu = M_Z$ chosen,
and ``new'' denotes new physics contributions.  In  Eqs.(\ref{eq:3_68},\ref{eq:3_69}), we have applied the $\ms$ renormalization prescription for $\hat{\alpha}(M_Z)$ and $\sin^2 \hat{\theta}_W (M_Z)$ proposed by 
\textcite{Marciano:1990dp}, which excludes  new-heavy-physics contributions in $A_{\gamma \gamma}^{{\rm new}}(q^2)$
 and $A_{\gamma Z}^{{\rm new}}(q^2)$. Consequently, these two self-energies are not included in the definitions
 of $S_Z$ and $S_W$. We recall that $\hat{s}^2 = 1-\hat{c}^2 = \sin^2 \hat{\theta}_W(M_Z)$ is the $\ms$ electroweak mixing parameter evaluated at the scale $\mu= M_Z$, $\Delta \hat{r}$ and 
$\Delta \hat{r}_W$ are the EWC in  Eq.(\ref{eq:3_4}) and  Eq.(\ref{eq:3_5}), respectively (cf. Section~\ref{sec:3.5}),
while $(\Delta \hat{r})_{{\rm ST}}$ and $(\Delta \hat{r}_W)_{{\rm ST}}$ are their values in the ST.
$\hat{\alpha}$ is the $\ms$ fine structure constant at $\mu = M_Z$ (cf. Section~\ref{sec:3.6}).

Alternatively, one defines $S \equiv S_Z$, $U \equiv S_W -S_Z$. $T$ and $U$ are primarily sensitive to
isodoublet mass splittings (generally, $U \ll T$), while $S$ probes contributions from mass-degenerate fermion doublets.

In conjunction with  Eqs.(\ref{eq:3_4},\ref{eq:3_5}), the modifications of $\Delta \hat{r}$ and $\Delta \hat{r}_W$
displayed in  Eqs.(\ref{eq:3_66},\ref{eq:3_67}), induce the linear shifts
\begin{align}
\hat{s}^2 &= \left(\hat{s}^2\right)_{{\rm ST}} + \frac{\hat{\alpha}}{4 (\hat{c}^2 -\hat{s}^2)} \left[ S - 4 \hat{s}^2 \hat{c}^2 T \right] \, , \label{eq:3_71} \\
M_W &= \left(M_W\right)_{{\rm ST}} \left[ 1 + \frac{\hat{\alpha} \hat{c}^2}{2 (\hat{c}^2 - \hat{s}^2)} T
+ \frac{\hat{\alpha}}{8 \hat{s}^2} U - \frac{\hat{\alpha}}{4  (\hat{c}^2 - \hat{s}^2)} S
\right] \, , \label{eq:3_72}
\end{align}
where EWC of ${\mathcal O}(\alpha^2)$ have been neglected.

Combining  Eq.(\ref{eq:3_71}) and  Eq.(\ref{eq:3_72}), one can solve for $S_W$:
\begin{align}
\frac{\hat{\alpha}}{4 \hat{s}^2} S_W = 2 B +C \, , \label{eq:3_73}
\end{align} 
where
\begin{align}
B  = \frac{M_W}{\left(M_W\right)_{{\rm ST}}} -1 \, ; \qquad
C  = \frac{\hat{s}^2 }{\left(\hat{s}^2\right)_{{\rm ST}}} -1 \, .
\label{eq:3_74}
\end{align}
In the case $U=0$, i.~e. $S_W = S_Z \equiv S$, we also have
\begin{align}
\hat{\alpha} \hat{c}^2 T = 2 \left[ B + \left(\hat{s}^2\right)_{{\rm ST}} C \right] \, ,
\label{eq:3_75}
\end{align}
where we have neglected a second-order term $C^2 \left(\left(\hat{s}^2\right)_{{\rm ST}}\!/\hat{s}^2\right)$
in the r.h.s. of the equation. In  Eqs.(\ref{eq:3_71}-\ref{eq:3_75}), $\left(M_W\right)_{{\rm ST}}$
and $\left(\hat{s}^2\right)_{{\rm ST}}$ are calculated using the EWC of the ST (cf. Section~\ref{sec:3.15}) and 
a chosen reference value for $M_H$, while $M_W$ is identified with the measured mass of the $W$-boson.
In turn, $\hat{s}^2$ is evaluated using the experimental value of $\sefl$, obtained from the $Z$-pole  asymmetries,
and applying  Eq.(\ref{eq:3_11p}).

In order to obtain the dependence of the neutral current observables on $S$ and $T$, one expresses the corresponding amplitudes in terms of $G_F$ and the ST EWC evaluated at the chosen reference value for $M_H$,
multiplies them by $\rho(0)^{{\rm new}} = 1 +\hat{\alpha} T$, and inserts the expression for $\hat{s}^2$ given in 
 Eq.(\ref{eq:3_71}). In particular, the weak charge $Q_W({\rm Cs})$, measured in atomic parity violation, is very insensitive to $T$ and thus provides a direct probe of $S$ \cite{Marciano:1990dp, Marciano:1991ix, Marciano:1993ep}.

A recent global analysis \cite{Baak:2011ze}  employs the reference values $M_{H, {\rm ref}} = 120$~GeV and 
$M_{t, {\rm ref}} = 173$~GeV and obtains
\be
S = 0.04 \pm 0.10  \, ; \quad
T = 0.05 \pm 0.11  \, ; \quad
U = 0.08 \pm 0.11  \, , \label{eq:3_76}
\ee
%
%The central values assume $M_H = 117$~GeV and the numbers in parentheses show the shifts in those values
%for $M_H = 300$~GeV. 
while, assuming $U=0$, the results are
\be
\left. S \right|_{U=0} = 0.07 \pm 0.09 \, ; \qquad \left. T \right|_{U=0} = 0.10 \pm 0.08 \, .
\ee
We see that the results in  Eq.(\ref{eq:3_76}) are in good agreement with the ST predictions
$S=T=U=0$. By comparison, a fourth generation of mass-degenerate fermions leads to 
$S = 4/ 6 \pi \approx 0.21$ (cf.~\textcite{Bertolini:1984dv}), while technicolor models roughly contribute $S \approx (0.05 {\rm \, to \, }  0.10) N_T N_D + 0.12$, where $N_T$ and $N_D$ are the number of technicolors and isodoublets,
respectively \cite{Marciano:1993ep}. Therefore, for one generation with $N_T = N_D =4$ one expects $S \approx 0.9 \mbox{ to } 1.7$, values
significantly larger than the $S$ result shown in  Eq.(\ref{eq:3_76}).

An alternative formulation of the $S$, $T$, $U$ analysis is based on the $\epsilon_i$ ($i=1,2,3,b$)
parameters, defined in terms of the physical quantities $M_W$, $\Gamma_l$,  $A_{{\rm FB}}^{(l)}$, and $\Gamma_{b \bar{b}}$ \cite{Altarelli:1993xx, Altarelli:1990zd, Altarelli:1991fk, Altarelli:1993sz, Altarelli:1993bh}.

The applications of the $S$, $T$, and $U$ formalism have focused mainly on ``new physics'' fermionic contributions to the self-energies. On the other hand, ``new physics'' bosonic contributions are also of general interest. However,  this poses a theoretical problem: as pointed out by \textcite{Degrassi:1993kn}, in contrast with the fermionic case, the bosonic contributions to the $S$, $T$, and $U$ parameters, defined in terms of the conventional self-energies (cf.  Eqs.(\ref{eq:3_68}-\ref{eq:3_70})), are gauge-dependent in the ST; furthermore, $T$ and $U$ are divergent unless a constraint is imposed among the gauge parameters. It is natural to expect that the same theoretical problems arise in the bosonic ``new physics'' contributions.  In order to circumvent this problem, \textcite{Degrassi:1993kn} proposed to replace the conventional self-energies in  Eqs.(\ref{eq:3_68}-\ref{eq:3_70}) by their pinch-technique counterparts
\cite{Cornwall:1981xx, Cornwall:1981zr, Cornwall:1989fg, Papavassiliou:1990xx, Degrassi:1992ue}, which are gauge-independent. Thus, this modification leads to a gauge independent  formulation of $S$, $T$, and $U$ in the bosonic sector.

%It is also important to note that $S_W$, $S_Z$, and $T$ (and, correspondingly, $S$, $T$, and $U$) are actually 
%parts of $\Delta \hat{r}$ and $\Delta \hat{r}_W$ (cf. Eqs.(\ref{eq:3_66},\ref{eq:3_67})), which also include vertex and box diagrams in gauge-independent %combinations. Thus, an alternative gauge-independent formalism, is to parametrize the ``new physics'' contributions by means of $\Delta \hat{r} = (\Delta %\hat{r})_{\rm ST} + (\Delta \hat{r})^{\rm new}$,
%$\Delta \hat{r}_W = (\Delta \hat{r}_W)_{\rm ST} + (\Delta \hat{r}_W)^{\rm new}$, together with an effective parameter $1 + \hat{\alpha} T_{\rm eff}$ that %multiplies the neutral current amplitudes.

\subsection{Supersymmetry}

In Section~{\ref{sec:3.16}} we pointed out that a recent global fit leads to values of the $S$, $T$, and $U$ parameters that are in good agreement with the ST expectations $S=T=U=0$. Thus, at present, the analysis of the precision electroweak data does not lead to clear signals of new physics beyond the ST.

However, there are powerful theoretical arguments that strongly suggest the presence of new physics. The most obvious one is that the ST does not incorporate gravity, one of the fundamental forces of nature. In fact, the unification of gravity with the ST in particular, and Quantum Mechanics in general, is one of the most important unsolved problems in theoretical particle physics. At present, there is a widespread belief among theorists that String Theory provides the most hopeful framework to achieve this major goal. On the other hand, String Theory leads to a landscape with an enormous number of possible vacua \cite{Bousso:2000zz, Susskind:2003zz}, without clear selection criteria, except perhaps for anthropic arguments.

Another powerful argument, involving radiative corrections (RC), is the Higgs boson mass hierarchy problem. This involves the important fact that the RC to $M_H^2$ are quadratically divergent. Thus, the relation of the physical, renormalized mass $M_H$, and the bare mass $M_H^0$, is of the form
\be
M_H^2 = \left(M_H^0\right)^2 +  O\left(\lambda, g^2, h_F^2 \right) \Lambda^2 + \cdots \, ,
\label{eq:3_n1}
\ee
where $g$ is the ${\rm SU}(2)_L$ gauge couplings, $\lambda$ the quartic Higgs self-coupling, $h_F = m_f/v$, $m_f$ the mass of fermion $f$, 
$v = \left(1/\sqrt{2} G_F \right)^{1/2} = 246$~GeV the vacuum expectation value of the Higgs field, and $\Lambda$ the cutoff introduced to regularize the ultraviolet (UV) divergence. The second term in  Eq.(\ref{eq:3_n1}) is the quadratically divergent part of the self-mass RC and the ellipses represent $\ln \Lambda$ contributions as well as finite terms\footnote{Rules of correspondence between the poles' positions in dimensional regularization and UV cutoffs in four dimensional calculations with $L$ loops were stated by \textcite{Veltman:1981zz} for quadratic divergences, and derived by \textcite{Ossola:2003ku}, on the basis of a heuristic argument, for quadratic and higher divergences.}.

Within the ST, the presence  of the ${\mathcal O}(\Lambda^2, \ln \Lambda) + \cdots$ terms in the r.h.s. of  Eq.(\ref{eq:3_n1}) does not cause difficulties: as in all renormalizable theories, they are canceled by the divergent part of the mass counterterm $-\delta M^2 = (M_H^0)^2 - M_H^2$.  In such an approach, $(M_H^0)^2$ and the RC are regarded as unobservable quantities and only $M_H^2$  has a physical meaning. However, if we assume that the ST is embedded in a larger theory that cuts off the momentum integral  in the RC at its own finite scale, $\Lambda$ acquires a physical meaning. Specifically, $\Lambda$ in  Eq.(\ref{eq:3_n1}) is then identified with the scale of the new physics. For example, if the new physics beyond the ST is gravity, $\Lambda$ is identified with the Planck mass: $\Lambda = M_P = G_N^{-1/2} = 1.2221 \times 10^{19}$~GeV, where $G_N$ is  Newton's gravitational constant.

To illustrate the effect of these considerations on  Eq.(\ref{eq:3_n1}), we consider a leading quadratically divergent contribution to the RC arising from the 
diagram $H \to \mbox{top loop} \to H$.  Employing  Eq.(8.6) in 
\textcite{Langacker:2010zza} 
with $m_t = 173.2$~GeV and $\nu=v=246$~GeV, we find that this 
diagram contributes  $\approx - 3.8 \times 10^{-2} \Lambda^2$. Using the gravity scale, $\Lambda = 1.221 \times 10^{19}$~GeV, one obtains a RC $\approx -5.6 \times 
10^{36}~{\rm GeV}^2$. Since, in absolute value, this is enormously larger than the expected value of $M_H^2$, there must be an extraordinarily fine-tunded 
cancellation between $(M_H^0)^2$ and the RC. As an illustration, if we assume $M_H = 125$~GeV, the level of the required fine-tuning is 
\begin{displaymath}
\frac{(M_H^0)^2 -5.6 \times 10^{36} {\rm GeV}^2} 
{ 5.6 \times 10^{36}~{\rm GeV}^2}  = \frac{(125)^2}{5.6\times 10^{36}} = 2.8 \times 10^{-33} \, , 
\end{displaymath}
namely $3$ parts in $10^{33}$! Such fine-tuning is generally regarded as unnatural. On the other hand, if we demand a relatively small level of fine-tuning, the same RC employed before leads to a value of $M_H$ rather close to $\Lambda$. For example, if we assume that the level of fine-tuning is $10 \%$, we have $M_H = 0.75 \times 10^{18}~{\rm GeV} \approx M_P/16$. This is usually referred to as the hierarchy problem. Namely, assuming a relatively small level of fine tuning, the quadratically divergent RC push the value of $M_H$ from the electroweak scale to a value within an order of magnitude of the gravitational scale.

The same problems occur when one considers the RC to the vacuum expectation value of the Higgs field $\langle 0|H|0\rangle$:
\be
v^2 = v^2_0 + O\left(\lambda, g^2, h_F^2 \right) \Lambda^2 + \cdots \, ,
\label{eq:3_n2}
\ee
where $v = 246~{\rm GeV}$ defines the electroweak scale. Again, if $\Lambda = M_P$, very large RC emerge, so that an unnaturally fine-tuned cancellation between $v_0$ and the RC must take place. On the other hand, if one demands a relatively small level of fine-tuning, the value of the weak scale $v$ moves close to $M_P$. 

One frequently invoked solution of the Higgs boson mass hierarchy problem is TeV scale supersymmetry. As is well-known, this theory, based on elegant symmetry principles, postulates that every fermion (boson) particle has boson (fermion) supersymmetric partners with the same quantum numbers and masses. Since mass-degenerate partners of known particles have not been found, it is clear that in nature supersymmetry is broken. 

A very important property of supersymmetry is that the quadratically divergent RC to $M_H^2$ arising  from the fermion and boson loops cancel each other, leaving only much smaller supersymmetry-breaking contributions. Thus, if  TeV scale supersymmetry is an approximate symmetry of nature, such cancellation would provide an elegant solution to the Higgs boson mass hierarchy problem, based on fundamental symmetry principles.

In fact, over the last several years, supersymmetric scenarios such as the Minimal-Supersymmetric Standard Model (MSSM) have emerged as leading candidates for theoretical frameworks beyond the ST. It involves five Higgs bosons: two neutral CP-even scalars $h$  and $H$ ($M_h < M_H$), one neutral CP-odd pseudoscalar $A$, and one charged pair $H^{\pm}$.

An interesting property is that, at the tree-level, $M_h \le M_Z$, which is ruled out by direct searches at $95 \%$ CL. This is also in contrast with the ST, where there is no tree-level upper limit on $M_H$, except for perturbativity and unitarity bounds. On the other hand, for large stop masses, there are sizable RC, dominated by top and stop loops, that significantly increase the upper bound for $M_h$. At present, the analysis yields $M_h \lesssim 135$~GeV \cite{Haber}. Thus, we see that RC indeed play a crucial role in ensuring the phenomenological consistency of the MSSM.

In the MSSM, supersymmetric contributions decouple if the superpartners' masses are much larger than $M_Z$. In that regime, the fits are of the same general quality as in the case of the ST. If some of them are of ${\mathcal O}(M_Z)$, the fits are worse, leading to constraints in the MSSM parameter space.

Another important result of supersymmetry is that the unification of gauge couplings is much more successful when they are extrapolated using the MSSM $\beta$ functions, with the couplings intersecting at $M_{{\rm GUT}}  \sim  3 \times 10^{16}$~GeV, than when employing the ST $\beta$ functions.
On the other hand, at present the agreement is not perfect: using $\hat{\alpha}(M_Z)$ and $\sin^2 \hat{\theta}_W (M_Z)$ as inputs, one finds the prediction $\alpha_s(M_Z) \approx 0.13$, which is slightly larger than the observed value $\approx 0.12$.

As discussed in Section~\ref{sec:3.20}, the possible contribution of supersymmetric partners of low mass may provide a natural explanation for the $\sim 3.5~\sigma$ discrepancy between the experimental and ST values of $a_\mu = (g_\mu -2)/2$.

Notwithstanding the impressive successes of supersymmetry, it is important to remember that the existence of supersymmetric partners, its most direct and compelling prediction, has not been established so far.

It is also important to note that a much more egregious hierarchy problem emerges in the analysis of the cosmological constant $\Lambda_c = 8 \pi G_N \rho$, where $\rho$ is the vacuum energy density of the universe. Assuming that the observed acceleration of the universe is due to $\Lambda_c$, the observed vacuum energy density is $\rho = {\mathcal O}(10^{-47}~{\rm GeV}^{4})$,  while estimates of the contribution to $\rho$ of elementary particles range roughly from ${\mathcal O}({\rm TeV}^4)$ in TeV supersymmetry to ${\mathcal O}((10^{19}~{\rm GeV})^4) = {\mathcal O}(10^{76}~{\rm GeV}^4)$ if the UV cutoff in the quartically divergent integrals is identified with $M_P$. Thus, there is mismatch of roughly $59$ to $123$ orders of magnitude between the estimates of $\Lambda_c$  from particle physics and the observed value! This implies that a cancellation between the bare cosmological constant $\Lambda_c^0$ and the contributions from elementary particles would require an extremely large and unnatural level of fine tuning. 
At the moment, it seems that there are no compelling explanations for the observed value of $\Lambda_c$, based on fundamental principles. In their absence, anthropic arguments are sometimes invoked: namely, the value of $\Lambda_c$ should be in the relatively small range that allows the formation of galaxies, a crucial requirement for the existence of life itself \cite{Weinberg:1989zz}.
Such anthropic arguments may serve, for example, as a selection criterion to choose among the multitude of vacua in the string landscape.  Nonetheless,
if a more fundamental explanation of the observed value of $\Lambda_c$ is not found, it seems clear that the requirement of natural fine-tuning faces a great challenge in the $\Lambda_c$ hierarchy problem.

\section*{Acknowledgments}

The authors are indebted to W.~J.~Marciano and M.~Passera for very interesting observations and valuable information. They also like to thank Fern Simes
for her help in the preparation of the manuscript.
The work of A. Sirlin was supported in part by the National Science Foundation Grant No. PHY-0758032.
The work of A.~Ferroglia was supported in part by the PSC-CUNY Award N. 64133-00 42 and by the National Science Foundation Grant No. PHY-1068317.

\bibliographystyle{apsrmp}
%\bibliography{nc,cds,sw,connes}

%\bibliography{rmp-sample}

\end{document}